\documentclass[11pt]{article}
\usepackage{cite}
\usepackage{amsmath,amsfonts,amssymb}
\pdfoutput=1
\usepackage[small,bf,hang]{caption}
\usepackage{slashed}
\input epsf.sty
\usepackage{epsfig}
\usepackage[titletoc,toc]{appendix}
\usepackage{xcolor}
\usepackage{caption}
\usepackage{subcaption}
\usepackage{tikz}

\graphicspath{ {./images/} }

\def\hybrid{
        \topmargin -20pt
        \oddsidemargin 0pt
        \headheight 0pt \headsep 0pt
        \textwidth 6.55in 
        \textheight 9.5in 
        \marginparwidth .875in
        \parskip 5pt plus 1pt \jot = 1.5ex}

\hybrid

 \csname
@addtoreset\endcsname{equation}{section}

\newcommand{\p}{\partial}

\def\moth{\mathsurround=0pt}
\newdimen\zo \zo=0pt

\def\tick{\leaders\hrule height 0.5ex depth 0pt \hskip 0.5pt}
\def\upboxfill{$\moth \setbox\zo\hbox{\tick}%
  \hskip 3pt\hbox to 0pt{$\tick$\hss}\hrulefill \hbox to 7.5pt{$\tick$\hss}$}

\def\dtick{\leaders\hrule height .34pt depth 0.5ex \hskip 0.5pt}
\def\downboxfill{$\moth \setbox\zo\hbox{\dtick}%
  \hskip 2pt\hbox to 0pt{$\dtick$\hss}\hrulefill \hbox to 2pt{$\dtick$\hss}$}

\def\bec{\begin{center}}
\def\ec{\end{center}}

\def\c{\gamma}

\def\vf{\varphi}

\def\s{\sigma}

\def\x{\xi}

\def\be{\begin{equation}}
\def\ee{\end{equation}}
\def\bea{\begin{eqnarray}}
\def\eea{\end{eqnarray}}
\def\ba{\begin{array}}
\def\ea{\end{array}}

\usepackage{hyperref}
\usepackage{comment}

\begin{document}

\begin{titlepage}

\rightline{\tt MIT-CTP-5530}
\hfill \today
\begin{center}
\vskip 0.5cm

{\Large \bf {Bootstrapping closed string field theory}
}

\vskip 0.5cm

\vskip 1.0cm
{\large {Atakan Hilmi F{\i}rat}}

{\em  \hskip -.1truecm
Center for Theoretical Physics \\
Massachusetts Institute of Technology\\
Cambridge, MA 02139, USA\\
\tt \href{firat@mit.edu}{firat@mit.edu} \vskip 5pt }

\vskip 1.5cm
{\bf Abstract}

\end{center}

\vskip 0.5cm

\noindent
\begin{narrower}
\noindent
The determination of the string vertices of closed string field theory is shown to be a conformal field theory problem solvable by combining insights from Liouville theory, hyperbolic geometry, and conformal bootstrap. We first demonstrate how Strebel differentials arise from hyperbolic string vertices by performing a WKB approximation to the associated Fuchsian equation, which we subsequently use it to derive a Polyakov-like conjecture for Strebel differentials. This result implies that the string vertices are generated by the interactions of $n$ zero momentum tachyons, or equivalently, a certain limit of suitably regularized on-shell Liouville action. We argue that the latter can be related to the interaction of three zero momentum tachyons on a generalized cubic vertex through classical conformal blocks. We test this claim for the quartic vertex and discuss its generalization to higher-string interactions.

\end{narrower}

\end{titlepage}

\baselineskip11pt

\tableofcontents

\baselineskip15pt

\section{Introduction}

Closed string field theory (CSFT) is a second-quantized formulation of closed string theory. It is supposed to provide a framework to explore solutions of string theory as well as to perform conventional field theoretical computations, such as mass renormalizations and vacuum shifts, see~\cite{Zwiebach:1992ie, deLacroix:2017lif, Erler:2019loq, Erbin:2021smf}. Despite the enormous progress for the latter in past few years, especially for superstrings~\cite{Pius:2014iaa, Pius:2014gza, Sen:2015uoa, Sen:2016gqt, Erler:2017pgf, DeLacroix:2018arq}, making advances for the former stays challenging. This is rather unsurprising: mass renormalizations and vacuum shifts don't really require a consistent choice of \textit{all} string vertices as one can work with a simple and convenient choice, such as $SL(2,\mathbb{C})$ vertices, order-by-order and argue that the physical results shouldn't depend on this arbitrary choice. 

On the other hand, this can't be done when seeking CSFT solutions: a consistent choice for string vertices is essential because they implement the gauge symmetry of CSFT through the geometric Batalin-Vilkovisky (BV) equation~\cite{Zwiebach:1992ie}. It is therefore imperative to choose a set of consistent string vertices and obtain the relevant geometric data on Riemann surfaces and their moduli spaces. This involves solving for the local coordinates around each puncture, and later, restricting to a sub-region $\mathcal{V}_{g,n}$ in the moduli space $\mathcal{M}_{g,n}$ where the moduli integration has to be performed. In CSFT, this has to be done for each genus $g$ Riemann surface with $n$ punctures.

Currently there are primarily two sets of string vertices that are shown to satisfy the geometric BV equation. These are minimal-area vertices (assuming their existence for higher genus)~\cite{Zwiebach:1990ni, Zwiebach:1990nh, ranganathan1992criterion,Wolf:1992bk,Headrick:2018dlw, Headrick:2018ncs, Naseer:2019zau} and hyperbolic vertices~\cite{Moosavian:2017qsp, Moosavian:2017sev, Costello:2019fuh, Cho:2019anu, Firat:2021ukc, Wang:2021aog, Ishibashi:2022qcz}. The former asks for minimal-area metrics of systole\footnote{Systole on a Riemann surface is defined as the shortest length of a non-contractible geodesic that is non-homotopic to a boundary component.} greater than $2 \pi$ (which is a convention) and uses the extremal properties of such metrics to show that the vertices satisfy the geometric BV equation. The local coordinates are defined by the semi-infinite flat cylinders appearing around the punctures due minimal-area property~\cite{ranganathan1992criterion, Wolf:1992bk}. Curves of length smaller than $2 \pi$ are prevented to appear in the geometry after sewing two vertices by the inclusion of stubs. The~\textit{vertex region} for the moduli integration is specified by restricting to surfaces whose heights of internal ring domains are smaller than $2 \pi$.

The hyperbolic vertices~\cite{Costello:2019fuh}, on the other hand, consider hyperbolic metrics (that is, metrics of constant negative curvature $K=-1$) on \textit{bordered} Riemann surfaces with geodesic borders of length $L$ whose systole is greater than $L$. The local coordinates are constructed by grafting a semi-infinite flat cylinder of circumference $L$ to each border. Curves of length smaller than $L$ are prevented to appear in the geometry after sewing two vertices by the collar lemma~\cite{Costello:2019fuh, buser2010geometry} as long as $0 < L \leq 2 \, \text{arcsinh} \, 1$. The vertex region contains surfaces whose systole is greater than $L$.

Our first result pertains on establishing the precise connection between the local coordinates of these two sets of vertices for genus zero surfaces, that is for punctured spheres, which has already been pointed out briefly in~\cite{Costello:2019fuh}. It is well-known that the classical minimal-area vertices arise from Strebel quadratic differentials whose critical graphs define the local coordinates~\cite{Saadi:1989tb, Kugo:1989aa, Zwiebach:1990ni}. For a comprehensive mathematical account see~\cite{strebel1984quadratic}. Remarkably, hyperbolic vertices reduce to minimal-area vertices defined by Strebel differentials in the $L \to \infty$ limit accompanied by an infinite scaling.

In this paper we demonstrate how Strebel differentials arise in the context of hyperbolic vertices. We do this by considering a (holomorphic) Fuchsian equation of the form
\begin{align} \label{eq:FuchsianIntro}
	\p^2 \psi 
	+ {1 \over 2} \, T(z) \, \psi =
	\p^2 \psi 
	+ {1 \over 2} \sum_{i=1}^n \bigg[ {\delta_i \over (z-\xi_i)^2} + {c_i \over z-\xi_i} \bigg] \psi = 0 \, ,
\end{align}
on a $n$-punctured sphere $\Sigma_{0,n}$ whose punctures are placed at $z = \xi_i$ for $i=1,\cdots,n$. The solutions of~\eqref{eq:FuchsianIntro} can be used to construct the local coordinates of hyperbolic vertices~\cite{Firat:2021ukc}. Here \textit{the classical weights} $\delta_i$ relate to the circumferences of the grafted cylinders $L_i \equiv 2\pi \lambda_i$ through
\begin{align}
	\delta_i 
	= {1 \over 2} + {1 \over 2} \left(  {L_i \over 2\pi} \right)^2 
	= {1 \over 2} + {\lambda_i^2 \over 2}  \, ,
\end{align}
and $c_i$'s are so-called \textit{accessory parameters} that have to be fixed by demanding hyperbolic $PSL(2,R)$ monodromy around each puncture, see section~\ref{sec:WKB} for more details. The hyperbolic CSFT demands taking the classical weights $\delta_i$ equal. This particular monodromy problem first appeared in~\cite{Hadasz:2003kp, Hadasz:2003he}. 

The equation~\eqref{eq:FuchsianIntro} under the WKB approximation $\lambda_i = \alpha_i \lambda \to \infty$ with fixed positive $\alpha_i \sim \mathcal{O}(1)$ shows that the Strebel differential $\varphi^{(S)} = \phi(z) dz^2$ on $\Sigma_{0,n}$ is related to $T(z)$ in~\eqref{eq:FuchsianIntro} by a limit
\begin{align} \label{eq:str}
	\phi(z) = -2\lim_{\lambda \to \infty} { T(z) \over \lambda^2} \,
	= \sum_{i=1}^n \left[ - {\alpha_i^2 \over (z-\xi_i)^2} + {\lim\limits_{\lambda \to \infty} \left( -2 c_i/\lambda^2 \right) \over z-\xi_i} \right] \, .
\end{align}
This limit also relates the accessory parameters of Strebel differentials $c_i^{(S)}$ to the accessory parameters of hyperbolic vertices $c_i$ through
\begin{align} \label{eq:StrebelWKB}
	c_i^{(S)} = - 2 \lim_{\lambda \to \infty} {c_i \over \lambda^2} \, .
\end{align}
Notice that the length of horizontal trajectories around the puncture $z = \xi_i$ is given by $2 \pi \alpha_i$.

On top of providing a conceptual unity among the classical vertices, the limit~\eqref{eq:StrebelWKB} provides an alternative expression for $c_i^{(S)}$ through the \textit{Polyakov conjecture}~\cite{Hadasz:2003kp, Hadasz:2003he,Hadasz:2005gk}. This conjecture relates a (suitably regularized and modified) on-shell Liouville action $S_{HJ}[\varphi]$, which we call~\textit{the on-shell Hadasz-Jask\'olski action} after~\cite{Hadasz:2003kp}, to the accessory parameters of the Fuchsian equation~\eqref{eq:FuchsianIntro} as
\begin{align} \label{eq:P}
c_j= - {\p  S_{HJ}[\varphi] \over \p \xi_j} \, ,
\end{align}
see~\eqref{eq:Lio5} for the definition of the action $S_{HJ}[\varphi]$. In section~\ref{sec:Polyakov}, we show that, through the geometric formulation of Liouville theory with suitable modifications and considering the WKB approximation above, a Polyakov-like conjecture for Strebel differentials is given by
\begin{align} \label{eq:PolyakovIntro}
	c_j^{(S)} =2 \, {\p \over \p \xi_j} \sum_{i=1}^n \alpha_i^2 \log r_i 
	\quad \text{with} \quad r_i = \left| {d z \over d w_i } \right|_{w_i = 0}
	\, ,
\end{align}
where $r_i$ is \textit{the mapping radius} of the map from unit disk $0< |w_i| \leq 1$ to the face of the critical graph the puncture at $z = \xi_i$ belongs. As a matter of fact, ~\eqref{eq:PolyakovIntro} has been \textit{rigorously} proved in geometric function theory~\cite{kuz1997methodsI, kuz1997methodsII, solynin1999moduli, solynin2009quadratic,solynin2020fingerprints, bakhtin2022generalized}, but appears to have been unnoticed in physics literature; here we rediscovered it. We test~\eqref{eq:PolyakovIntro} against some of the known cases of Strebel differentials in section~\ref{sec:Polyakov}.

The quantity on the right-hand side of~\eqref{eq:PolyakovIntro} comes from taking the WKB limit  of $S_{HJ}[\varphi]$ and is a sum of a well-known quantity associated with simply-connected domains called the~\textit{reduced modulus}~\cite{strebel1984quadratic}. It is beneficial to define the total modulus, which we simply call the \textit{modulus},\footnote{Our definition of the modulus differ by $2 \pi$ from the conventional definition in~\cite{strebel1984quadratic}.}
\begin{align} \label{eq:Modulus}
	\mathcal{S}_{0,n}(\xi_i; \alpha_i) \equiv \sum_{i=1}^n \alpha_i^2 \log r_i \, ,
\end{align}
where $r_i$ is now understood as the mapping radius of an arbitrary punctured simply-connected domain on the Riemann sphere. This is a functional such domains. Dividing the Riemann sphere into $n$ non-overlapping domains and evaluating $\mathcal{S}_{0,n}$, it can be shown that the division defined by the critical graph of the Strebel differential maximizes the modulus~\cite{strebel1984quadratic}. Notice that the modulus, as we defined it, is \textit{not} an ambigious quantity since we always use the local coordinates specified by the non-overlapping domains. We denote the modulus associated with Strebel differentials by $\mathcal{S}^\ast_{0,n}(\xi_i;\alpha_i)$. Take note that this is a real function of $\xi_i$ and \textit{not} holomorphic in them.

The quantity $\mathcal{S}^\ast_{0,n}$ is closely related to the closed string tachyon potential~\cite{Belopolsky:1994sk, Belopolsky:1994bj}
\begin{align}
	V(t, \cdots) = -{1 \over 2} \, t^2 - \sum_{i=1}^\infty {v_n \over n!} \, t^n + \cdots \, ,
\end{align}
where dots represent the interactions of the zero momentum tachyon $t$ with other fields. The coefficients $v_n$ are given by the following integral over the vertex region $\mathcal{V}_{0,n}$
\begin{align} \label{eq:Tachyon}
	v_n 
	= (-1)^n {2 \over \pi^{n-3} } \int_{\mathcal{V}_{0,n}}
	\left( \prod_{i=1}^{n-3} d^2 \xi_i \right)
	\prod_{j=1}^n {1 \over r_j^2}
	=
	(-1)^n {2 \over \pi^{n-3} } \int_{\mathcal{V}_{0,n}}
	\left( \prod_{i=1}^{n-3} d^2 \xi_i \right)
	e^{- 2 \mathcal{S}^\ast_{0,n}(\xi_i; \alpha_i = 1)} \, .
\end{align}
The integration is performed over the $n-3$ unfixed positions of punctures after $3$ of them fixed using the global conformal symmetry. In the second line we write the product of the mapping radii $r_i$ in terms of $\mathcal{S}^\ast_{0,n}$. So the relation~\eqref{eq:PolyakovIntro}, combined with the fact that the accessory parameters are sufficient to characterize the vertices of CSFT~\cite{Belopolsky:1994bj}, shows that \textit{the (integrand of) off-shell interaction of $n$ zero momentum tachyons generates the coefficients of the remaining interactions in CSFT}. In other words, the modulus $\mathcal{S}^\ast_{0,n}(\xi_i, \alpha_i=1)$ fully determines string vertices.

The accessory parameters $c_i^{(S)}$ for $n \geq 4$ were previously obtained by numerically solving complicated integral equations~\cite{Belopolsky:1994bj, Moeller:2004yy,Moeller:2006cw, Moeller:2007mu}, or very recently, using machine learning~\cite{Erbin:2022rgx}. The conjecture~\eqref{eq:PolyakovIntro} provides an alternative path forward. The question of classical string vertices in CSFT reduces to evaluating $\mathcal{S}^\ast_{0,n}$ as a function of moduli $\xi_i$ for $n \geq 3$ without using the local coordinates. There is a closed-form expression for $n=3$ for the general case and it is given by
\begin{align} \label{eq:BaseCase}
	&\mathcal{S}^\ast_{0,3}(\alpha_1, \alpha_2, \alpha_3) = 
	\log \bigg[
	|2\alpha_1|^{2 \alpha_1^2} \; |2\alpha_2|^{2 \alpha_2^2} \; |2\alpha_3|^{2 \alpha_3^2} \;
	|\alpha_1 + \alpha_2 + \alpha_3|^{-{1 \over 2}  (\alpha_1 + \alpha_2 + \alpha_3)^2 } \\
	& \quad
	|-\alpha_1 + \alpha_2 + \alpha_3|^{-{1 \over 2} (-\alpha_1 + \alpha_2 + \alpha_3)^2 }
	|\alpha_1 - \alpha_2 + \alpha_3|^{- {1 \over 2}  (\alpha_1 - \alpha_2 + \alpha_3)^2 }
	|\alpha_1 + \alpha_2 - \alpha_3|^{- {1 \over 2}  (\alpha_1 + \alpha_2 - \alpha_3)^2 }
	\bigg] \nonumber \, ,
\end{align} 
assuming the positions of three punctures are fixed to $0,1, \infty$.\footnote{We fix the positions of last three punctures to $0,1,\infty$ unless stated otherwise, so the dependence on $\xi_i$ should be understood after such operation. For $n=3$, this eliminates the dependence on $\xi_i$ so we don't indicate it. Also when $\alpha_i = 1$ for all $i =1, \cdots, n$ (the case relevant for CSFT), we suppress the dependence of  $\alpha_i$ in $\mathcal{S}^\ast_{0,n}$.}This is derived by taking an appropriate limit of the mapping radii of the hyperbolic three-vertex of~\cite{Firat:2021ukc}, see section~\ref{sec:recursion}. We point out that $\mathcal{S}^\ast_{0,3}(\alpha_1, \alpha_2, \alpha_3)$ is totally symmetric in its arguments and it has an eerie resemblance to the DOZZ three-point function of Liouville theory. We argue that this is not a coincide in section~\ref{sec:recursion}.

Evaluating $\mathcal{S}^\ast_{0,n}(\xi_i; \alpha_i)$ has a rich and fruitful history in geometric function theory and it goes under the name of \textit{the extremal decomposition of Riemann sphere}, see~\cite{kuz1997methodsI, kuz1997methodsII, solynin1999moduli, solynin2009quadratic, solynin2020fingerprints, bakhtin2022generalized, jenkins1954recent,kuz1982problem,fedorov1982maximum, emelyanov2004problem}. Actually, $\mathcal{S}^\ast_{0,4}(\xi_i)$ is claimed to have a closed-form expression in term of Jacobi elliptic functions~\cite{kuz1982problem,fedorov1982maximum}, and the partial results are known for $n \geq 5$, although these expressions seem to be implicit at best. There may be value of investigating them further to derive a closed-form expression for $\mathcal{S}^\ast_{0,n}(\xi_i; \alpha_i)$, however this problem appears to be intractable/impractical to the author.

Regardless, we can import insights from Liouville theory to construct $\mathcal{S}^\ast_{0,n}(\xi_i; \alpha_i)$ as an expansion using the \textit{classical conformal blocks}~\cite{Zamolodchikov:1995aa,zamolodchikov1987conformal, francesco2012conformal} in the spirit of conformal bootstrap. Recall that the Virasoro conformal blocks
\begin{align} \label{eq:ConfBlock}
\mathcal{F}_{1 + 6 Q^2 , \, \Delta } 
\begin{bmatrix}
\Delta_3 & \Delta_2\\
\Delta_4 & \Delta_1
\end{bmatrix} 
(\xi) 
=
\xi^{\Delta - \Delta_2 - \Delta_1}\left[
1 + {(\Delta + \Delta_3  - \Delta_4) (\Delta + \Delta_2  - \Delta_1) \over 2 \Delta} \, \xi + \cdots 
\right] \, ,
\end{align}
are the universal functions representing the (holomorphic) part of the four-point function fixed by the conformal symmetry in conformal field theory (CFT). The conformal block $\mathcal{F}$ is a function of cross-ratio $\xi$, depends on the central charge of CFT $c \equiv 1 + 6Q^2$ as well as on the conformal weights $\Delta_i$ of the external primaries for $i=1,2,3,4$ and the conformal weight $\Delta$ of the so-called intermediate primary. The four-point function is given in terms of conformal blocks and three-point functions between two of the external operators and the intermediate operator. The explicit closed-form expressions for the conformal blocks are not known, but there exists efficient algorithms à la Zamolodchikov to compute them as a series expansion for (almost) all $\xi$~\cite{zamolodchikov1987conformal}. Observe that our notation for the blocks underlines how the four-point function is decomposed by its columns.

The classical conformal blocks are the semi-classical limit ($c \to \infty, Q \to \infty$) of the Virasoro conformal blocks~\eqref{eq:ConfBlock} with ``heavy'' weights
\begin{align}
	\Delta_i =  Q^2 \, {\delta_i \over 2} \, ,
	\quad \quad
	\Delta =  Q^2  \, {\delta \over 2},
\end{align}
for which $\delta_i, \delta \sim \mathcal{O}(1)$ and they are related to~\eqref{eq:ConfBlock} through exponentiation~\cite{Belavin:1984vu} 
\begin{align} \label{eq:ClassConfBlock}
	\mathcal{F}_{1 + 6 Q^2 , \Delta } 
	\begin{bmatrix}
	\Delta_3 & \Delta_2\\
	\Delta_4 & \Delta_1
	\end{bmatrix} 
	(\xi)
	= \exp\left(
	Q^2 \, \widetilde{f}_{\delta/2}
	\begin{bmatrix}
	\delta_3/2 & \delta_2/2 \\
	\delta_4/2 & \delta_1/2
	\end{bmatrix} (\xi)
	\right)\, .
\end{align}
This exponential behavior is non-trivial~\cite{zamolodchikov1986two, zamolodchikov1987conformal} and only recently proven in~\cite{becsken2020semi}. As alluded through the equations~\eqref{eq:StrebelWKB}-\eqref{eq:Modulus}, the modulus $\mathcal{S}^\ast_{0,n}(\xi_i;\alpha_i)$ is intimately related to the on-shell Liouville action. We explain in section~\ref{sec:recursion} how the latter can be understood as a correlator of the heavy primaries in the semi-classical limit of the DOZZ formulation of Liouville theory~\cite{ Zamolodchikov:1995aa, Teschner:2001rv, Dorn:1994xn}. Conformal bootstrap then suggests that it can be determined using the on-shell Liouville action with three hyperbolic singularities derived in~\cite{Hadasz:2003he} and classical conformal blocks~\eqref{eq:ClassConfBlock}. 

On top of this, we consider the WKB  limit to restrict our considerations to the modulus $\mathcal{S}^\ast_{0,n}$. Here, the WKB limit is in the sense of taking $\lambda \to \infty$ while keeping $\alpha, \alpha_i \sim \mathcal{O}(1)$ fixed so that
\begin{align}
	\delta_i = {1 \over 2} + {\alpha_i^2 \lambda^2 \over 2} \to {\alpha_i^2 \lambda^2 \over 2}  ,
	\quad \quad
	\delta = {1 \over 2} + {\alpha^2 \lambda^2 \over 2}  \to {\alpha^2 \lambda^2 \over 2} \, ,
\end{align}
and this results in
\begin{align} \label{eq:them}
	f_{\alpha^2}
	\begin{bmatrix}
	\alpha_3^2 & \alpha_2^2\\
	\alpha_4^2 & \alpha_1^2
	\end{bmatrix} (\xi)
	\equiv \lim_{\lambda \to \infty} {1 \over \lambda^2} \, 
	\widetilde{f}_{\delta/2}
	\begin{bmatrix}
	\delta_3/2 & \delta_2/2 \\
	\delta_4/2 & \delta_1/2
	\end{bmatrix} (\xi) \, .
\end{align}
A quick inspection of the expansion of $\widetilde{f}$ shows that this limit exists and is different from zero (see appendix~\ref{app:ConfBlocks}). For brevity, we also call the functions~\eqref{eq:them} classical conformal blocks.

For instance, the procedure described above entails the following relation for $\mathcal{S}^\ast_{0,4} (\xi, \alpha_i = 1)$:
\begin{align}
	\exp\left( -{Q^2 \lambda^2 \over 2} \, \mathcal{S}^\ast_{0,4} (\xi) \right) 
	\sim
	&\int_0^\infty d \alpha 
	\exp \bigg[
	-{Q^2 \lambda^2 \over 2} \, \mathcal{S}^\ast_{0,3} (\alpha,1,1) 
	-{Q^2 \lambda^2 \over 2} \, \mathcal{S}^\ast_{0,3} (1,1, \alpha) \\
	& \hspace{1in}
	+Q^2 \lambda^2 \,
	 f_{\alpha^2}
	\begin{bmatrix}
	1 & 1\\
	1 & 1
	\end{bmatrix} (\xi)
	+Q^2 \lambda^2 \,
	\overline{f_{\alpha^2}}
	\begin{bmatrix}
	1 & 1\\
	1 & 1
	\end{bmatrix} (\overline{\xi}) 
	\bigg]\nonumber \, ,
\end{align}
without using the symmetry of $\mathcal{S}^\ast_{0,3} $ to make the overall structure apparent. Here the bar indicates complex conjugation. Since we consider the semi-classical and WKB limits, the saddle point at $\alpha = \alpha_s$ dominates the integral. The saddle point $\alpha_s$ is found by solving
\begin{align} \label{eq:Saddle}
	{\p \over \p \alpha} \bigg[
	-{1 \over 2}\, \mathcal{S}^\ast_{0,3} (\alpha,1,1) 
	-{1 \over 2}\, \mathcal{S}^\ast_{0,3} (1,1,\alpha)
	+ f_{\alpha^2}
	\begin{bmatrix}
	1 & 1\\
	1 & 1
	\end{bmatrix} (\xi)
	+
	 \overline{f_{\alpha^2}}
	\begin{bmatrix}
	1 & 1\\
	1 & 1
	\end{bmatrix} (\overline{\xi}) 
	\bigg]_{\alpha = \alpha_s} = 0 ,
\end{align}
and it results in the expression, at the leading order,
\begin{align}
	\mathcal{S}^\ast_{0,4} (\xi) =
	\mathcal{S}^\ast_{0,3} (\alpha_s(\xi,\overline{\xi}),1,1) 
	+ \mathcal{S}^\ast_{0,3} (1,1,\alpha_s(\xi,\overline{\xi}))
	- 2 f_{\alpha_s^2(\xi,\overline{\xi})}
	\begin{bmatrix}
	1 & 1\\
	1 & 1
	\end{bmatrix} (\xi)
	-
	2 \overline{ f_{\alpha_s^2(\xi,\overline{\xi})} }
	\begin{bmatrix}
	1 & 1\\
	1 & 1
	\end{bmatrix} (\overline{\xi}) \, .
\end{align}
The function $\mathcal{S}^\ast_{0,4} (\xi)$ can be used to find the accessory parameters as a function of the moduli $\xi$ using~\eqref{eq:PolyakovIntro}. From this, we additionally get the following for the accessory parameter associated with the cross-ratio $\xi$
\begin{align} \label{eq:IntroAlt}
c^{(S)} = 2 \, {\p \mathcal{S}^\ast_{0,4} (\xi) \over \p \xi} 
= - 4 \, {\p  f_{\alpha^2} \over \p \xi}
\begin{bmatrix}
1 & 1\\
1 & 1
\end{bmatrix} (\xi) \Bigg|_{\alpha = \alpha_s(\xi,\overline{\xi})} \, ,
\end{align}
by noticing any term that multiplies a derivative of $\alpha_s(\xi,\overline{\xi})$ drops out by the saddle point equation~\eqref{eq:Saddle} and all that remains is the derivative of the classical conformal block with respect to~$\xi$.

There are few comments in order for these relations. First, $2 \pi \alpha_s$ is the length of the $s$-channel geodesic.\footnote{This is the geodesic separating the punctures at $z=0,\xi$ from those at $z=1, \infty$.} This is justified by the semi-classical expectation for the resulting geometry on $\Sigma_{0,n}$. Second, there is nothing special about $s$-channel decomposition and the similar procedure can be repeated for the $t$-and $u$-channels. Consequently, this leads to crossing equations for $\mathcal{S}^\ast_{0,4} (\xi_i)$. We repeat the procedure for each channel and show that they pass various consistency checks in section~\ref{sec:recursion}. 

In fact, it is possible to express the boundary of the vertex region $\p \mathcal{V}_{0,4}$ in terms of classical conformal blocks by investigating the constant $\alpha_s, \alpha_t$ and $\alpha_u$ curves in the moduli space $\mathcal{M}_{0,4}$. For instance, we find
\begin{align} \label{eq:DV}
	\left| \exp 
	{\p f_{\alpha^2} \over \p \alpha }
	\begin{bmatrix}
	1 & 1\\
	1 & 1
	\end{bmatrix} (\xi)
	\right|_{\alpha = 1} =  {4 \over 3 \sqrt{3} }  \, ,
\end{align}
for the curve that separates the vertex region $\mathcal{V}_{0,4}$ from the $s$-channel Feynman region $\mathcal{F}_{s}$ (see figure~\ref{fig:vertex}). The expressions for $t$-and $u$-channels are similar. It is somewhat amusing that a famous number in string field theory appeared on the right-hand side of this expression~\cite{Sen:1999nx}. As a corollary, we also find the holomorphic map that relates the Schwinger parameter $\mathfrak{q}$ of a propagator in a given channel to the complex moduli $\xi$, see equation~\eqref{eq:Sch}.

The arguments for $n=4$ and $\alpha_i =1$ sketched above have obvious generalizations to $n \geq 5$, which we briefly discuss in section~\ref{sec:recursion}. Hence, it appears that the classical conformal blocks play a central role in hyperbolic CSFT: it is possible to specify \textit{all} background-independent data of CSFT in terms of them, reducing it to a problem in CFT. This is unexpected. Moreover, our framework simplifies considerably when the vertices are taken to be defined by Strebel differentials, indicating that they may be the ``canonical'' choice as far as the classical hyperbolic vertices are concerned. Beyond possibly providing an analytic tool to generate higher string vertices we think that the framework sketched here may point out a deeper relation between CSFT and Liouville theory. We collect some of these speculations in conclusion.

The rest of the paper is organized as follows. In section~\ref{sec:WKB}, we establish the precise connection between classical hyperbolic and minimal-area vertices, first pointed out in~\cite{Costello:2019fuh}. In section~\ref{sec:Polyakov}, we construct a version of Liouville theory whose on-shell action plays the role of a generating function for the accessory parameters of hyperbolic string vertices. We also consider the WKB limit of the resulting relation to argue for a Polyakov-like conjecture for Strebel differentials. In section~\ref{sec:recursion} we propose an ``operator formalism'' and explicitly bootstrap the string interactions for $n=4$ and $\alpha_i =1$. We conclude the paper in section~\ref{sec:conc}. In appendices~\ref{app:direct} and~\ref{app:DOZZ} we provide a brief review of the arguments in~\cite{Hadasz:2003kp,Hadasz:2003he} respectively and in appendix~\ref{app:ConfBlocks} we review the classical conformal blocks and our computations of them thereof.

\section{Hyperbolic string vertices and Strebel differentials} \label{sec:WKB}

In this section we establish the relation between the classical hyperbolic and minimal-area string vertices. To that end, we use the connection between hyperbolic vertices and a certain hyperbolic monodromy problem of the Fuchsian equation first investigated in~\cite{Firat:2021ukc}, which we review in subsection~\ref{sec:WKBrev}. The contents of this subsection are known in the literature~\cite{Firat:2021ukc,Hadasz:2003he,Hadasz:2003kp}, but here we additionally provide an argument for the existence and uniqueness of the solutions to the hyperbolic monodromy problem based on the uniformization theorem. In subsection~\ref{sec:WKBsub}, we solve the relevant Fuchsian equation using the WKB approximation. This allows us to relate hyperbolic vertices and Strebel differentials. We demonstrate how the latter is just a particular limit of the former.

\subsection{Hyperbolic string vertices and the Fuchsian equation} \label{sec:WKBrev}

Consider the (holomorphic) Fuchsian equation\footnote{Similar results hold for the anti-holomorphic Fuchsian equation.}
\begin{align} \label{eq:Fuchsian}
	\p^2 \psi + {1 \over 2} \, T(z) \psi = 0 \, ,
\end{align}
on an $n$-punctured Riemann sphere $\Sigma_{0,n}$. Here $T(z)$ is taken to be
\begin{align} \label{eq:StressEnergy} 
	T(z) = \sum_{i=1}^{n} \left[
	{\delta_i \over (z-\xi_i)^2} + {c_i \over z-\xi_i}
	\right],
	\hspace{0.5in}
	\delta_i = {1 \over 2} + {\lambda_i^2 \over 2}
	\, ,
\end{align}
where $\x_i$ are the position of punctures, $\delta_i \geq 1/2$ are the \textit{classical weights}, and $c_i \in \mathbb{C}$ are the \textit{accessory parameters} that should be chosen so that the solutions to equation~\eqref{eq:Fuchsian} have a real hyperbolic monodromy around each puncture, as this leads to the local coordinates of hyperbolic vertices~\cite{Hadasz:2003kp, Hadasz:2003he,Hadasz:2005gk,Firat:2021ukc}. The punctures at $z=\xi_i$ are sometimes called~\textit{hyperbolic singularities} for this reason. Finding the accessory parameters $c_i$'s as functions of $\xi_i$ is the \textit{hyperbolic monodromy problem}.

The Fuchsian equation~\eqref{eq:Fuchsian} is invariant under the conformal transformation $z \to \widetilde{z}$ as long as the objects $\psi(z)$ and $T(z)$ transform as
\begin{align} \label{eq:StressEnergyTransform}
	\psi(z) =  \left( {\p \widetilde{z} \over \p z} \right)^{-1/2} \widetilde{\psi}(\widetilde{z})\,, 
	\hspace{0.5in}
	T(z) = \left( {\p \widetilde{z} \over \p z} \right)^2 \widetilde{T}(\widetilde{z}) + \{\widetilde{z},z\} \, .
\end{align}
Here $\{\cdot, \cdot\}$ is the Schwarzian derivative
\begin{align}
	\{\widetilde{z},z\} \equiv { \p^3 \widetilde{z} \over \p \widetilde{z} } - {3 \over 2} \left( { \p^2 \widetilde{z} \over \p \widetilde{z} } \right)^2 \, .
\end{align}
Since ~\eqref{eq:Fuchsian} is invariant under conformal transformations, we can consider it patch-by-patch on any Riemann surface. Hence we can use it to construct the local coordinates of quantum hyperbolic vertices after placing a restriction on the values of $\lambda_i$~\cite{Costello:2019fuh}. However, we restrict ourselves to genus zero Riemann surfaces, that is punctured Riemann spheres, as we are just interested in classical string vertices at this moment.

Assuming no punctures are at $z=\infty$ and demanding regularity there for the Fuchsian equation~\eqref{eq:Fuchsian} on $\Sigma_{0,n}$ constrains the accessory parameters to satisfy
\begin{align} \label{eq:Conditions}
\sum_{i=1}^n c_i = 0,
\quad \quad
\sum_{i=1}^n (\delta_i + c_i \, \xi_i) = 0,
\quad \quad
\sum_{i=1}^n (2 \delta_i \xi_i + c_i \, \xi_i^2) = 0,
\end{align} 
and forbids having regular terms in $(z-\xi_i)$ in~\eqref{eq:StressEnergy}. These constraints can be argued by inverting~\eqref{eq:StressEnergy} by $\widetilde{z} = 1/z$ and subsequently using~\eqref{eq:StressEnergyTransform}. Note that the Fuchsian equation~\eqref{eq:Fuchsian} is the hypergeometric equation when $n=3$ and it has used to obtain the local coordinates for the hyperbolic three-string vertex~\cite{Hadasz:2003he,Firat:2021ukc}. However for $n \geq 4$, $n-3$ accessory parameters remain to be fixed.

It can be shown that a (possibly singular) hyperbolic metric on $\Sigma_{0,n}$ can be obtained using the solutions of~\eqref{eq:Fuchsian} that realize a real hyperbolic monodromy around each puncture~\cite{Hadasz:2003he,Firat:2021ukc}. Taking two linearly independent solutions $\psi^{\pm}(z)$ to~\eqref{eq:Fuchsian} with a diagonal real hyperbolic monodromy around $z=\xi_i$ and normalizing them so that their Wronskian is $W(\psi^-, \psi^+) = 1$, the single-valued hyperbolic metric takes the form
\begin{align} \label{eq:SingHypMet}
	ds^2 = {-4 \; |dz|^2  \over ( \overline{\psi^-(z)} \psi^+(z) - \overline{\psi^+(z)} \psi^-(z) )^2}
	= {\lambda_i^2 \; |\p \rho_i(z) |^2 \over |\rho_i(z)|^2 \, \sin^2(\lambda_i \log|\rho_i(z)|)}  |dz|^2  \, .
\end{align}
In the second line we have defined \textit{the scaled ratio}
\begin{align} \label{eq:ScaledRatio}
	\rho_i(z) \equiv \left( {\psi^+(z) \over \psi^-(z) }\right)^{1 \over i \lambda_i}.
\end{align}
The branch choice for the imaginary exponentiation is irrelevant for the metric in~\eqref{eq:SingHypMet}, as it only shifts the argument of $\sin$ by an integer multiple of $2 \pi$. So we pick a convenient branch for the scaled ratio when needed. We note that the local coordinates are proportional to the scaled ratio~\cite{Firat:2021ukc}.

Geometrically, the metric~\eqref{eq:SingHypMet} looks like a hyperbolic metric on an $n$-bordered sphere with geodesic borders of length $2 \pi \lambda_i$, together with semi-infinite series of hyperbolic cylinders grafted at each border. A priori, the metric~\eqref{eq:SingHypMet} may be singular away from the singularities of these hyperbolic cylinders, but we argue below that this never happens. This is because there is a unique set of accessory parameters that solves the monodromy problem given $\xi_i$, which is exactly those arising from the hyperbolic metric on a bordered sphere with grafted hyperbolic cylinders.

We first establish the existence of the accessory parameters $c_i$  solving the hyperbolic monodromy problem for~\eqref{eq:Fuchsian}. In order to accomplish this, recall that the Riemann sphere with $n$ borders can be equipped with the hyperbolic metric with geodesic borders of length $2 \pi \lambda_i$ by the uniformization theorem. One can then construct the metric mentioned in the paragraph above by grafting a semi-infinite series of hyperbolic cylinders at each border, resulting in a singular hyperbolic metric on $\Sigma_{0,n}$. Say this metric is of the form
\begin{align} \label{eq:HyperbolicMetric}
	ds^2 =e^{\vf(z,\overline{z})} |dz|^2 \, ,
\end{align}
on $\Sigma_{0,n}$ with the punctures are placed at $z=\xi_i$ for $i=1, \cdots, n$.\footnote{It is always possible to consider this metric for given positions of punctures due to the homeomorphism established in~\cite{mondello2011riemann}, see Theorem 5.4. Also see remarks in~\cite{Costello:2019fuh}.} Since~\eqref{eq:HyperbolicMetric} is a hyperbolic metric it satisfies \textit{Liouville's equation}
\begin{align} \label{eq:LiouvillesEquation}
	K = -1 \iff \p \overline{\p} \vf(z,\overline{z}) = {1 \over 2} e^{\vf(z,\overline{z})} \, ,
\end{align}
when it is not singular. Given the metric~\eqref{eq:HyperbolicMetric} we define
\begin{align} \label{eq:StressEnergyPlugged}
	T_{\vf} (z) \equiv -{1 \over 2} (\p \vf)^2 + \p^2 \vf  = - 2 e^{\vf/2} \p^2 e^{-\vf/2} \, .
\end{align}
This is holomorphic (i.e. $\overline{\p} \, T_{\vf} = 0$) as a consequence of Liouville's equation~\eqref{eq:LiouvillesEquation}. Clearly, by taking $T(z) = T_{\vf} (z)$ in~\eqref{eq:Fuchsian}, we can write $e^{\vf(z,\overline{z})} $ in terms of \textit{particular} solutions to~\eqref{eq:Fuchsian} as in~\eqref{eq:SingHypMet}. Moreover, this metric has to take the following form as $z \to \xi_i$
\begin{align} \label{eq:LeadingTerm}
ds^2 = {\lambda_i^2 \, |dz|^2 \over |z-\xi_i|^2\, \sin^2(\lambda_i \log|z-\xi_i| + \theta_i)}  \left( 1 + \mathcal{O}(z-\xi_i) \right)  \, ,
\end{align}
by having a semi-infinite series of hyperbolic cylinders around the puncture at $z=\xi_i$. Here $\theta_i$ is an unspecified phase factor. Combining two facts above shows that
\begin{align} \label{eq:LeadingTerm2}
\rho_i(z) \sim ( z-\xi_i) + \mathcal{O}((z-\xi_i)^2) 
\implies \psi^{\pm} (z) \sim  (z-\xi_i)^{1/2 \pm i \lambda_i/2} \left( 1 + \mathcal{O}(z-\xi_i) \right) \, ,
\end{align}
around the puncture $z=\xi_i$ for some normalized basis of solutions to~\eqref{eq:Fuchsian}. Observe that this argument doesn't fix the overall coefficients of $\psi^{\pm}(z)$ (which we indicate with $\sim$) and this is related to having $\theta_i$ in~\eqref{eq:LeadingTerm}.

Nonetheless, there are few things to notice here. First, this asymptotic implies
\begin{align}
	\p ^2\psi^\pm (z) +{1 \over 2} {\delta_i \over (z-\xi_i)^2} \psi^\pm (z)
	= \mathcal{O}\left( {1 \over z- \xi_i} \right) \, .
\end{align}
That is, there exists a double pole of residue $\delta_i$ at $z=\xi_i$ in $T(z)$ of~\eqref{eq:Fuchsian}. Second, the leading term in~\eqref{eq:LeadingTerm2} produces a real hyperbolic monodromy around the puncture $z=\xi_i$, since
\begin{align} \label{eq:Monodromy}
	z-\xi_i \to e^{2 \pi i} (z-\xi_i) \implies 
	\begin{bmatrix}
	\psi^+(z) \\
	\psi^-(z)
	\end{bmatrix}
	\to
	\begin{bmatrix}
	-e^{-\pi \lambda_i} & 0 \\
	0 & -e^{ \pi \lambda_i}
	\end{bmatrix}
		\begin{bmatrix}
	\psi^+(z) \\
	\psi^-(z)
	\end{bmatrix} \, .
\end{align}
These statements remain true after a real change of basis for $\psi^\pm(z)$, which is the most general transformation preserving the form of the metric~\eqref{eq:SingHypMet}. In particular, it is possible to relate the basis of solutions that have~\eqref{eq:LeadingTerm2} for different punctures as a linear combination of each other by the fact that either solutions are supposed to describe the metric~\eqref{eq:LiouvillesEquation}. Considering this fact and repeating the same logic for each puncture, we must have
\begin{align} \label{eq:Fuchsian2}
	\p^2 \psi(z) + {1 \over 2} \sum_{i=1}^{n} \left[
	{\delta_i \over (z-\xi_i)^2} + {c_i \over z-\xi_i}
	\right] \psi(z)  = 0 \, ,
\end{align}
with $c_i$ are certain complex numbers. Again, we are supposed to have a trivial monodromy around $z=\infty$ for the solutions. This imposes the constraints~\eqref{eq:Conditions}. Furthermore, we are made sure that the there exists a real hyperbolic monodromy around each puncture by~\eqref{eq:Monodromy} for the solutions that lead to the same metric, albeit it is not going to be in the diagonal form like it does in~\eqref{eq:Monodromy} in general. This shows that there exists a set of accessory parameters $c_i$ such that the solutions to the equation~\eqref{eq:Fuchsian} can realize a real hyperbolic monodromy around each puncture.

The accessory parameters $c_i$ are also unique. To show this, suppose that there are two distinct sets of accessory parameters $c_i$, $c_i'$ solving the hyperbolic monodromy problem for given $\xi_i$. Then the equation~\eqref{eq:Fuchsian2} is satisfied for the same solution $\psi(z)$ that realize real hyperbolic monodromies and the difference of the equation with two distinct accessory parameters gives
\begin{align}
	\psi(z) \sum_{i=1}^n {c_i - c_i' \over z-\xi_i} = 0 \, ,
\end{align}
which holds for any $z$. This is only possible when $c_i = c_i'$, showing that the accessory parameters are unique. We conclude that the metric in~\eqref{eq:SingHypMet} shouldn't contain any singularity beyond those associated with the grafted hyperbolic cylinders: it would be inconsistent with the uniqueness of the accessory parameters that lead to their construction.

The reasoning above shows that the Fuchsian equation~\eqref{eq:Fuchsian2} can be used to find the local coordinates of any classical hyperbolic vertex. All one has to do is to find a basis of solution to~\eqref{eq:Fuchsian2} that has a diagonal monodromy around a puncture~\eqref{eq:Monodromy} and make sure that the monodromy around the remaining punctures are real hyperbolic by adjusting the appropriate constants multiplying $\psi^\pm(z)$. Subsequently, the local coordinates are given by the scaled ratio $\rho_i(z)$ up to a multiplicative constant. Geometrically, this corresponds to taking out semi-infinite series of hyperbolic cylinders around punctures and replacing them with semi-infinite flat cylinders. This is an example of \textit{Thurston metric}~\cite{Costello:2019fuh}. This procedure has been explicitly done for $n=3$ in~\cite{Firat:2021ukc}.

Even after we somehow obtain the accessory parameters $c_i$ as a function of $\xi_i$, we still have to derive the connection formulas for the solutions to~\eqref{eq:Fuchsian2} so that the hyperbolic real monodromies can be implemented. These formulas are not known explicitly for~\eqref{eq:Fuchsian2} when $n \geq 4$. In the next subsection, we consider a scenario for which the knowledge of these connection formulas is not necessary to obtain the local coordinates and this will lead us to the construction underlined by Strebel differentials.

\subsection{The WKB approximation of the Fuchsian equation} \label{sec:WKBsub}

Rather than trying to solve~\eqref{eq:Fuchsian2} exactly for general classical weights $\delta_i$, let us consider the limit where $\delta_i$ are taken to infinity while none of them being parametrically small. That is, we take
\begin{align} \label{eq:Lim}
	\lambda_i = \alpha_i \lambda \quad \text{with} \quad \lambda \to \infty \, ,
\end{align}
while $\alpha_i \sim \mathcal{O}(1)$ is fixed. We call~\eqref{eq:Lim} the~\textit{WKB limit}, as it corresponds to the WKB limit of the equation~\eqref{eq:Fuchsian2}. The WKB limit is interesting for two reasons. First, classical hyperbolic string vertices reduce to those coming from classical minimal-area vertices, which are described by Strebel differentials~\cite{Costello:2019fuh}. The argument for this is simple: the area of the bordered surface stays finite by the Gauss-Bonnet theorem as $L_i = 2\pi \lambda_i \to \infty$ and the hyperbolic part of the Thurston metric reduces to a measure-zero graph on the Riemann sphere as result. The flat cylinders become the faces of such graph and the Riemann surface naturally gets endowed with a Strebel differential. The graph in question is the critical graph of this Strebel differential.

The second reason is that we can consider the WKB approximation to~\eqref{eq:Fuchsian2} and establish an explicit relation between the local coordinates of two sets of vertices. In particular, the WKB approximation connects the accessory parameters of the Fuchsian equation~\eqref{eq:Fuchsian2} solving the hyperbolic monodromy problem to the accessory parameters of Strebel differentials. Later, we use this relation to export insights to the latter from the former that have non-trivial consequences.

Motivated by above, we begin by deriving the scaling of the accessory parameters $c_i$ of~\eqref{eq:Fuchsian} in the WKB limit. We claim they scale like the classical weights $\delta_i$, that is
\begin{align} \label{eq:AccessoryParameterScaling}
	c_i = c_i^{(2)} \lambda^2 + \cdots \, ,
\end{align}
where the terms of order $\lambda^n$ with $n>2$ are absent and $c_i^{(2)} \sim \mathcal{O}(1)$ are some complex numbers and dots stand for the terms scale slower than $\lambda^2$ as $\lambda \to \infty$. This can be argued as follows. Say the leading behavior of $c_i$'s are given by $c_i = c_i^{(n)} \lambda^n + \cdots$ where $n \in \mathbb{R}$. Since the double pole at $z = \xi_i$ suppose to exist for real hyperbolic monodromies regardless of how large the classical weights are, we must have $\forall \, \lambda>0, \; \exists \, \delta >0$ such that for $0 < |z - \xi_i| < \delta$
\begin{align}
	{\alpha_i^2 \lambda^2 \over 2 |z-\xi_i|^2} > {\lambda^n |c_i^{(n)}| \over  |z-\xi_i|} 
	\implies
	{\alpha_i^2 \over 2 |z- \xi_i| \lambda^{n-2} } > |c_i^{(n)}| \, .
\end{align}
That is, the first order pole shouldn't overwhelm the second order pole as $\lambda \to \infty$ in an open set around $z = \xi_i$. Clearly this necessitates $c_i^{(n)} = 0$ for $n>2$, leaving us with~\eqref{eq:AccessoryParameterScaling} as the most general possibility.

Next, let us consider the Fuchsian equation~\eqref{eq:Fuchsian2} as $\lambda \to \infty$, which takes the form of
\begin{align} \label{eq:FuchsianWKB}
\p^2 \psi(z) + {\lambda^2 \over 4} \sum_{i=1}^{n} \left[
{\alpha_i^2 \over (z-\xi_i)^2} + {2 c_i^{(2)}\over z-\xi_i}
\right] \psi(z) + \cdots  = 0\, .
\end{align}
As $\lambda \to \infty$, it is natural to attempt to solve~\eqref{eq:FuchsianWKB} using the WKB approximation.~So make the ansatz
\begin{align} \label{eq:Spin1/2WKB}
	\psi(z)  = \exp\left(\lambda \int^z S(z') dz' + \cdots \right) \, .
\end{align}
For now we keep the lower bound on the integration arbitrary. This implies
\begin{align}
	S(z)^2 +  {1 \over 4} \sum_{i=1}^{n} \left[
	{\alpha_i^2 \over (z-\xi_i)^2} + {2 c_i^{(2)}\over z-\xi_i}
	\right] = 0 \implies S(z) = \pm {1 \over 2} \sqrt{ \sum_{i=1}^{n} \left[
		{-\alpha_i^2 \over (z-\xi_i)^2} + {-2 c_i^{(2)}\over z-\xi_i}
		\right]} \, ,
\end{align}
and we find two linearly independent solutions to~\eqref{eq:FuchsianWKB} are
\begin{align}
	\psi^{\pm}(z) = \exp\left(\pm {\lambda \over 2} \int^z \sqrt{ \sum_{i=1}^{n} \left[
		{-\alpha_i^2 \over (z'-\xi_i)^2} + {-2 c_i^{(2)}\over z'-\xi_i}
		\right]} dz' + \cdots \right) \, .
\end{align}
Note that the solutions $\psi^\pm(z)$ have the expected real hyperbolic monodromy behavior around the punctures given in~\eqref{eq:Monodromy}. Placing the same lower bound on the integrals for both solutions in order to normalize their Wronskian as before, we find the scaled ratio~\eqref{eq:ScaledRatio} associated with the puncture at $z=\xi_j$ is given by
\begin{align} \label{eq:ScaledRatioWKB}
	\rho_j(z) = \exp\left( - {i \over \alpha_j} \int^z \sqrt{ \sum_{i=1}^{n} \left[
		{-\alpha_i^2 \over (z'-\xi_i)^2} + {-2 c_i^{(2)}\over z'-\xi_i}
		\right]} dz' + \cdots \right) \, .
\end{align}

In the strict limit of $\lambda \to \infty$ the dots above disappear. As we mentioned earlier,~\eqref{eq:ScaledRatioWKB} is related to the local coordinates $w_j$ around the punctures up to a multiplicative constant. This implies
\begin{align} \label{eq:StrebelNatural}
	-\alpha_j^2 {d w_j^2 \over w_j^2} = - \alpha_j^2 {d \rho_j^2 \over \rho_j^2} = \sum_{i=1}^{n}  \left[
	{-\alpha_i^2 \over (z-\xi_i)^2} + {-2 c_i^{(2)}\over z-\xi_i}
	\right] dz^2
	= - 2 \lim_{\lambda \to \infty} {T(z) \over \lambda^2}  dz^2  \, ,
\end{align}
for the local coordinates $w_j = w_j(z)$ around the puncture $z=\xi_j$ for $j=1, \cdots, n$. Hence the Strebel differential $\varphi^{(S)} $ on $\Sigma_{0,n}$ is equal to
\begin{align} \label{eq:Str}
	\varphi^{(S)} = \sum_{i=1}^{n}  \left[
	 {-\alpha_i^2 \over (z-\xi_i)^2} + {-2 c_i^{(2)}\over z-\xi_i}
	\right] dz^2 \, ,
\end{align}
as its expression is given by~\eqref{eq:StrebelNatural} in the local coordinates $0<|w_j|\leq1$ and it is appropriately sewn at $|w_j| = 1$ (i.e. on the critical graph)~\cite{strebel1984quadratic} as explained earlier. The differential $\varphi^{(S)}$ indeed transforms as a quadratic differential under conformal transformations, even though it comes from $T(z)$ that has an anomalous term in its transformation~\eqref{eq:StressEnergyTransform}. This is because the anomalous part is independent of $\lambda$, see~\eqref{eq:StrebelNatural}. We note in passing that the critical graph of a Strebel differential is an example of \textit{Stokes curves} from the perspective of the theory of Fuchsian equations~\cite{takei2017wkb}.

As a corollary to~\eqref{eq:StrebelNatural}, we see that the accessory parameters of Strebel differentials $c^{(S)}_i$ are related to the accessory parameters of the Fuchsian equation~\eqref{eq:Fuchsian2} that realize a real hyperbolic monodromy around each puncture via
\begin{align} \label{eq:WKBRES}
\boxed{
	c^{(S)}_i = -2  \lim_{\lambda \to \infty} {c_i \over \lambda^2} \, ,
}
\end{align} 
and they satisfy
\begin{align}  \label{eq:StrebelCond}
\sum_{i=1}^n c^{(S)}_i = 0,
\quad
\sum_{i=1}^n (-\alpha_i^2 + c^{(S)}_i  \xi_i) = 0,
\quad
\sum_{i=1}^n (- 2\alpha_i^2 \xi_i + c^{(S)}_i  \xi_i^2) = 0,
\end{align} 
by taking the WKB limit of~\eqref{eq:Conditions}. These are the expected constraints for Strebel differentials. In passing, we remark that the reasoning here can be viewed as an alternative argument for why Strebel differentials exist and are unique on a given punctured Riemann sphere: Thurston metrics (without internal flat cylinders) exist and are unique by the uniformization theorem, so too Strebel differentials.\footnote{The uniqueness of Thurston metrics with internal flat cylinders hasn't been rigorously established, see~\cite{Costello:2019fuh}.}

Lastly, take note that the connection formulas are no longer necessary to construct the local coordinates using~\eqref{eq:ScaledRatioWKB} in the WKB limit, as the monodromy behavior around each puncture is correctly accounted, see~\eqref{eq:ScaledRatioWKB}. However, this issue is replaced with the problem of placing a lower bound on the integral in~\eqref{eq:ScaledRatioWKB} so that $|w_i| = 1$ when the coordinate patches overlap. This lower bound should be a point on the critical graph~\cite{Erbin:2022rgx}. Luckily, this is not as hard as finding the connection formulas for Fuchsian equations. It can be easily done after solving the accessory parameters $c_i^{(S)}$ and finding the zeros of the resulting quadratic differential~\eqref{eq:ScaledRatioWKB}. They always lie on the critical graph and can be used as a lower bound for the integral~\eqref{eq:ScaledRatioWKB}.

\section{The Polyakov conjecture for Strebel differentials} \label{sec:Polyakov}

In the previous section, we have related the accessory parameters of Strebel differentials to those of the Fuchsian equation solving the hyperbolic monodromy problem. It turns out that the latter can be generated using an on-shell Liouville action by the \textit{Polyakov conjecture}. In this section, we argue for the Polyakov conjecture for hyperbolic singularities and what it entails for Strebel differentials and CSFT. The contents here are partially based on~\cite{Hadasz:2003kp, Hadasz:2003he, Hadasz:2005gk} but we provide an alternative argument based on a slight modification to the geometric formulation of Liouville theory that formalizes the conjecture.

We begin by reviewing Liouville theory in subsection~\ref{sec:Liouville}. For deeper exposition reader should refer to numerous excellent reviews~\cite{seiberg1990notes, Teschner:2001rv, Nakayama:2004vk, Harlow:2011ny, erbin2015notes,Ribault:2014hia}. Here we slightly modify the conventional Liouville theory by making the cosmological constant position-dependent to formally argue for the Polyakov conjecture for hyperbolic singularities in subsection~\ref{sec:PC} by considering the semi-classical limit of this modified theory. In subsection~\eqref{sec:PCSD}, we take the WKB limit and argue for the Polyakov conjecture for Strebel differentials. Lastly, we check the symmetry properties of the conjecture as well as benchmark against to the cases of 3-and 4-punctured spheres in subsection~\eqref{sec:Test}.

\subsection{Liouville theory and the position-dependent cosmological constant} \label{sec:Liouville}

The action of Liouville theory on an $n$-punctured sphere $\Sigma_{0,n}$ coupled to a reference metric $\tilde{g}$ is given by~\cite{Harlow:2011ny}
\begin{align} \label{eq:Lio}
S_L'[\phi] = {1 \over 4\pi} \int \sqrt{\tilde{g}} \; dx \wedge dy  \left[
\tilde{g}^{a b} \p_a \phi \p_b \phi + Q \tilde{R} \phi + 4 \pi \mu e^{2 b \phi}
\right] \, ,
\end{align}
Here $Q$ is the \textit{background charge} and the tilded quantities refer to the reference metric $\tilde g$. The last term in the action is the \textit{cosmological constant} term with $\mu \geq 0$, and $b$ is a parameter that would eventually be related to the background charge $Q$. The exponential operator is defined by the normal ordering using the distances defined by the reference metric $\tilde g$.

Liouville theory is invariant under the diffeomorphisms of the reference metric $\tilde g$ as well as the Weyl invariance of the form
\begin{subequations}
\begin{align}
	&\tilde{g}'_{ab}(x,y) = e^{2 \omega(x,y) } \tilde{g}_{ab}(x,y) \, , \label{eq:EmergentWeyl2} \\
	&\phi'(x,y) = \phi(x,y) - Q  \omega(x,y) \label{eq:EmergentWeyl1} \, .
\end{align}
\end{subequations}
This combination leaves the physical metric
\begin{align} \label{eq:Physmet}
g_{ab} = e^{{2 \over Q} \phi} \tilde{g}_{ab} \, ,
\end{align}
invariant. Classically, the transformation~\eqref{eq:EmergentWeyl1} reduces to the conformal transformation of $\phi$ when~\eqref{eq:EmergentWeyl2} is a diffeomorphism of $g'_{ab}$.  Liouville theory is a classical CFT with the choice $Q = 1/b$ as a result. Of course, these statements get modified in the quantum theory.

Before we fix the reference metric let us do something unusual: give $\mu$ position-dependence by demanding $\mu = 0$ for some non-overlapping simply-connected regions $H_i$ around $n$ punctures at $z=\xi_i$ for $i = 1, \cdots, n$ and a positive constant $\mu > 0$ elsewhere. That is,
\begin{align} \label{eq:cc}
\mu(x,y) = 
\begin{cases}
0  & \quad \text{for} \quad (x,y) \in H_i \\
\mu &\quad \text{for} \quad (x,y) \in \mathbb{CP}^1 \setminus \bigcup\limits_{i=1}^n H_i \equiv R \neq \emptyset
\end{cases} \, ,
\end{align}
with $H_i \cap H_j = \emptyset$ (except possibly at their boundary) when $i \neq j$. Here $R \neq \emptyset$ denotes the rest of the surface with the simply connected domains $H_i$ cut out. Having a position-dependent $\mu$ doesn't change the dynamics of Liouville theory, in particular it being a CFT\footnote{This is classically true, but there may be subtleties for the quantum theory. We are not going to delve into them and~\textit{assume} such a CFT makes sense, at least semi-classically.}. We simply treat $\mu$ as a background field that is restricted to be of the form~\eqref{eq:cc}. Notice $\mu$ has to be a scalar $\mu(x, y) = \mu' (x',y')$ under diffeomorphisms so that to keep the Liouville action~\eqref{eq:Lio} invariant. We indicate this modification by $S_L'[\phi; \mu]$.

Now let us set the reference metric to be flat, $\tilde{g}_{ab} = \delta_{ab}$, and switch to the complex coordinates $z = x+ iy$. The action~\eqref{eq:Lio} becomes\footnote{We only consider the holomorphic part of Liouville theory. The anti-holomorphic part behaves analogously. The measure of the action is defined by $d^2 z \equiv i/2 \, dz \wedge d\overline{z} = dx \wedge dy$.}
\begin{align} \label{eq:Lio1}
S_L'[\phi; \mu] =
&
{1 \over \pi} \left[ \; \int\limits_R  d^2z \left( \p \phi \, \overline{\p} \phi + \pi \mu e^{2 b \phi} \right) +
\sum_{i=1}^n \int\limits_{H_i} d^2 z \, \p \phi \, \overline{\p} \phi \; \right] 
\, . 
\end{align}
The subscript of the integrals indicate the regions where $\mu$ is zero and different from zero. The resulting equation of motion is given by
\begin{subequations} \label{eq:EOM}
\begin{align}
	\p \overline{\p} \phi = \pi \mu b e^{2 b \phi} &\quad \text{for} \quad z \in R = \mathbb{CP}^1 \setminus \bigcup\limits_{i=1}^n H_i, \\
	\p \overline{\p} \phi = 0 &\quad \text{for} \quad z \in H_i.
\end{align}
\end{subequations}
Now the motivation behind considering $\mu = \mu(z, \overline{z})$ like in~\eqref{eq:cc} is apparent: the physical metric~\eqref{eq:Physmet} can be a Thurston metric. That is, unlike Liouville theory where the equation of motion (Liouville's equation) imposes the physical metric to be of negative constant curvature~\textit{everywhere} on the punctured Riemann sphere, we may have disjoint regions on the surface determined by $\mu$ where the metric is flat now.

In order to Thurston metrics to solve~\eqref{eq:EOM} we need to impose correct asymptotic behavior. This involves demanding regularity at $z=\infty$ and making sure that the physical metric describes a flat semi-infinite cylinder around the punctures $z= \xi_i$. Begin with demanding regularity at $z= \infty$. Assuming no punctures are placed at $z=\infty$ we must have
\begin{align} \label{eq:BC1}
\phi \to - 2 Q \log|z| + \cdots \quad \text{as} \quad |z| \to \infty 
\implies
ds^2 = e^{2\phi/Q} |dz|^2  \to {|dz|^2 \over |z|^4} 
\, ,
\end{align}
up to a multiplicative constant. This is desired because we have $ds^2 \to |dz|^2$ up to a multiplicative constant after the inversion $z \to 1/z$ so that the physical metric isn't singular at $z= \infty$. Similarly, having a semi-infinite flat cylinder around the puncture at $z=\xi_i$ imposes
\begin{align} \label{eq:BC2}
\phi \to - Q \log|z-\xi_i| + \cdots \quad \text{as} \quad |z-\xi_i| \to 0 
\implies
ds^2 = e^{2\phi/Q} |dz|^2 \to {|dz|^2 \over |z-\xi_i|^2}
\, ,
\end{align}
up to a multiplicative constant.

So we modify the action~\eqref{eq:Lio1} by including boundary terms in order to impose the asymptotics~\eqref{eq:BC1} and~\eqref{eq:BC2}:
\begin{align} \label{eq:Lio2}
S_L'[\phi; \mu] =
&\lim_{\epsilon \to 0} \Bigg[
{1 \over \pi}\int\limits_{R^{1/\epsilon} } d^2z \left[ \p \phi \, \overline{\p} \phi + \pi \mu e^{2 b \phi} \right] 
+{\epsilon \, Q \over \pi} \int\limits_{|z|=1/\epsilon} |dz| \phi - 2 Q^2 \log \epsilon \nonumber \\
&\quad \quad \quad
+ \sum_{i=1}^n \bigg[
{ 1 \over \pi} \int\limits_{H_i^\epsilon} d^2 z \, \p \phi \, \overline{\p} \phi 
+ 
{Q \over 2 \pi \epsilon}\int\limits_{|z - \xi_i| = \epsilon} |dz| \phi
+ {Q^2 \over 2} \log \epsilon
\bigg] \Bigg]
\, . 
\end{align}
Here $\epsilon$ on the sets indicate the original sets after excluding a small circle around either $z = \infty$ or punctures $z=\xi_i$, whose orientations are induced by $R^{1 / \epsilon} \cup H_i^\epsilon$. More specifically,
\begin{align}
H_i^\epsilon \equiv H_i \setminus \{ z \in H_i \; | \; |z-\xi_i| < \epsilon \} \, ,
\quad \quad \quad 
R^{1 / \epsilon}  \equiv R \setminus\left\{ z \in R \; | \; |z| > {1 \over \epsilon} \right\} \, .
\end{align}
Using the same $\epsilon> 0$ doesn't make a difference: we can always make $\epsilon$ small enough so that the circle around infinity excludes all $H_i$. The boundary terms impose the asymptotic behaviors~\eqref{eq:BC1} and~\eqref{eq:BC2} through
\begin{align}
	{1 \over 2 \epsilon} \, {\p \phi \over \p |z| }  + Q = 0 \quad \text{at} \quad |z|={1 \over \epsilon} ,
	\hspace{0.5in} \text{and} \hspace{0.5in}
	\epsilon \,  {\p \phi \over \p |z- \xi_i| } + {Q \over 2} = 0 \quad \text{at} \quad |z - \xi_i| = \epsilon
	\, .
\end{align}
from the variation of $\phi$. We have also included the constant terms to~\eqref{eq:Lio2} to guarantee that the on-shell action doesn't diverge due to the metric having the asymptotic behaviors~\eqref{eq:BC1} and~\eqref{eq:BC2}.

Let us begin considering the quantum theory and its operators. Possibly the most important operator is the stress-energy tensor $T_L$ of Liouville theory given by
\begin{align} \label{eq:LioT}
T_L(z) = - (\p \phi)^2 + Q \p^2 \phi \, ,
\end{align}
which results in the central charge $c = 1 + 6 Q^2 $. Another important set of local operators is the exponential operators defined by
\begin{align} \label{eq:ExpOp}
V_\beta (z, \overline{z}) = e^{2 \beta \phi(z,\overline{z})} \, .
\end{align}
Classically they transform under the conformal transformation $z \to \widetilde{z}$ as
\begin{align}
V_\beta(z, \overline{z}) 
= \left({\p \tilde{z} \over \p z}\right)^{\beta Q} \left({\p \overline{\tilde{z}} \over \p \overline{z}}\right)^{\beta Q} V_\beta(\tilde{z}, \overline{\tilde{z}}) \, ,
\end{align}
by~\eqref{eq:EmergentWeyl1}, so they are primaries. The weights of these operators read $\beta Q$ classically, but this gets modified in quantum theory. Instead we have $\Delta = \beta \; (Q-\beta) $: the addition of $-\beta^2$ is coming due to normal ordering. In particular, $V_b(z,\overline{z})$ appearing in the action~\eqref{eq:Lio} must have its weight equal to $1$ in order to define a conformally invariant quantum theory (up to a possible $c$-number anomaly). This modifies the relation between $Q$ and $b$ to be
\begin{align}
	Q = b + {1 \over b} \, .
\end{align}
Take note that none of these arguments required  $\mu$ to be a constant.

Next, let us consider the partition function based on the action~\eqref{eq:Lio2}
\begin{align} \label{eq:PartFunc1}
Z[\mu] = \int [d \phi] \, e^{- S_L'[\phi; \mu]} \, ,
\end{align}
where we integrate over all conformal factors $\phi$ over $\Sigma_{0,n}$ that have the asymptotic behavior~\eqref{eq:BC1} and~\eqref{eq:BC2}. The measure won't be needed for our formal arguments so we don't attempt to define it.

Let us separate the term in $S_L'[\phi; \mu]$ that imposes regularity at the punctures and reinterpret~\eqref{eq:PartFunc1} as a correlator of these ``puncture'' operators. That is, we have
\begin{align}
\exp\left[ {Q \over 2 \pi \epsilon}\int_{|z - \xi_i| = \epsilon} |dz| \phi \right] = 
\exp\left( Q \; \phi(\xi_i, \overline{\xi_i}) \right) = V_{Q/2}(\xi_i, \overline{\xi_i}) \, ,
\end{align}
in $Z[\mu]$ and we interpret adding a puncture at $z=\xi_i$ to be inserting $V_{Q/2}(\xi_i,\overline{\xi_i})$, akin to the parabolic singularities (i.e cusps)~\cite{Hadasz:2005gk}. Notice that this operator has conformal weight $\Delta = Q^2 / 4$, since $\beta = Q/2$.

It is also possible to consider the correlation functions of more exotic insertions, especially given that $\mu$ is position-dependent like in~\eqref{eq:cc}. One of these exotic insertions is the mapping radius associated with the simply connected region $H_i$ sans the puncture
\begin{align}
r_i[H_i] \equiv \left| {dz \over dw_i} \right|_{w_i = 0} \, ,
\end{align}
obtained from the conformal map $z= z(w_i)$, $\xi_i = z(0)$ from an unit disk $0 < |w_i| \leq 1$ to $H_i$. The mapping radii of the regions $H_i$ is in general some complicated functionals of $\mu$ as the latter carries the same information as $H_i$, so $r_i[\mu] = r_i[H_i]$. Nevertheless, it is perfectly well-defined and one can insert such objects in correlators, acting as a background field in effect. Observe that the mapping radius changes as a primary of weight $-1/2$ under the conformal transformation $z \to \widetilde{z}$
\begin{align}
r_i[\mu] = \left({\p \tilde{z} \over \p z}\right)^{-1/2} 
\left({\p \overline{\tilde{z}} \over \p \overline{z}}\right)^{-1/2}
\widetilde{r_i}[\widetilde{\mu}] \, .
\end{align}
This fact holds true in quantum theory since $\mu$ is non-dynamical. 

Given our remarks above, we specialize to the following correlator
\begin{align} \label{eq:PI}
	\left< \Sigma_{0,n} \right> \equiv
	\left< \prod_{i=1}^n V_{Q/2}(\xi_i, \overline{\xi_i}) \; r_i[H_i]^{-Q^2 \lambda_i^2 /2} \right>
	= \int [d \phi]\, e^{- S_L[\phi; \mu]} \, .
\end{align}
Here $\lambda_i$ are some collections of real numbers. Above, we have collected the integrand of the path integral into the exponential by defining
\begin{align} \label{eq:Lio3}
S_L[\phi; \mu] = S_L'[\phi;\mu] + {Q^2 \over 2} \sum_{i=1}^n \lambda_i^2 \log r_i[H_i] \, ,
\end{align}

A final thing to note that the puncture operators, together with the insertions of mapping radii as above, have the conformal weights greater than $Q^2/4$
\begin{align} \label{eq:FinalWeights1}
\Delta_i = 
{Q^2 \over 2} \left({1 \over 2} + {\lambda_i^2 \over 2} \right)=
\beta (Q-\beta)
\quad \text{where} \quad
\beta = {Q \over 2} (1+i \lambda)
 \, .
\end{align}
So instead of calling them ``puncture'' operators it is more appropriate to call them ``hole'' operators, as they are going to be closely related to having flat cylinders in the geometry in the semi-classical limit. In fact, we can understand the justification behind the inclusion of the mapping radii factors based on these weights---it is natural to expect the primaries of weights~\eqref{eq:FinalWeights1} to describe holes on the Riemann surface in the semi-classical limit (claimed and checked in~\cite{Hadasz:2005gk, Hadasz:2006rb}) based on their elliptic and parabolic counterparts. It appears straightforward to achieve this by simply making $\mu$ position-dependent like in~\eqref{eq:cc} and inserting mapping radii. This was the main motivation behind our modification of Liouville theory.

\subsection{The Polyakov conjecture for hyperbolic singularities} \label{sec:PC}

Let us consider the semi-classical limit of $S_L[\phi;\mu]$. Per usual, the semi classical limit is $c \to \infty$ (or $Q \to \infty, b \to 0$), while keeping $K \equiv - 4 \pi \mu b^2$ and the physical metric~\eqref{eq:Physmet} fixed. As a result the classical $\phi_c \equiv 2 b \phi$ has a fixed limit and the action $S_L[\phi;\mu]$ evaluates to
\begin{align} \label{eq:Lio4}
&S_L[\phi_c; K] =
{1 \over 2b^2} \lim\limits_{\epsilon \to 0} \Bigg[
{1 \over 2\pi}\int\limits_{R^{1/\epsilon}} d^2z \left[ \p \phi_c \, \overline{\p} \phi_c- K e^{\phi_c} \right] 
+{\epsilon \over \pi} \int\limits_{|z|=1/\epsilon} |dz| \phi_c - 4 \log \epsilon \nonumber \\
&\quad 
+ \sum_{i=1}^n \bigg[
{ 1 \over 2\pi} \int\limits_{H_i^\epsilon} d^2 z \, \p \phi_c \, \overline{\p} \phi_c
+ 
{1 \over 2 \pi \epsilon}\int\limits_{|z - \xi_i| = \epsilon} |dz| \phi_c
+ \log \epsilon 
\bigg] 
+ \sum_{i=1}^n \lambda_i^2 \log r_i[H_i]
+ \mathcal{O} (b^2) \Bigg]
\, . 
\end{align}
From~\eqref{eq:EOM}, we see that $K$ is the Gaussian curvature of the physical metric $ds^2 = e^{\phi_c} |dz|^2$ on the region $R$ on-shell. We set the profile of $\mu$ so that $K = -1$ on $R$.

Let us investigate the stationary points of~\eqref{eq:Lio4} as we can use them to evaluate the correlator~\eqref{eq:PI} semi-classically. First of all, we remark that \textit{not} every profile of $\mu$ results in a stationary point of the action~\eqref{eq:Lio4}---only certain profiles do. This is due to the existence of the Thurston metric discussed in section~\ref{sec:WKB}. As we mentioned there, Thurston metrics have hyperbolic ($R$) and flat portions ($H_i$). A priori, not every choice of $R$ and $H_i$ can be endowed with a Thurston metric. So we restrict our attention to profiles for which a stationary point of the action~\eqref{eq:Lio4} exist. As we shall see, the explicit knowledge of these profiles won't be needed for our arguments, their existence is sufficient.

Denoting a stationary point of the action $S_L[\phi_c; K] $ by $\varphi$, the correlator~\eqref{eq:PI} evaluates to
\begin{align} \label{eq:PartFunc2}
\left<\Sigma_{0,n} \right> \sim \exp\left[ {- {Q^2 \over 2} S_{HJ}[\varphi]} \right] \, ,
\end{align}
in the semi-classical limit. Here $\sim$ indicates that the equality holds up to an overall normalization.  Above we have defined the \textit{on-shell Hadasz-Jask\'olski (HJ) action} $S_{HJ}[\varphi]$ after its originators~\cite{Hadasz:2003kp}:
\begin{align} \label{eq:Lio5}
&S_{HJ}[\varphi] =
\lim\limits_{\epsilon \to 0} \Bigg[
{1 \over 2\pi}\int\limits_{R^{1/\epsilon}} d^2z \left[ \p \varphi \, \overline{\p} \varphi + e^{\varphi} \right] 
+{\epsilon \over \pi} \int\limits_{|z|=1/\epsilon} |dz| \varphi - 4 \log \epsilon \nonumber \\
&\quad \quad \quad
+ \sum_{i=1}^n \bigg[
{ 1 \over 2\pi} \int\limits_{H_i^\epsilon} d^2 z \, \p \varphi \, \overline{\p} \varphi
+ 
{1 \over 2 \pi \epsilon}\int\limits_{|z - \xi_i| = \epsilon} |dz| \, \varphi
+ \log \epsilon 
\bigg] 
+ \sum_{i=1}^n \lambda_i^2 \log r_i[H_i] \Bigg]
\, . 
\end{align}

Let us consider the conformal Ward identity for~\eqref{eq:PI} and take the semi-classical limit  to argue for the Polyakov conjecture. The conformal Ward identity reads
\begin{align} \label{eq:ConfWard1}
\left< T_L(z) \Sigma_{0,n} \right> =
\sum_{i=1}^n \left[ {\Delta_i \over (z-\xi_i)^2} + {1 \over z-\xi_i} {\p \over \p \xi_i}\right]
\left< \Sigma_{0,n} \right> \, ,
\end{align}
with $\Delta_i \sim Q^2$ are given as in~\eqref{eq:FinalWeights1}. We can evaluate $\left< \Sigma_{0,n} \right> $ as $c \to \infty$ and the result is in~\eqref{eq:PartFunc2}. As we mentioned before, the saddle points exist in certain profiles of $\mu$, so we adjust its shape appropriately to make sure a saddle point appears. In particular we make it so that the lengths of the geodesic \textit{seams} where hyperbolic and flat metrics touch are $2 \pi \lambda_i$.

Similarly we can evaluate $\left< T_L(z) \Sigma_{0,n} \right> $ as $c \to \infty$. The result is
\begin{align} \label{eq:TSigma1}
\left< T_L(z) \Sigma_{0,n} \right> \sim 
T_L^{cl}(z)
\exp\left[-{Q^2 \over 2} S_{HJ}[\varphi]  \right] \, ,
\end{align}
as the correlator is expected to factorize in this limit. Here $T_L^{cl}(z) = \langle T_L(z) \rangle$ is the classical stress-energy tensor~\eqref{eq:LioT} evaluated at the saddle-point $\varphi$:
\begin{align} \label{eq:fact}
	T_L^{cl}(z) \equiv {Q^2 \over 2} \left[-{1 \over 2} (\p \varphi)^2 + \p^2 \varphi \right]
	= {Q^2 \over 2} \sum_{i=1}^n \left[ {\delta_i \over (z-\xi_i)^2} + {c_i \over z-\xi_i} \right]
	  \, .
\end{align}
The second equality above has been argued above~\eqref{eq:Fuchsian2}. We would like to solve for the accessory parameters $c_i$. Using \eqref{eq:PartFunc2}, \eqref{eq:ConfWard1}, \eqref{eq:TSigma1}, and \eqref{eq:fact} together we have
\begin{align} \label{eq:Polyakov1}
	c_i = - {\p  S_{HJ}[\varphi] \over \p \xi_i} \, .
\end{align}
This was the main result of~\cite{Hadasz:2003kp} and it is \textit{the Polyakov conjecture for hyperbolic singularities}. Of course, the arguments here were mostly formal, nonetheless it can be shown that~\eqref{eq:Polyakov1} holds true by a direct computation like in~\cite{Hadasz:2003kp}, which we briefly summarize in appendix~\ref{app:direct}. So even though certain properties haven't justified rigorously (such as the normalization factors in~\eqref{eq:PartFunc2} and~\eqref{eq:TSigma1} as well as the factorization for the latter), it appears that the final result doesn't get affected.

Let us perform some post-derivation justifications for~\eqref{eq:Polyakov1}. First, take note that the accessory parameters are constrained according to~\eqref{eq:Conditions} as a consequence of the regularity of the stress-energy tensor~\eqref{eq:LioT} at $z= \infty$. Next, let us try to make sense of the mapping radii terms in~\eqref{eq:Lio5} by considering the metric on $H_i$. The (flat) metric on $H_i$ is given by
\begin{align} \label{eq:Flat}
ds^2 = e^\varphi |dz|^2 =  \lambda_i^2 \, \bigg| {d \rho_i \over \rho_i} \bigg|^2 
\implies \varphi = \log\left[{\lambda_i^2 \over |\rho_i(z)|^2 } \bigg|{d \rho_i(z) \over dz  }\bigg|^2 \right]
\, ,
\end{align}
with $\rho_i$ is the scaled ratio~\eqref{eq:ScaledRatio}. Note that the overall scale of $\rho_i$ drops out, so we can replace it with the local coordinates $0 < |w_i| \leq 1$ which shows
\begin{align}
\int\limits_{H_i^\epsilon} d^2z \; e^\varphi = 
\int\limits_{H_i^\epsilon} d^2z \; {\lambda_i^2 \over |w_i(z)|^2 } \bigg|{d w_i(z) \over dz  }\bigg|^2 =
\lambda_i^2 \int\limits_{w_i^{-1}(H_i^\epsilon)} {d^2w_i  \over |w_i|^2} =
2 \pi \lambda_i^2 \int_{\epsilon/r_i[H_i]}^1 {d r \over r} + \mathcal{O}(\epsilon) \, .
\end{align}
This subsequently implies
\begin{align} \label{eq:red}
\lim_{\epsilon \to 0} \left[
{1 \over 2 \pi}\int_{H_i^\epsilon} d^2z e^\varphi  + \lambda_i^2 \log \epsilon	\right]
= \lambda_i^2 \log r_i[H_i] \, .
\end{align}
So the mapping radii terms in~\eqref{eq:Lio5} seem to provide the cosmological constant terms that were missing for the regions $H_i$ in $S_{HJ}[\varphi]$. From this perspective it is quite natural to include them. In fact, the equation~\eqref{eq:red} shows that the sum of them is \textit{the reduced area} by definition, see~\cite{Saadi:1989tb}.

\subsection{The WKB limit of the Polyakov conjecture} \label{sec:PCSD}

Combining the results~\eqref{eq:WKBRES} and~\eqref{eq:Polyakov1} we argue for a Polyakov-like conjecture for Strebel differentials in this subsection. Begin with $S_{HJ}[\varphi]$ in~\eqref{eq:Lio5} and focus on the part that has a dependence on $\xi_i$. This is given by, 
\begin{align} \label{eq:GetRidofeps1}
&S_{HJ}[\varphi] = 
{1 \over 2 \pi}\int\limits_{R} d^2z \left[ \p \varphi \, \overline{\p} \varphi + e^{\varphi} \right] 
+  \lim_{\epsilon \to 0}  \sum_{i=1}^n  \bigg[
{ 1 \over 2 \pi} \int\limits_{H_i^\epsilon} d^2 z \, \p \varphi \, \overline{\p} \varphi 
+ \lambda_i^2 \log r_i[H_i]
\bigg] + \cdots
\, . 
\end{align}
Notice that all boundary and the constant $\log \epsilon$ terms drop out as $\epsilon \to 0$ since the leading order asymptotics are independent of $\xi_i$.

Focus on the terms integrated over $R$ in~\eqref{eq:GetRidofeps1}. We claim that these terms vanish upon differentiating with respect to $\xi_i$ at the leading order (that is, $\mathcal{O}(\lambda^2)$) in the WKB limit. In order to argue for this first recall that the Gauss-Bonnet theorem implies
\begin{align} \label{eq:GB0}
n-3 = {1 \over 2 \pi} \int\limits_{R} dA = {1 \over 2 \pi}  \int\limits_{R} d^2z \, e^\varphi \, ,
\end{align}
given the hyperbolic metric on $R$ is $ds^2 = e^\varphi |dz|^2$. This holds as $\lambda \to \infty$ and it stays independent of the position of punctures, hence giving a vanishing derivative with respect to $\xi_i$.

Furthermore, since $\varphi$ is determined by~\eqref{eq:SingHypMet} we see that
\begin{align} \label{eq:GB}
n-3 = {\lambda_i^2 \over 2 \pi} \int\limits_{R} d^2z \, {|\p \rho_i|^2 \over |\rho_i|^2 \, \sin^2\left(\lambda_i \log |\rho_i| \right)} \, .
\end{align}
This must hold as $\lambda \to \infty$ and this is only possible if $\log|\rho_i|$ vanishes identically to prevent infinite oscillation (the rest of the integrand does not oscillate). Therefore
\begin{align}
|\rho_i(z) | \to 1 \quad \text{as} \quad \lambda \to \infty \, ,
\end{align}
for $z \in R$, otherwise the finite limit can't exist. This argument holds for each $i = 1, \cdots, n$, so the hyperbolic part of the surface reduces to the critical graph $\mathbb{CG} $ of a Strebel differential  $\varphi^{(S)}$~\eqref{eq:Str}:
\begin{align}
\mathbb{CG} = \big\{ z \in \mathbb{CP}^1 \big| \lim_{\lambda \to \infty}|\rho_i(z)| = 1 
\quad \text{for} \quad i = 1, \cdots, n \big\} 
= \bigcup_{i=1}^n \p H_i \, ,
\end{align}
Clearly $\mathbb{CG}$ describes the collection of geodesic seams, so in fact we can take
\begin{align} \label{eq:SeamLimit}
	\lambda_i \log |\rho_i(z)| \to {\pi \over 2} \quad \text{as} \quad \lambda \to \infty \, ,
\end{align}
for $z \in \mathbb{CG}$ after adjusting the branches for $\rho_i$ so that the geodesic seams are at $|\rho_i| = \exp(-\pi/2\lambda_i)$.

Now note that
\begin{align}
\p \varphi = -\lambda_i  \cot(\lambda_i \log |\rho_i|)  { \p \rho_i \over \rho_i} + \mathcal{O}(\lambda^0) \, , 
\end{align}
using~\eqref{eq:SingHypMet} as $\lambda \to \infty$. This shows
\begin{align} \label{eq:Expan}
{1 \over 2 \pi}\int\limits_{R} d^2z \; \p \varphi \, \overline{\p} \varphi 
&= {\lambda_i^2 \over 2 \pi} \int\limits_{R} d^2z\cot^2(\lambda_i \log |\rho_i|) { |\p \rho_i|^2 \over |\rho_i|^2}  + \cdots \, .
\end{align}
Dots above indicate the terms that can't have possible $\mathcal{O}( \lambda^2)$ contributions. They contain contributions $\lambda_i \cot (\lambda \log |\rho_i(z)|)$ coming from the cross-terms of the product $\p \varphi \overline{\p} \varphi$. In the view of~\eqref{eq:SeamLimit} they can't give $\mathcal{O}( \lambda^2)$ contributions upon integration, hence they are subleading in $\lambda$.

Looking at the leading term in~\eqref{eq:Expan} we observe
\begin{align}
	{\lambda_i^2 \over 2 \pi} \int\limits_{R} d^2z\cot^2(\lambda_i \log |\rho_i|) { |\p \rho_i|^2 \over |\rho_i|^2}  \leq 
	{\lambda_i^2 \over 2 \pi} \int\limits_{R} d^2z \, { |\p \rho_i|^2 \over |\rho_i|^2 \sin^2(\lambda_i \log|\rho_i|)} = n-3 \, .
\end{align}
So this term cannot be $\mathcal{O}( \lambda^2)$ as well. Combining with~\eqref{eq:GB0}, we conclude that the contributions from $R$ terms in the Polyakov conjecture are subleading in $\lambda$.

Next, look at the $H_i^\epsilon$ contributions as $\lambda_i \to \infty$. But this is clearly subleading in $\lambda$ as can be seen from identities~\eqref{eq:Coll}. So their derivatives with respect to $\xi_i$ are subleading in $\lambda$ as well. Thus we are only left with the modulus
\begin{align} \label{eq:Reducing}
S_{HJ}[\varphi] = \lambda^2 \sum_{i=1}^n \alpha_i^2 \log r_i[H_i] + \cdots =  \lambda^2 \mathcal{S}_{0,n}^\ast(\xi_i; \alpha_i) + \cdots \,. 
\end{align}
at the leading order in $\lambda$. We put a star on $\mathcal{S}_{0,n}$ as the mapping radii appearing in $S_{HJ}[\varphi]$ are now those associated with Strebel differentials since the regions $H_i$ become the faces of critical graphs. 

Considering we have~\eqref{eq:WKBRES} and~\eqref{eq:Polyakov1}, we immediately deduce that the accessory parameters of Strebel differentials satisfy
\begin{align} \label{eq:StrebelPolyakov1}
\boxed{
	c_j^{(S)} = 2 \, {\p \over \p \xi_j} \sum_{i=1}^n \alpha_i^2 \log  r_i[H_i] = 2 \, {\p \mathcal{S}_{0,n}^\ast \over \p \xi_j} (\xi_i; \alpha_i).
}
\end{align}
We call the equation~\eqref{eq:StrebelPolyakov1} \textit{the Polyakov conjecture for Strebel differentials} and it is one of the central results of this paper. In the subsequent subsection we are going to benchmark this conjecture against the known Strebel differentials. 

We have already commented in the introduction that this relation is proven \textit{rigorously} in geometric function theory, see~\cite{kuz1997methodsI, kuz1997methodsII, solynin1999moduli, solynin2009quadratic, solynin2020fingerprints, bakhtin2022generalized, jenkins1954recent,kuz1982problem,fedorov1982maximum, emelyanov2004problem}. Here we argued in the spirit of Liouville theory which forms a bridge between these previously unrelated sub-fields. As we shall see in the next section, Liouville theory perspective to~\eqref{eq:StrebelPolyakov1} allows us to bootstrap the modulus $\mathcal{S}_{0,n}^\ast$ .

Let us close this subsection by emphasizing the implications of the appearance of $\mathcal{S}_{0,n}^\ast$ in~\eqref{eq:StrebelPolyakov1}. The modulus is related to the interaction of zero momentum tachyons through~\eqref{eq:Tachyon} and by~\eqref{eq:StrebelPolyakov1} it follows that all CSFT vertices are determined by the interactions of tachyons. Second, given a punctured Riemann surface and any arbitrary set of simply connected regions containing just a single puncture (call them $D_i$) we have the inequality~\cite{strebel1984quadratic}
\begin{align}
\sum_{i=1}^n \alpha_i^2 \log  r_i[D_i] \leq \sum_{i=1}^n \alpha_i^2 \log  r_i[H_i] =\mathcal{S}_{0,n}^\ast \, ,
\end{align}
where $r_i[D_i]$ is the mapping radius associated with the region $D_i$. This equality is saturated when $D_i = H_i$ for $i=1,\cdots,n$. This is intriguing: the relation~\eqref{eq:StrebelPolyakov1} conjures that the change of the maximum of $\mathcal{S}_{0,n}$ determines the Strebel differential. In fact, this was the idea behind the rigorous proof of~\eqref{eq:StrebelPolyakov1}~\cite{solynin1999moduli}.

Finally, let us comment on the fact that the mapping radius is an ambigious quantity: it is only sensible upon introducing a local coordinate vanishing at the puncture~\cite{Belopolsky:1994bj}. Here the natural set of local coordinates are provided by the shapes of $H_i$. So the modulus $\mathcal{S}_{0,n}$ is unambiguous for each punctured Riemann sphere. This doesn't necessarily mean comparing the modulus for distinct surfaces is unambiguous. However, we compare $\mathcal{S}_{0,n}^\ast$ in~\eqref{eq:StrebelPolyakov1}, the modulus determined by the choice of local coordinates made according to $H_i$, i.e. the faces of the critical graph of a Strebel differential. So this is sensible to do, as there is a unique choice for them for each distinct surface. Therefore $\mathcal{S}_{0,n}^\ast$ is a well-defined function over $\mathcal{M}_{0,n}$.

\subsection{Testing the conjecture} \label{sec:Test}

We establish the symmetry properties of the conjecture~\eqref{eq:StrebelPolyakov1} and demonstrate that they are consistent with our expectations in this subsection. We are interested in how Strebel differentials~\eqref{eq:str} transform under a general \textit{active} global conformal transformation
\begin{align} \label{eq:GC}
z \to \widetilde{z} = f(z) ={a z + b \over c z +d}
\quad \text{where} \quad a,b,c,d \in \mathbb{C}
\quad ad - bc =1 \, .
\end{align}
Here the transformation is active in the sense that we move the punctures themselves, rather than a mere coordinate transformation. After somewhat easy calculation we see\footnote{We drop $(S)$ on the accessory parameters for Strebel differentials as long as there is no chance for confusion.}
\begin{align}
\widetilde{\phi} (\widetilde{z} ) = 
\sum_{i=1}^n \left[
{-\alpha_i^2 \over (\widetilde{z}-f(\xi_i))^2} + {(c \xi_i + d)^2 \left(c_i - {2 c \alpha_i^2 \over c \xi_i +d}\right) \over \widetilde{z}-f(\xi_i)}
\right]
=
\sum_{i=1}^{n} \left[
{-\alpha_i^2 \over (\widetilde{z}-\widetilde{\xi_i})^2} + {\widetilde{c_i} \over \widetilde{z}-\widetilde{\xi_i}} \right] \, .
\end{align}
This implies that we must have
\begin{align} \label{eq:TP}
\xi_i \to \widetilde{\xi_i} = f(\xi_i) = {a \, \xi_i + b \over c \, \xi_i +d} \implies c_i  \to \widetilde{c_i} = (c \, \xi_i + d)^2 \left(c_i - {2 c \, \alpha_i^2 \over c \, \xi_i +d}\right) \, .
\end{align}
Note that having a global conformal transformation was important here since it is well-defined everywhere on the Riemann sphere and the representation~\eqref{eq:str} stays valid.

The transformation property~\eqref{eq:TP} is consistent with the conjecture~\eqref{eq:StrebelPolyakov1}, since the modulus~\eqref{eq:Modulus} transforms under a generic conformal transformation $z \to \widetilde{z}$ as
\begin{align} \label{eq:ModTrans}
\widetilde{S_{0,n}} = S_{0,n} + \sum_{i=1}^n \alpha_i^2 \log\left|{\p \widetilde{z} \over \p z} \right|_{z = \xi_i} \, ,
\end{align}
and for the \textit{global} conformal transformation~\eqref{eq:GC} this entails
\begin{align} \label{eq:MG}
\widetilde{S_{0,n}} = S_{0,n} - 2 \sum_{j=1}^n \alpha_i^2 \log \left|{ c \, \xi_i + d } \right| \, ,
\end{align}
from which we see
\begin{align} \label{eq:Anomaly1}
\widetilde{c_i} =2 {\p \widetilde{S_{0,n}^\ast} \over \p \widetilde{\xi_i}} 
&= 2 {\p \xi_k \over \p \widetilde{\xi_i} } {\p \over \p \xi_k}
\left[
S_{0,n}^\ast - 2 \sum_{j=1}^n \alpha_j^2 \log \left|{ c \, \xi_j + d } \right|
\right]  =  (c \xi_i + d)^2 \left[ c_i - {2 c \alpha_i^2 \over c \xi_i +d}
\right] \, .
\end{align}
The Polyakov conjecture for Strebel differentials is indeed consistent with the transformations~\eqref{eq:TP} and~\eqref{eq:MG}. We note in passing that the conjecture holds true if a moduli is complex conjugated, since $S_{0,n} \in \mathbb{R}$, as well as after permuting the punctures and their associated accessory parameters.

Now let us begin collecting evidence for the conjecture~\eqref{eq:StrebelPolyakov1}. Begin with the three-punctured sphere whose generic Strebel differential is given by
\begin{align} \label{eq:GW}
\varphi^{(S)} = \phi(z) \, dz^2 = {-\alpha_1^2 \over (z - \xi_1)^2} +  {-\alpha_2^2 \over (z - \xi_2)^2} +  {-\alpha_3^2 \over (z - \xi_3)^2} + {c_1 \over z - \xi_1} + {c_2 \over z - \xi_2} + {c_3 \over z - \xi_3} \, .
\end{align}
These satisfy~\eqref{eq:StrebelCond} which fix $c_i$ to be
\begin{align}
	c_1 &=  {\alpha_3^2 - \alpha_1^2 - \alpha_2^2 \over \xi_1 - \xi_2} +
	{\alpha_2^2 - \alpha_1^2 - \alpha_3^2 \over \xi_1 - \xi_3} \, ,
\end{align}
with $c_2, c_3$ are similarly given after cyclically permuting $1,2,3$. In fact a global conformal transformation can fix the positions of $\xi_i$ and turn $c_i$'s independent of $\xi_i$. However we opt out doing this to demonstrate that~\eqref{eq:StrebelPolyakov1} still holds true to test the consistency of the conjecture.

From reverse engineering, it is possible to read out that the Polyakov conjecture~\eqref{eq:StrebelPolyakov1} is satisfied with the following choice for $S_{0,3}^\ast(\xi_i;\alpha_i)$:
\begin{align} \label{eq:3-answer1}
S_{0,3}^\ast(\xi_i;\alpha_i) &= (\alpha_1^2 + \alpha_2^2 - \alpha_3^2 ) \log|\xi_1 - \xi_2| +
(\alpha_1^2 + \alpha_3^2 - \alpha_2^2 )\log|\xi_1 - \xi_3| \nonumber \\
& \quad \quad \quad
+ (\alpha_2^2 + \alpha_3^2 - \alpha_1^2 )\log|\xi_2 - \xi_3| 
+ s(\alpha_1, \alpha_2, \alpha_3)\, .
\end{align}
Here $s$ is a function of $\alpha_i$'s and it is independent of $\xi_i$. It is undetermined by the Polyakov conjecture.

Let us see how $S_{0,3}^\ast(\xi_i;\alpha_i)$ can be determined using the mapping radii. We can do this as follows. First, place the punctures $\xi_1, \xi_2, \xi_3$ at $0,1,\infty$ respectively. This is done by the (active) global conformal transformation
\begin{align}
z \to \tilde{z} = {\xi_2 - \xi_3 \over \xi_2 - \xi_1} {z - \xi_1 \over z - \xi_3},
\hspace{0.5in}
{d \tilde{z} \over d z } = - {(\xi_1-\xi_3)(\xi_2 - \xi_3) \over (\xi_1 - \xi_2) (z- \xi_3)^2}
\, .
\end{align}
where we also included its derivative above. After this transformation, the modulus is just a function of $\alpha_i$, that is $\widetilde{\mathcal{S}_{0,3}^\ast}(\alpha_i) = s(\alpha_1,\alpha_2,\alpha_3)$. But recall that the modulus transforms like in~\eqref{eq:ModTrans}  and we have just evaluated both terms on the right-hand side. These show $\mathcal{S}_{0,3}(\xi_i;\alpha_i)$ indeed given by~\eqref{eq:3-answer1}. In the hindsight, it is not surprising that the global conformal symmetry has entirely fixed the form of the conjecture here: there were no undetermined accessory parameters. Calculating $\widetilde{\mathcal{S}_{0,3}} = s(\alpha_1,\alpha_2,\alpha_3)$ takes a non-trivial effort, which we do in section~\ref{sec:recursion}.

Now let us consider 4-punctured spheres with $\alpha_i = 1$. This is the first check of~\eqref{eq:StrebelPolyakov1} beyond the symmetry considerations. The accessory parameters in this case are only obtained numerically in the past, see~\cite{Belopolsky:1994bj,Moeller:2004yy,Erbin:2022rgx}. However, if the positions of punctures are restricted to $\{\xi,0,1,\infty\}$ with $\xi \in \mathbb{R}$ it is possible to obtain a closed-form expression for the accessory parameter $c$ associated with $\xi$ is given by~\cite{Erbin:2022rgx}
\begin{align} \label{eq:Arctan}
c \, (\xi = \overline{\xi}) 
= \begin{cases} \vspace{0.1in}
	{-2 \over \xi (\xi-1)}  & \quad \text{for} \quad \xi \leq 0 \\ \vspace{0.1in}
	{4\xi -2 \over \xi (\xi-1)}  & \quad \text{for} \quad  0 \leq \xi \leq 1 \\
	{2 \over \xi (\xi-1)}  & \quad \text{for} \quad  1 \leq \xi
\end{cases} \, .
\end{align}
The rest of the accessory parameters are fixed by demanding having a puncture at $z= \infty$.

We only consider $0 \leq \xi \leq 1$ as the remaining cases are related to this case by an appropriate conformal transformation. The Strebel differential reads
\begin{align} \label{eq:RQ}
	\varphi^{(S)} = \phi(z)\,  dz^2 = - {(z^2 + \xi - 2 z \xi)^2 \over z^2 (z-1)^2 (z-\xi)^2} dz^2 \, .
\end{align}
We need to compute the modulus $\mathcal{S}_{0,4}^\ast(\xi)$ of $\varphi^{(S)}$. In order to do this we use the following formula for the mapping radii~\cite{Belopolsky:1994bj,Moeller:2004yy,Erbin:2022rgx}
\begin{align} \label{eq:map_rad}
\log r_i = \lim_{\epsilon \to 0} \left( \mathrm{Im} \int_{\xi_i + \epsilon}^{z_c} \sqrt{\phi(z')} \, dz' + \log | \epsilon | \right) \, .
\end{align}
Here $\epsilon \in \mathbb{C}$ lies on the linear path between $\xi_i$ and $z_c$, a zero of~\eqref{eq:RQ}. The sign of the square root has to be chosen such that limit exists. For more details see~\cite{Erbin:2022rgx}.

It is possible to evaluate the integral~\eqref{eq:map_rad} analytically and we find
\begin{align} \label{eq:Rmap}
	r_1 = (1-\xi)\xi, \quad \quad r_2 = \xi, \quad\quad  r_3 = 1-\xi, \quad \quad r_4 = 1 \, .
\end{align}
We take the conventions for which $r_1, r_2, r_3, r_4$ are the mapping radii associated with $z= \xi,0,1, \infty$ respectively.\footnote{This is different from the conventions used in~\cite{Moeller:2004yy,Erbin:2022rgx}.} All of them are positive for $0 \leq \xi \leq 1$ as expected. This implies that the modulus is
\begin{align} \label{eq:MapRe}
	\mathcal{S}_{0,4}^\ast(\xi = \overline{\xi}) =\log\left( |1-\xi|^2 |\xi|^2 \right) \, .
\end{align}
Take note that we use the absolute values in this expression to make the modulus symmetric in either using $\xi$ or its complex conjugate $\overline{\xi}$. Then we see
\begin{align}
	2{\p \mathcal{S}_{0,4}^\ast \over \p \xi}(\xi = \overline{\xi}) =
	2 \left[ {-1\over 1- \xi} + {1 \over \xi} \right]
	= { 4\xi - 2 \over \xi (\xi-1)} = c \, (\xi = \overline{\xi}) \, ,
\end{align}
which is consistent with the Polyakov conjecture~\eqref{eq:StrebelPolyakov1}.

\begin{figure}[h]
	\includegraphics[width=0.49\textwidth]{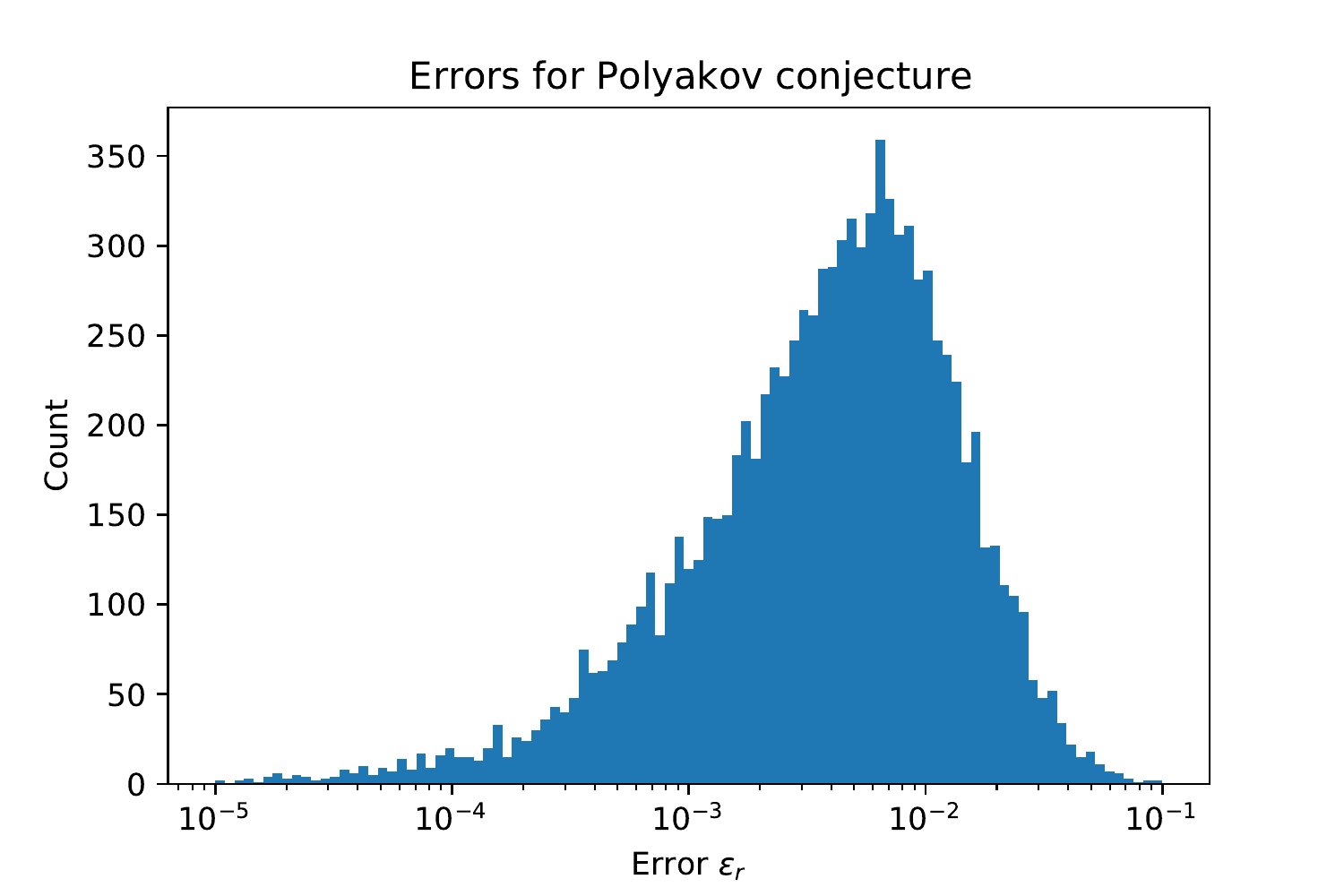}
	\includegraphics[width=0.49\textwidth]{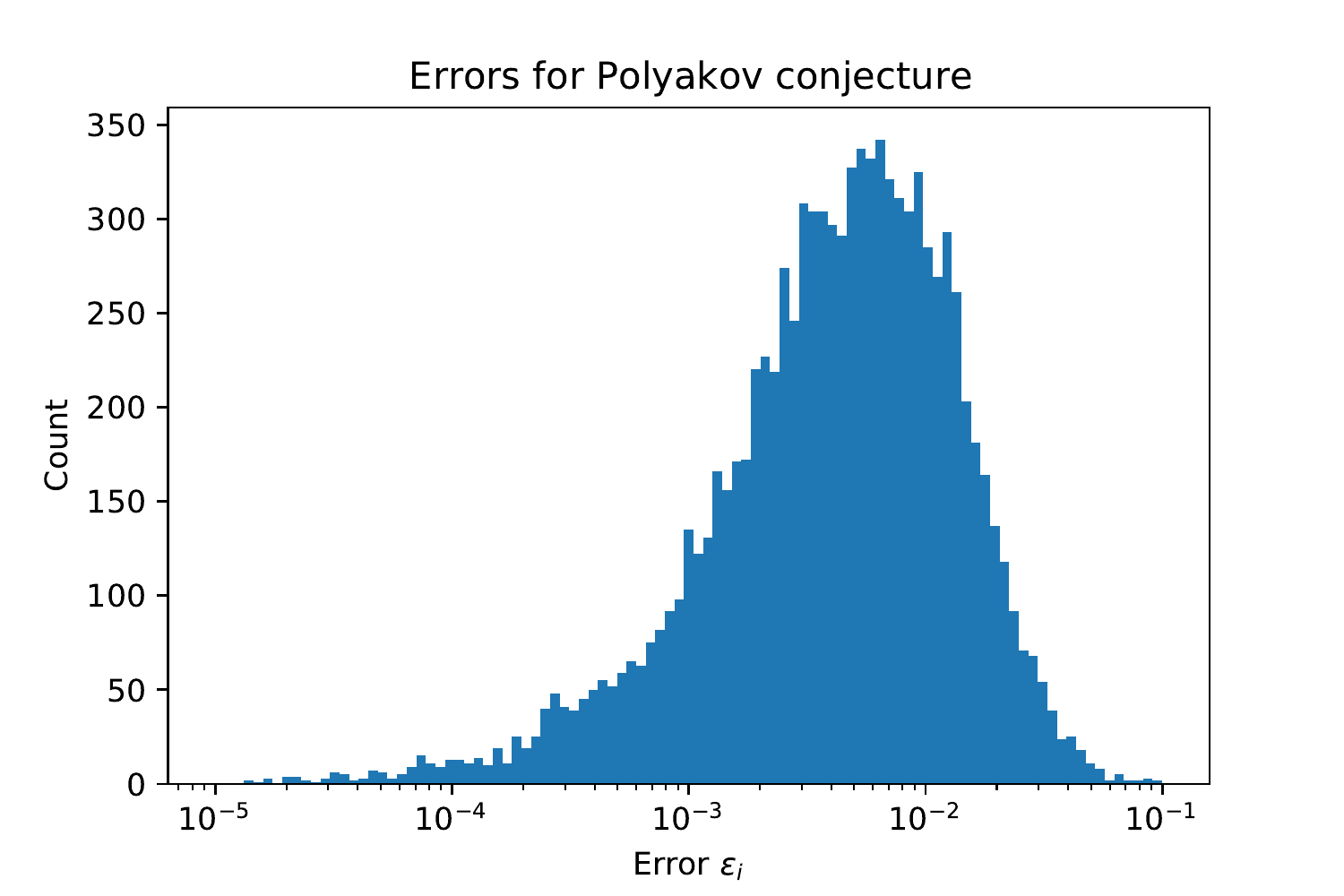}
	\caption{\label{fig:errors}The distribution of~\eqref{eq:errors} for randomly sampled $\approx 10^4$ points in $\mathcal{M}_{0,4}$. We only show $\epsilon < 0.1$. While there were few points for which the error was sizable, they got smaller upon changing the precision of the evaluation of~\eqref{eq:map_rad}. So we think this was an issue of the implementation in~\cite{Erbin:2022rgx}.}
\end{figure}

Finally, let us check the Polyakov conjecture~\eqref{eq:StrebelPolyakov1} for 4-punctured spheres with $\alpha_i = 1$ for generic $\xi \in \mathbb{C}$. Unlike $\xi \in \mathbb{R}$, only numerical solutions are available and we use the results from~\cite{Erbin:2022rgx}. There, the non-trivial accessory parameter $c$ and the modulus $\mathcal{S}_{0,4}^\ast(\xi)$ are obtained as artificial neural networks. This allows us to differentiate the latter and compare it with the former. More precisely, we consider the absolute errors
\begin{align} \label{eq:errors}
	\epsilon_r (\xi, \overline{\xi})  = \left| \mathrm{Re} \left( c(\xi, \overline{\xi})  - 2 {\p \mathcal{S}_{0,4}^\ast(\xi, \overline{\xi})  \over \p \xi} \right) \right| \, ,
	\hspace{0.4in}
	\epsilon_i (\xi, \overline{\xi})  = \left| \mathrm{Im} \left( c(\xi, \overline{\xi})  - 2 {\p \mathcal{S}_{0,4}^\ast(\xi, \overline{\xi})  \over \p \xi} \right) \right|
	\, ,
\end{align}
and check whether $\epsilon_r  \approx \epsilon_i \approx 0 $ over $\mathcal{M}_{0,4}$ by randomly sampling points.\footnote{We sampled points from the circle of radius $1.7$ centered at $\xi = 0.5$ and excised circles of radius $0.2$ around $\xi = 0,1$.}. The results are shown in figure~\ref{fig:errors}. The errors are small and have the expected order of magnitude from~\cite{Erbin:2022rgx}.

Unfortunately, the accessory parameters, even as a fit, for higher-punctured spheres are not available (although see~\cite{Moeller:2006cw,Moeller:2007mu}). So instead of viewing the Polyakov conjecture~\eqref{eq:StrebelPolyakov1} as something Strebel differentials happen to satisfy, we are going to look at it as a tool to generate Strebel differentials. In particular, we look for relations among $\mathcal{S}_{0,n}^\ast$ for different $n$ starting next section.

\section{Classical bootstrap for string vertices} \label{sec:recursion}

Given that $\mathcal{S}_{0,n}^\ast$ generates the accessory parameters for Strebel differentials through~\eqref{eq:StrebelPolyakov1}, we now focus on ways to compute it. As we mentioned, the direct computation of the modulus appears to be as complicated as computing the accessory parameters. However, it is apparent that $\mathcal{S}_{0,n}^\ast$ comes with an index $n$, so it is natural to ask whether it is possible to compute it \textit{recursively}. In this section, we claim that this is indeed possible akin to conformal bootstrap~\cite{Zamolodchikov:1995aa}. We primarily focus on $n=4$ with equal weights to show how this can be done and comment on $n \geq 5$ at the end.

In order to do this we are going to use the DOZZ formulation of Liouville theory~\cite{Zamolodchikov:1995aa,Dorn:1994xn}. It is already shown in~\cite{Hadasz:2003he} that the DOZZ three-point function is related to~\eqref{eq:Lio5} with three hyperbolic singularities in the semi-classical limit, which we review in appendix~\ref{app:DOZZ}. Given such a connection, we consider the WKB limit of~\eqref{eq:Lim} and what it entails for the same quantity in subsection~\ref{sec:S03}. This simply produces~\eqref{eq:BaseCase} for $\mathcal{S}_{0,3}^\ast(\alpha_1,\alpha_2,\alpha_3)$. We remark on its significance.

Given the close connection between the DOZZ three-point function and the on-shell HJ action with three hyperbolic singularities, together with the explicit WKB limit for the latter, we argue that $\mathcal{S}_{0,4}^\ast (\xi)$ can be constructed using techniques from conformal bootstrap in subsection~\ref{sec:OpForm} by thinking it as a 4-point function in an ``operator formalism'' and considering its semi-classical limit. The semi-classical limit of the conformal blocks plays a crucial role in this construction, which we provide a brief review in appendix~\ref{app:ConfBlocks}. In the subsequent subsection we numerically show that the crossing symmetry is satisfied for different ways of constructing $\mathcal{S}_{0,4}^\ast (\xi)$ .

In the penultimate subsection, we describe how the vertex $\mathcal{V}_{0,4}$ and Feynman regions $\mathcal{F}_{0,4} = \mathcal{M}_{0,4} \setminus \mathcal{V}_{0,4}$ fits into this framework. In particular, we determine the Schwinger parameter of a string propagator $\mathfrak{q}$ in terms of the cross-ratio $\xi$ in $\mathcal{F}_{0,4}$. This allows us to describe the boundary of the vertex region $\p \mathcal{V}_{0,4} $ \textit{analytically}. Hence we obtain a \textit{complete} analytic characterization of 4-string contact interactions in terms of classical conformal blocks and $\mathcal{S}_{0,3}^\ast(\alpha_1,\alpha_2,\alpha_3)$. In the last subsection we discuss the generalization to higher-punctured spheres.

\subsection{Computing $\mathcal{S}_{0,3}^\ast(\alpha_1,\alpha_2,\alpha_3)$} \label{sec:S03}

In this subsection we compute $\mathcal{S}_{0,3}^\ast(\alpha_i)$ using the generalized hyperbolic three-vertex of~\cite{Firat:2021ukc} by taking the WKB limit~\eqref{eq:Lim} and demonstrate that it is given by~\eqref{eq:BaseCase}. Begin with considering the (logarithm of) mapping radii of the generalized hyperbolic-three vertex
\begin{align} \label{eq:3maprad}
	\log r_i[H_i] =  {\pi \over 2\lambda_i} - {v_i \over \lambda_i} \, ,
\end{align}
with $v_i$' are given in terms of the function (see (1.5) in~\cite{Firat:2021ukc})
\begin{align} \label{eq:v_func}
v\left(\lambda_{1}, \lambda_{2}, \lambda_{3}\right) &\equiv \frac{1}{2i} \; \text{log}\left[\frac{\Gamma\left(-i \lambda_{1}\right)^{2}}{\Gamma\left(i \lambda_{1}\right)^{2}} \frac{\gamma\left(\frac{1}{2}(1+i \lambda_{1}+i \lambda_{2}+i \lambda_{3})\right) \gamma\left(\frac{1}{2}(1+i \lambda_{1}-i \lambda_{2}+i \lambda_{3})\right)}{\gamma\left(\frac{1}{2}(1-i \lambda_{1}-i \lambda_{2}+i \lambda_{3})\right) \gamma\left(\frac{1}{2}(1-i \lambda_{1}+i \lambda_{2}+i \lambda_{3})\right)} \right] \, ,
\end{align}
and its respective permutations. For example $v_1 = v(\lambda_1, \lambda_2, \lambda_3)$ and the expressions for $v_2, v_3$ are  analogous. Notice there was $\ell \in \mathbb{Z}_{<0}$ in~\cite{Firat:2021ukc}, however we don't include it by adjusting the branches so that the geodesic seams correspond $|w_i|=1$ in local coordinates. Here
\begin{align}
\gamma(x) &\equiv \frac{\Gamma(x)}{\Gamma(1-x)} \, .
\end{align}

We are interested in the WKB limit~\eqref{eq:Lim} of~\eqref{eq:v_func}. Using the identities
\begin{align} \label{eq:Stirling}
	\log \Gamma(1+ i x) \approx ix \, (\log ix - 1) \, ,
	\hspace{0.5in}
	\log \Gamma\left({1 \over 2}+ {i x \over 2}\right) \approx 
	{ix \over 2} \left(\log {ix \over 2} - 1\right) 
	\, ,
\end{align}
for $x\in\mathbb{R}$ as $x \to \pm \infty$ we see that
\begin{align}
v_1  &= \lambda \left[
-2 \alpha_1 \log  |2 \alpha_1| + {1 \over 2} \sum_{\sigma_2, \sigma_3 = \pm 1} (\alpha_1 + \sigma_2 \alpha_2 + \sigma_3 \alpha_3) \log |\alpha_1 + \sigma_2 \alpha_2 + \sigma_3 \alpha_3|
\right]
\, ,
\end{align}
after some manipulations. This, together with~\eqref{eq:3maprad}, shows that the modulus is indeed given by
\begin{align} \label{eq:BaseCase1}
&\mathcal{S}^\ast_{0,3}(\alpha_1, \alpha_2, \alpha_3) = \sum_{i=1}^3 \alpha_i^2 \log r_i [H_i] =
\log \bigg[
|2\alpha_1|^{2 \alpha_1^2} |2\alpha_2|^{2 \alpha_2^2} |2\alpha_3|^{2 \alpha_3^2}
|\alpha_1 + \alpha_2 + \alpha_3|^{-{1 \over 2}  (\alpha_1 + \alpha_2 + \alpha_3)^2 } \nonumber\\
&
|-\alpha_1 + \alpha_2 + \alpha_3|^{-{1 \over 2} (-\alpha_1 + \alpha_2 + \alpha_3)^2 }
|\alpha_1 - \alpha_2 + \alpha_3|^{- {1 \over 2}  (\alpha_1 - \alpha_2 + \alpha_3)^2 }
|\alpha_1 + \alpha_2 - \alpha_3|^{- {1 \over 2}  (\alpha_1 + \alpha_2 - \alpha_3)^2 }
\bigg]  \, ,
\end{align} 
This is a totally symmetric function of $\alpha_i$'s and positive as expected. Note that $\pi / 2 \lambda_i$ terms in~\eqref{eq:3maprad} are subleading, so they won't contribute to the modulus in the WKB limit (i.e. the branch choice was irrelevant). Notice we can use scaling to eliminate the dependence on one of the variables.

It is possible to derive~\eqref{eq:BaseCase1} using alternative methods, here we list some of them. One can derive it starting from~\eqref{eq:LambdaDer} and more or less following a similar logic above. Alternatively, it can be derived by evaluating the integrals~\eqref{eq:map_rad} for the Strebel differential~\eqref{eq:GW}, although this seems awfully hard to do. In a completely different way, one can use geometric function theory, just as reported in~\cite{jenkins1954recent}. Take note that the expression~\eqref{eq:BaseCase1} is given in a different form in this paper, we checked that they are equivalent nonetheless.

We remark that there are essentially $3$ distinct behaviors for the critical graphs of Strebel differentials with three second order poles~\cite{Saadi:1989tb} which is reflected in~\eqref{eq:BaseCase1}: $\alpha_1 < \alpha_2 + \alpha_3$\footnote{We also have $\alpha_2 < \alpha_1 + \alpha_3$ and $\alpha_3 < \alpha_1 + \alpha_2$.}, $\alpha_1 = \alpha_2 + \alpha_3$ and $\alpha_1 > \alpha_2 + \alpha_3$ which we call type I, II and III vertices respectively and collectively call them the \textit{generalized cubic vertex}. For the type I vertex, three strings (i.e. the faces of the critical graph) touch each others but not to themselves. The symmetric cubic vertex (that is $\alpha_1 = \alpha_2 = \alpha_3 = 1$) is type I and its modulus is given by
\begin{align}
\mathcal{S}_{0,3}^\ast(1,1,1) = \log {64 \over 81 \sqrt{3}} = \log \left({4 \over 3 \sqrt{3}}\right)^3 \, ,
\end{align}
consistent with its conventional derivation~\cite{Rastelli:2000iu}.

The type II vertex is the light-cone vertex. We observe that there is a term~($|-\alpha_1 + \alpha_2 + \alpha_3|^{-(-\alpha_1 + \alpha_2 + \alpha_3)^2/2}$) in the modulus~\eqref{eq:BaseCase1} that may have a non-trivial limit. However, we have
\begin{align} \label{eq:lcexp}
\mathcal{S}_{0,3}^\ast(\alpha_2 + \alpha_3  + \epsilon, \alpha_2 ,\alpha_3)
= \log \left[ |\alpha_2|^{-2\alpha_2} |\alpha_3|^{-2\alpha_3} |\alpha_2 + \alpha_3|^{2(\alpha_2 + \alpha_3)} \right] \epsilon + \mathcal{O}(\epsilon^2 \log \epsilon) \, ,
\end{align}
by taking $\alpha_1 = \alpha_2 +\alpha_3 + \epsilon$. This effectively shows that such terms don't pose a treat and the modulus vanishes as expected for the type II vertices. Lastly, we have type III vertices for which one string touch itself while the remaining ones only touch the first string. We remark that the modulus~\eqref{eq:BaseCase1} is continuous. In fact, its first derivative also exists and is continuous by the expansion~\eqref{eq:lcexp}. However, it can be shown that the second derivative of the modulus diverges for type II vertices. 

Observe the striking resemblance of~\eqref{eq:BaseCase1} to the DOZZ formula~\eqref{eq:DOZZ}---as if the functions $\Upsilon(x)$ is replaced by $|x|^{x^2/2}$. This is somewhat expected from~\eqref{eq:DOZZexp} together with the WKB limit. It is well-known that the DOZZ formula (and the conformal blocks) can be used to express the $N$-point functions of Liouville theory. So it is natural to ask whether this is also possible for Strebel differentials in the vein of~\cite{Hadasz:2005gk,Hadasz:2006rb}. In the next subsection, we investigate this question.

\subsection{The operator formalism and computing $\mathcal{S}_{0,4}^\ast(\xi)$} \label{sec:OpForm}

We now turn our attention to computing $\mathcal{S}_{0,4}^\ast(\xi; \alpha_i)$ characterizing the four-string vertex. Remember that we have defined the ``hole'' operator below~\eqref{eq:FinalWeights1} by
\begin{align} \label{eq:HoleOp}
	\mathcal{H}_\lambda(\xi_i, \overline{\xi_i}) \equiv V_{Q/2} (\xi_i, \overline{\xi_i}) \, r_i [H_i]^{-Q^2 \lambda^2/2} \, ,
\end{align}
where $V_{Q/2} (\xi_i, \overline{\xi_i}) $ was given in~\eqref{eq:ExpOp} and $r_i[H_i]$ was the mapping radius for the region $H_i$ dressing $V_{Q/2} (\xi_i, \overline{\xi_i}) $ to generate conformal weights $\Delta > Q^2 /4$. We are going to set up an ``operator'' formalism and consider the correlators given in~\eqref{eq:PI} from this heuristic perspective.

We begin with the zero-, one-and two-point functions of the hole operators~\eqref{eq:HoleOp}. The associated surfaces have non-negative Euler characteristics, so they don't endow hyperbolic metrics and it is more natural to take $\mu < 0$ for the first two and $\mu = 0$ for the latter. The zero-point function may give a number which we don't need for our purposes, but the one-point function on the sphere vanishes as $\langle V_{Q/2} (\xi_i, \overline{\xi_i}) \rangle = 0$ .

For the two-point function, we evaluate
\begin{align} \label{eq:2point}
	\langle \mathcal{H}_{\lambda'}(\infty, \infty) \,
	\mathcal{H}_{\lambda}(0, 0) \rangle = 
	 r_{\lambda'}^{-Q^2 \lambda'^2/2} r_\lambda^{-Q^2 \lambda^2/2} \, .
\end{align}
Here $r_\lambda, r_{\lambda'}$ are the mapping radii associated with the holes around $z=0$ and $z=\infty$ respectively. This is essentially due to the operator product expansion~\cite{polchinski1998string}
\begin{align}
	V_{Q/2} (\infty, \infty) \, V_{Q/2} (0,0) = \lim_{z \to \infty} \, |z|^{2 \, Q^2} \, V_{-Q/2} (z, \overline{z})\, V_{Q/2} (0,0) = 1 + \cdots \, . 
\end{align}

A priori there is no relation between $\lambda$ and $\lambda'$ in~\eqref{eq:2point}. However, it is natural to demand that the two-point function to describe a flat cylinder of length $s \in \left[0, \infty \right]$ of circumference $2 \pi \lambda$ in the semi-classical limit for our bootstrap program as we shall see.~\footnote{The length of the cylinder is actually given by $\lambda s$ for the metric under consideration. However, we are still going to call $s$ the length of the cylinder.} These types of 2-point functions are only possible if we take $\lambda = \lambda'$. So we instead demand
\begin{align} \label{eq:2point1}
\langle \mathcal{H}_{\lambda'}(\infty, \infty) \,
\mathcal{H}_{\lambda}(0, 0) \rangle = 
e^{Q^2 \lambda^2 s / 2} \delta(\lambda - \lambda') \, ,
\end{align}
using the relation
\begin{align}
	s = \log {1/r_{\lambda'} \over r_\lambda} = - \log  r_\lambda r_{\lambda'} \, ,
\end{align}
which relates the length $s$  of the flat cylinder between two holes around $z=0,\infty$ to their mapping radii. Note that $s=0$ is a possibility for which the holes around $z=0, \infty$ touch each other.

We demand that the three-point function is given by the (symmetric) DOZZ formula~\eqref{eq:DOZZ},\eqref{eq:SymmDOZZ}
\begin{align}
	\langle \mathcal{H}_{\lambda_3}(\infty, \infty) \,
	\mathcal{H}_{\lambda_2}(1, 1)  \,
	\mathcal{H}_{\lambda_1}(0, 0) \rangle 
	= \widetilde{C} (\lambda_1, \lambda_2, \lambda_3) \, .
\end{align}
Using this formula for our purposes is justified, especially in the semi-classical limit (see~\eqref{eq:DOZZexp}), given that the geometric formulation of Liouville theory for the same problem described in the previous section. 

We can now express the four-point function of the hole operators as follows
\begin{align} \label{eq:4point}
	&\left\langle \mathcal{H}_{\lambda_4}(\infty, \infty) \,
	\mathcal{H}_{\lambda_3}(1, 1) \,
	\mathcal{H}_{\lambda_2}(0,0)  \,
	\mathcal{H}_{\lambda_1}(\xi, \overline{\xi})\right\rangle \nonumber \\
	& \quad \quad \quad
	= \int\limits_{{Q \over 2} (1 + i \mathbb{R}^+) } {d \lambda} \;
	\widetilde{C}(\lambda, \lambda_3, \lambda_4) \;
	e^{Q^2 \lambda^2 s/2} \;
	\widetilde{C}(\lambda_1, \lambda_2, -\lambda)
	\left| \mathcal{F}_{1 + 6 Q^2 , \Delta } 
	\begin{bmatrix}
	\Delta_3 & \Delta_2\\
	\Delta_4 & \Delta_1
	\end{bmatrix} 
	(\xi)\right|^2 \, ,
\end{align}
in the vein of conformal bootstrap. As we mentioned in the introduction, the functions $\mathcal{F}$ are the conformal blocks, which are reviewed in appendix~\ref{app:ConfBlocks}. They are entirely determined by the Virasoro algebra. Although no explicit closed-form expression exists, it has a well-defined semi-classical limit given by the classical conformal blocks~\eqref{eq:ClassConfBlock}. For now we consider the $s$-channel decomposition as in~\eqref{eq:4point}. We comment on different ways of decomposing and how they relate to each other in the next subsection.

The remaining factors in~\eqref{eq:4point} also have well-defined semi-classical limits. For example, the three-point function is given by~\eqref{eq:DOZZexp}, while the left-hand side is given by~\eqref{eq:PartFunc2} as $Q \to \infty$, i.e.
\begin{align}
	\left\langle \mathcal{H}_{\lambda_4}(\infty, \infty)
	\mathcal{H}_{\lambda_3}(1, 1)
	\mathcal{H}_{\lambda_2}(0,0) 
	\mathcal{H}_{\lambda_1}(\xi, \overline{\xi}) \right\rangle
	\sim \exp \left[ {- {1 \over 2}Q^2 { S_{HJ}^{(4)}(\xi;\lambda_i) }} \right] \, ,
\end{align}
where $ S_{HJ}^{(n)}(\xi_i;\lambda_i)$ stands for the on-shell HJ action with $n$ hyperbolic singularities. Instead of using $\varphi$, we use $\lambda_i$ for its arguments. Together they imply
\begin{align} \label{eq:Hyp}
	 \exp \left[ {- {1 \over 2} Q^2 { S_{HJ}^{(4)}(\xi;\lambda_i)}} \right]
	 &\sim \int\limits_{0}^\infty { d \lambda}
	 \exp \Bigg[ -{1 \over 2} Q^2 \bigg(
	 { S_{HJ}^{(3)}(\lambda, \lambda_3, \lambda_4) }
	 -  {\lambda^2 s}
	 + { S_{HJ}^{(3)}(\lambda_1, \lambda_2, \lambda) } \nonumber \\
	 & \hspace{0.75in}
	 -2 \widetilde{f}_{\delta/2}
	 \begin{bmatrix}
	 \delta_3/2 & \delta_2/2\\
	 \delta_4/2 & \delta_1/2
	 \end{bmatrix} (\xi)
	 -2\overline{\widetilde{f}_{\delta/2}}
	 \begin{bmatrix}
	 \delta_3/2 & \delta_2/2\\
	 \delta_4/2 & \delta_1/2
	 \end{bmatrix} (\overline{\xi})
	 \bigg)
	 \Bigg] \, .
\end{align}
after keeping the leading terms in $Q$ and using the invariance of $S_{HJ}^{(3)}$ under $\lambda_i \to - \lambda_i$. The explicit expression for $S_{HJ}^{(3)}$ is given in~\eqref{eq:LambdaDer}. We can develop~\eqref{eq:Hyp} to construct the local coordinates for the classical hyperbolic vertices. However, this would be rather involved and make the whole procedure more convoluted than needed. So we additionally take the WKB limit~\eqref{eq:Lim} to simplify $S_{HJ}^{(3)}(\lambda_s, \lambda_3, \lambda_4) $ and $S_{HJ}^{(4)}(\xi;\lambda_i)$. In any case, this is just a technical simplification and not a conceptual one, see~\cite{Hadasz:2005gk,Hadasz:2006rb,Harrison:2022frl} where such a limit is absent. 

So we begin our exposition with the equation
\begin{align} \label{eq:Stre}
\exp \left[ {- {1 \over 2} Q^2 \lambda^2 { \mathcal{S}_{0,4}^\ast (\xi;\alpha_i)}} \right]
&\sim \int\limits_{0}^\infty { d \alpha} \;
\exp \Bigg[ -{1 \over 2} Q^2 \lambda^2 \bigg(
{\mathcal{S}_{0,3}^\ast (\alpha, \alpha_3, \alpha_4) }
-  \alpha^2 s
+ { \mathcal{S}_{0,3}^\ast (\alpha_1, \alpha_2, \alpha) } \nonumber \\
& \hspace{1.25in}
-2 f_{\alpha^2}
\begin{bmatrix}
\alpha_3^2 & \alpha^2_2\\
\alpha^2_4 & \alpha^2_1
\end{bmatrix} (\xi)
-2 \overline{f_{\alpha^2}}
\begin{bmatrix}
\alpha^2_3 & \alpha^2_2\\
\alpha^2_4 & \alpha^2_1
\end{bmatrix} (\overline{\xi})
\bigg)
\Bigg] \, ,
\end{align}
ignoring the terms subleading in $\lambda$. Because of the semi-classical and WKB limit, the integral in~\eqref{eq:Hyp} is dominated by the saddle point at $\alpha= \alpha_s$
\begin{align} \label{eq:SaddlePoint}
	{\p \over \p \alpha} \, \Bigg[
	{\mathcal{S}_{0,3}^\ast (\alpha, \alpha_3, \alpha_4) }
	- \alpha^2 s
	+ { \mathcal{S}_{0,3}^\ast (\alpha_1, \alpha_2, \alpha) } 
	-2 f_{\alpha^2}
	\begin{bmatrix}
	\alpha_3^2 & \alpha^2_2\\
	\alpha^2_4 & \alpha^2_1
	\end{bmatrix} (\xi)
	-2 \overline{f_{\alpha^2}}
	\begin{bmatrix}
	\alpha^2_3 & \alpha^2_2\\
	\alpha^2_4 & \alpha^2_1
	\end{bmatrix} (\overline{\xi})
	\Bigg]_{\alpha = \alpha_s} = 0 \, ,
\end{align}
and we find, only keeping the leading order in the limits,
\begin{align} \label{eq:mod}
\mathcal{S}_{0,4}^\ast(\xi;\alpha_i) 
=
{ \mathcal{S}_{0,3}^\ast (\alpha_s, \alpha_3, \alpha_4) }
-  {\alpha_s^2 s}
+ { \mathcal{S}_{0,3}^\ast (\alpha_1, \alpha_2, \alpha_s) } 
-2 f_{\alpha_s^2}
\begin{bmatrix}
\alpha_3^2 & \alpha^2_2\\
\alpha^2_4 & \alpha^2_1
\end{bmatrix} (\xi)
-2 \overline{f_{\alpha_s^2}}
\begin{bmatrix}
\alpha^2_3 & \alpha^2_2\\
\alpha^2_4 & \alpha^2_1
\end{bmatrix} (\overline{\xi})  \, .
\end{align}

Let us remark on the significance of the last two equations, beginning with the interpretation of the saddle point $\alpha_s = \alpha_s(\xi, \overline{\xi})$. Since it is related to the weight of the internal hole operator, $2 \pi \alpha_s$ is the length of the $s$-channel geodesic. In fact, there is a ring domain of height $s$ and circumference $2 \pi \alpha_s$ because the term containing $s$ comes from the normalization of the two-point function and it describes a flat cylinder in the semi-classical limit, see~\eqref{eq:2point1}. This way, not only we can describe the Strebel differentials ($s=0$) relevant to the vertex region, but also the \textit{Jenkins-Strebel differentials} ($s\geq0$) describing string Feynman diagrams. We call the quadratic differentials with at most second order poles with residues equal to $-1$ and whose systole is greater than or equal to $2\pi$ with measure zero critical trajectory \textit{Zwiebach differentials}~\cite{Zwiebach:1990nh}. These are the quadratic differentials relevant for the classical CSFT. We set $s=0$ for the rest of this subsection and focus on the cases with $s>0$ in subsection~\ref{sec:Feynman}. 

Given the modulus $\mathcal {S}_{0,3}^\ast$ (see~\eqref{eq:BaseCase1}) and the expansion of the classical conformal block in $\xi$ (see~\eqref{eq:xi}), the saddle point $\alpha_s$ can be found solving~\eqref{eq:SaddlePoint} perturbatively in $\xi$. Let us show how to do this in the case of $\alpha_i = 1 $. We need the following two derivatives with respect to $\alpha$:
\begin{subequations} \label{eq:Der}
\begin{align}
	&{\p f_{\alpha^2}  \over \p \alpha} (\xi) =
	{\alpha \over 2} \log \xi
	+{ \alpha \over 4} \xi
	+ \left[-{1 \over 8 \alpha^3} + {13 \alpha \over 128} \right] \xi^2 
	+ \left[ - {1 \over 8 \alpha^3} + {23 \alpha \over 384} \right] \xi^3 
	+ \mathcal{O}(\xi^4) \, , \\
	&{\p \mathcal {S}_{0,3}^\ast \over \p \alpha} (\alpha,1,1) =
	\log \left[ {\alpha-2 \over \alpha+2} \right]^2 
	+ \alpha  \log {16 \alpha^2 \over |4 - \alpha^2 |}
	= \alpha \log {4 \alpha^2 \over e^2} + {\alpha^3 \over 12} + {\alpha^5 \over 160 } 
	+ \mathcal{O}(\alpha^7) \, .
\end{align}
\end{subequations}
Strictly speaking, this second expansion holds only for $0 \leq \alpha \leq 2$ as there is a discontinuity at $\alpha = 2$. This subtlety isn't relevant for the $4-$punctured spheres with $\alpha_i = 1$ as there can't be a situation for which $\alpha_s > 2$. On the other hand, there may be additional subtlety for the higher-string interactions.\footnote{This can be overcame by performing the expansion around $\alpha =2$ instead. This would produce an expansion applicable to any situation, however it complicates the subsequent ansatze slightly.}

Let us find the saddle point $\alpha_s$ using last two equality. Make an ansatz of the form
\begin{align} \label{eq:Ansatz}
	\alpha_s(\xi, \overline{\xi}) &= 
	|\xi|^{1/2} \, ( a_0 + a_1 \cos(\theta) + a_2 \cos(2\theta) + a_3 \cos(3\theta) + \cdots ) \nonumber \\
	&+ |\xi|^{3/2}\, ( b_0 + b_1 \cos(\theta) + b_2 \cos(2\theta) + b_3 \cos(3\theta)  + \cdots ) \nonumber \\
	&+ |\xi|^{5/2} \, ( c_0 + c_1 \cos(\theta) + c_2 \cos(2\theta) + c_3 \cos(3\theta) + \cdots ) + \cdots \, ,
\end{align}
with $\xi = |\xi| e^{i \theta}$. This is an expansion in both $|\xi|$ and $\theta$. Double series of this form is expected in general since the saddle point $\alpha_s$ is expected to be a non-holomorphic function of $\xi$. The leading power of $|\xi|$ series is dictated by $\log$'s in~\eqref{eq:Der} and the fact that $\alpha_s \to 0$ as $\xi \to 0$ for the $s$-channel geodesic to shrink.

Ignoring the dependence of $\alpha_s$ on the angle $\theta$, the saddle point equation~\eqref{eq:SaddlePoint} takes the form
\begin{align}
	2 \alpha_s \log {4 \alpha_s^2 \over e^2} + {\alpha_s^3 \over 6} + {\alpha_s^5 \over 80 }  + \cdots
	- 2 \alpha_s \log | \xi | + \cdots = 0 \, .
\end{align}
Here the first dots represent the terms of order $\alpha_s^7$ and higher and the second dots represent the terms that depend on $\theta$. We need to plug the ansatz~\eqref{eq:Ansatz} excluding $\theta$-dependent terms, giving
\begin{align}
	\alpha_s(\xi, \overline{\xi})  &= |\xi|^{1/2} \left( {e \over 2} + \cdots \right)  + |\xi|^{3/2} \left( -{e^3 \over 192} + \cdots \right) + |\xi|^{5/2} \left( {7 e^5 \over 184 320} + \cdots \right) + \cdots \, .
\end{align}
This procedure can be repeated for higher powers of $|\xi|$.

Now, let us add back the $\theta$ dependence by considering the higher order terms in the expansion of classical conformal blocks. The full saddle point equation is
\begin{align}
&2 \alpha_s \log {4 \alpha_s^2 \over e^2} + {\alpha_s^3 \over 6} + {\alpha_s^5 \over 80 }  + \cdots - 2 \alpha_s \log | \xi |  \nonumber\\
& \hspace{0.3in}
- \alpha_s |\xi| \cos(\theta)
- \left[-{1 \over 2 \alpha_s^3} + {13 \alpha_s \over 32} \right] |\xi|^2 \cos(2\theta)
- \left[ - {1 \over 2 \alpha_s^3} + {23 \alpha_s \over 96 } \right] |\xi|^3 \cos(3\theta) 
+ \cdots = 0 \, .
\end{align}
This can be solved by plugging the ansatz~\eqref{eq:Ansatz} and our final result for $\alpha_s $ is
\begin{align} \label{eq:Res}
\alpha_s(\xi, \overline{\xi}) &= 
|\xi|^{1/2} \left(  {e \over 2}  -  {1 \over e^3} \cos(2\theta) + \cdots \right) + |\xi|^{3/2} \left( -{e^3 \over 192}+ {e \over 8} \cos(\theta) - {1 \over 96 e} \cos(2\theta) - {1 \over 4 e^3} \cos(3\theta) + \cdots \right) \nonumber \\
&+ |\xi|^{5/2} \left( {7 e^5 \over 184 320} -{e^3 \over 256} \cos(\theta) + { 6133 e \over 92160} \cos(2\theta) -{1 \over 128 e } \cos(3\theta) + \cdots \right) + \cdots \, .
\end{align}
This expansion can be repeated to higher orders as well. We have worked up to~$\mathcal{O}(|\xi|^{7\over2}, \cos(5\theta) )$, but for brevity we only report the terms that have shown above.

We can run following numerical checks for the result~\eqref{eq:Res}. First, we have a closed-form expression for the lengths when $\xi \in \mathbb{R}$~\cite{Erbin:2022rgx}. This corresponds to taking either $\theta = 0$ or $\theta = \pi$ in~\eqref{eq:Res}. When $\theta=0$ ($\xi>0$), we have
\begin{align}
	2 \pi \alpha_s(\xi=\overline{\xi}) = 8 |\xi|^{1/2} + {4 \over 3} |\xi|^{3/2} + {3 \over 5} |\xi|^{5/2} +\cdots
	\approx 8.14 |\xi|^{1/2} + 1.36 |\xi|^{3/2} + 0.62 |\xi|^{5/2} + \cdots \, ,
\end{align}
Similarly for $\theta=\pi$ ($\xi<0$) we have
\begin{align}
2 \pi \alpha_s(\xi=\overline{\xi}) &= 8 |\xi|^{1/2} - {8 \over 3} |\xi|^{3/2} + {8 \over 5} |\xi|^{5/2} +\cdots \approx 8.14 |\xi|^{1/2} -2.71 |\xi|^{3/2} + 1.64 |\xi|^{5/2} + \cdots \, .
\end{align}
We observed that~\eqref{eq:Res} appears to converge to the exact result, albeit slowly, by increasing the order of the expansion. Given $\alpha_s$, it is possible to find the modulus $\mathcal{S}_{0,4}^\ast (\xi) $ and it is given by
\begin{align} \label{eq:S4xi}
	S_{0,4}^\ast (\xi) = 2 \log |\xi|
	&+ |\xi| \left( - {e^2 \over 4} -{1 \over e^2} \cos(2\theta) + \cdots \right)  \\
	&+ |\xi|^2 \left({e^4 \over 384} - {e^2 \over 4} \cos(\theta) - {7 \over 48} \cos(2\theta) - {1 \over 2 e^2 } \cos(3\theta)  + \cdots \right) \nonumber\\
	&+ |\xi|^3\left(-{e^6 \over 46080} + {e^4 \over 384} \cos(\theta) - {947 \over 11520} \cos(2\theta) - {7 \over 48 e^2 } \cos(3\theta)  + \cdots \right) + \cdots \, .\nonumber
\end{align}
Again, we can compare it with the exact results for $\xi \in \mathbb{R}$~\eqref{eq:MapRe} and get a reasonable argument for the coefficients. For $\xi > 0$, this is
\begin{align}
	S_{0,4}^\ast (\xi = \overline{\xi}) = 
	2 \log \xi - 2 \xi - \xi^2 -{2 \over 3} \xi^3 + \cdots
	\approx 2 \log \xi -1.99 \xi -0.99 \xi^2 -0.67 \xi^3 + \cdots \, ,
\end{align}
and for $\xi <0$
\begin{align}
S_{0,4}^\ast (\xi = \overline{\xi}) = 
2 \log (-\xi) + 2 \xi +\xi^2 + {2 \over 3} \xi^3 + \cdots
\approx 2 \log (-\xi) + 1.99 \xi +0.99 \xi^2 + 0.65 \xi^3 + \cdots \,  .
\end{align}

Given the Polyakov conjecture~\eqref{eq:StrebelPolyakov1}, it is also possible to find the accessory parameter. For this, we have to take the derivative of~\eqref{eq:S4xi} with respect to $\xi$. It will be handy to use the following identity:
\begin{align} \label{eq:derid}
{\p \over \p \xi} (|\xi|^m \cos(n\theta)) 
&= {|\xi|^{m-1} \over 2}  \bigg[  {m+n \over 2} \cos((n-1)\theta) +{m-n \over 2} \cos((n+1)\theta) \\
&\hspace{1in}+ i \left(  {m+n \over 2} \sin((n-1)\theta) - {m-n \over 2} \sin((n+1)\theta) \right)
\bigg] \nonumber \, .
\end{align}
for $m \geq 1, n \geq 0$. Note that the real part is even in $\theta$ while the imaginary part is odd, which is consistent with the conjugation of the accessory parameter. Using~\eqref{eq:derid}, the accessory parameter is given by
\begin{subequations} \label{eq:acc}
\begin{align}
	\mathrm{Re} (c) = {2 \over |\xi|} \cos(\theta)  &+\left[ - \left( {3 \over 2 e^2} + {e^2 \over 4} \right) \cos(\theta) + \cdots \right] \\
	&+|\xi| \left[-{3 e^2 \over 16} + \left(- {7 \over 24 } + {e^4 \over 192} \right) \cos(\theta) - \left({5 \over 4 e^2} +{e^2 \over 16} \right) \cos(2\theta) + \cdots \right] + \cdots \nonumber \, ,
\end{align}
and
\begin{align}
\mathrm{Im} (c) = {2 \over |\xi|} \sin(\theta) &+\left[ \left(- {3 \over 2 e^2} + {e^2 \over 4} \right) \sin(\theta) + \cdots \right] \\
&+|\xi| \left[- \left( {7 \over 24 } + {e^4 \over 192} \right) \sin(\theta) + \left(-{5 \over 4 e^2} +{e^2 \over 16} \right) \sin(2\theta) + \cdots \right] + \cdots \nonumber \, .
\end{align}
\end{subequations}
Again, comparing the real part with the exact solution in~\eqref{eq:Arctan} gives a reasonable agreement. We opt out to report this.

As we mentioned above, the expansion in $\xi$ can be pushed to higher orders. However, for our purposes, it is better to do this using the expansion of the classical conformal blocks in the elliptic nome $q=q(\xi)$ defined by
\begin{align} \label{eq:nome}
q(\xi) \equiv \exp \left[ - \pi {K(1-\xi) \over K(\xi) }\right] = {\xi \over 16} + {\xi^2 \over 32} + \mathcal{O}(\xi^3) \, ,
\hspace{0.5in}
K(\xi) = \int_{0}^1 {dt \over \sqrt{(1-t^2)(1- \xi t^2) } } \, ,
\end{align}
as this would improve the convergence and allow us to focus on the vertex region rather than being restricted to the region around the degeneration. Here $K(\xi)$ is the complete elliptic integral of the first kind. The classical conformal blocks in $q$-expansion for $\alpha_i=1$ is given in~\eqref{eq:qexp}.

So we instead make the ansatz
\begin{align} \label{eq:Ansatzq}
\alpha_s(\xi, \overline{\xi}) &= 
|q|^{1/2} \, ( a_0 + a_2 \cos(2t) + a_4 \cos(4t) + a_6 \cos(6t) + \cdots ) \nonumber \\
&+ |q|^{3/2} \, ( b_0 + b_2 \cos(2t) + b_4 \cos(4t) + b_6 \cos(6t)  + \cdots ) \nonumber \\
&+ |q|^{5/2} \, ( c_0 + c_2 \cos(2t) + c_4 \cos(4t) + c_6 \cos(6t) + \cdots ) + \cdots \, ,
\end{align}
with $q = |q| e^{it}$. Because of the symmetry of $q$, the cosine terms with odd arguments are absent. Plugging the ansatz in the saddle point equation~\eqref{eq:SaddlePoint} and solving perturbatively, we obtain
\begin{align} \label{eq:Resq}
\alpha_s(\xi, \overline{\xi}) &= 
|q|^{1/2} \left( 2 e - {4 \over e^3} \cos(2t) - {58 \over e^7}\cos(4t) - {2948 \over 3 e^{11}} \cos(6t)  + \cdots \right) \nonumber \\
&+ |q|^{3/2} \left( -{e^3 \over 3}- {2 \over 3 e} \cos(2t) + {19 \over 3 e^5}\cos(4t) - {270 \over  e^9} \cos(6t) + \cdots \right) \nonumber \\
&+ |q|^{5/2} \left( {7 e^5 \over 180} + {13 e \over 90} \cos(2t) + {9 e \over 20 e^3}\cos(4t) + {13097 \over 270 e^7} \cos(6t)+ \cdots \right) + \cdots \, .
\end{align}
We work up to order $\mathcal{O}(|q|^{7 \over 2}, \cos(12t))$, but we only report to orders shown above for brevity.

\begin{figure}[h]
	\centering
	\includegraphics[width=0.75\textwidth]{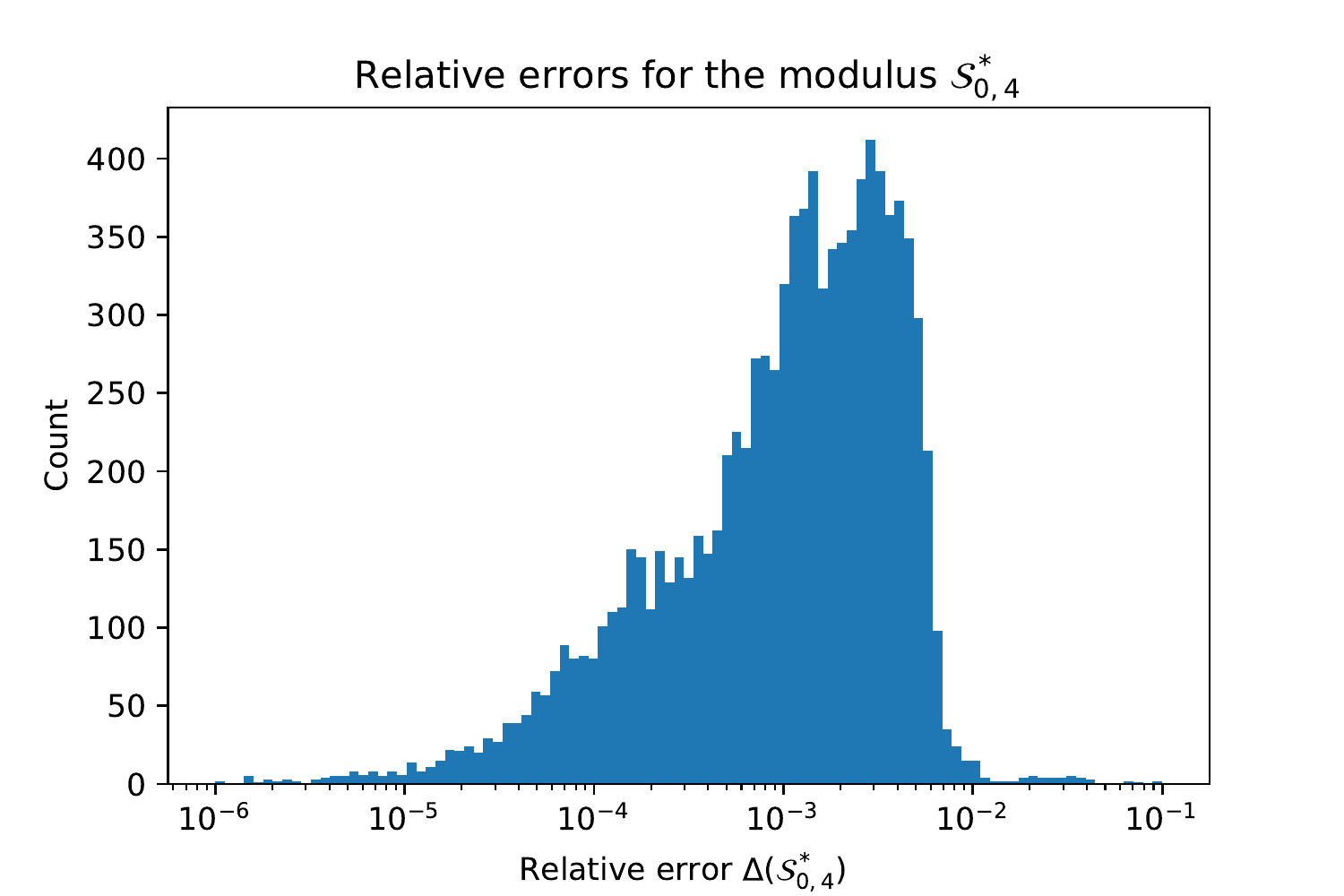}
	\caption{\label{fig:mod_errors}The distribution of the relative errors~\eqref{eq:relerror} for randomly sampled $10^4$ points in $\mathcal{M}_{0,4}$.}
\end{figure}

Inserting~\eqref{eq:Resq} to~\eqref{eq:mod} we further obtain
\begin{align} \label{eq:modulusq}
	&\mathcal{S}_{0,4}^\ast(\xi) =
	2 \log |\xi | + 2 \log |1- \xi| 
	+ 8 \log \left|{2 \over \pi} K(\xi)\right| + \widehat{\mathcal{S}}_{0,4}^\ast(q(\xi)) \, , \\
	&\widehat{\mathcal{S}}_{0,4}^\ast(q) = |q| \left[-4 e^2 - {16 \over e^2} \cos(2t) -{72 \over e^4} \cos(4t) - {2080 \over 3 e^6} \cos(6t) + \cdots \right] \nonumber \\
	&\hspace{0.5in}+ |q|^2 \left[ {2 e^4 \over 3} - {16 \over 3} \cos(2t) + {104 \over 3 e^4} \cos(4t) - {896 \over 9 e^8} \cos(6 t)+ \cdots \right] \nonumber \\
	&\hspace{0.5in} + |q|^3 \left[- {4 e^6 \over 45} -{32 e^2 \over 45} \cos(2t) - {344 \over 45 e^2} \cos(4 t) + {448 \over 5 e^6} \cos(6t) + \cdots \right] + \cdots \, . \nonumber
\end{align}
In figure~\ref{fig:mod_errors} we show how this expansion compares against the modulus $\mathcal{S}_{0,4}^{NN}(\xi)$ generated using the neural networks (NN) of~\cite{Erbin:2022rgx}. Notice we have chosen to plot the relative error between these two results defined by 
\begin{align} \label{eq:relerror}
	\Delta(\mathcal{S}_{0,4}^\ast(\xi)) = \left|1 - {\mathcal{S}_{0,4}^\ast(\xi) \over \mathcal{S}_{0,4}^{NN}(\xi)} \right| \, .
\end{align}
We observe that the errors are quite small and have the expected order of magnitude from~\cite{Erbin:2022rgx}.

We can generate the accessory parameter using the replacement rule similar to~\eqref{eq:acc}, with $\xi$ is replaced with $q$ and $\theta$ is replaced with $t$. Progressing analogously we get
\begin{align} \label{eq:C}
	c(\xi, \overline{\xi}) = {2 \over \xi (1-\xi) } \left[{2E(\xi) \over K(\xi) } - 1\right]
	+ {\pi^2 q \,  \widetilde{c} (q(\xi), \overline{q}(\overline{\xi})) \over 2 \xi (1-\xi)  K(\xi)^2 },
	\quad \quad
	\widetilde{c} (q, \overline{q}) \equiv {\p \widehat{\mathcal{S}}_{0,4}^\ast \over \p q} \, .
\end{align}
Here $E(\xi)$ is the complete elliptic integral of the second kind defined as
\begin{align}
E(\xi) = \int_0^1 dt \sqrt{1- \xi t^2 \over 1-t^2} \, ,
\end{align}
coming from the derivatives
\begin{align}
	{d K(\xi) \over d \xi} = {E(\xi) - (1-\xi) K(\xi) \over 2 \xi (1-\xi)},
	\hspace{0.5in}
	{d q(\xi) \over d \xi} = {\pi^2 q(\xi) \over 4\xi (1-\xi) K(\xi)^2} \, .
\end{align}
The real and imaginary parts of $\widetilde{c}$ in~\eqref{eq:C} is given by
\begin{subequations}
\begin{align}
	\mathrm{Re}(\widetilde{c})
	&= \left[\left( -{12 \over e^2} - 2 e^2 \right) \cos(t)+ \left(-{90 \over e^6} + {4 \over e^2} \right) \cos(3t) + \left(-{3640 \over 3 e^{10} } + {54 \over e^6} \right) \cos(5t)  + \cdots \right] \\
	&+ |q| \left[ \left(- {16 \over 3} + {2 e^4 \over 3} \right) \cos(t) + {52 \over e^4} \cos(3t) + \left(-{1792 \over 9 e^8} - {52 \over 3 e^4} \right) \cos(5t) + \cdots \right] + \cdots \nonumber \, ,
\end{align}
and
\begin{align}
	\mathrm{Im}(\widetilde{c})
	&= \left[ \left(-{12 \over e^2} + 2 e^2 \right) \sin(t) + \left( - {90 \over e^6} - {4 \over e^2} \right) \sin(3t) - \left({3640 \over 3 e^{10}} + {54 \over e^6} \right) \sin(5t) + \cdots \right] \\
	&+ |q| \left[ - \left({16 \over 3} + {2e^4 \over 3} \right) \sin(t) + {52 \over e^4} \sin(3t) + \left(-{1792 \over 9e^8} + {52 \over 3 e^4} \right) \sin(5t) + \cdots
	\right] + \cdots \, . \nonumber
\end{align}
\end{subequations}

\begin{figure}[h]
	\centering
	\includegraphics[width=0.49\textwidth]{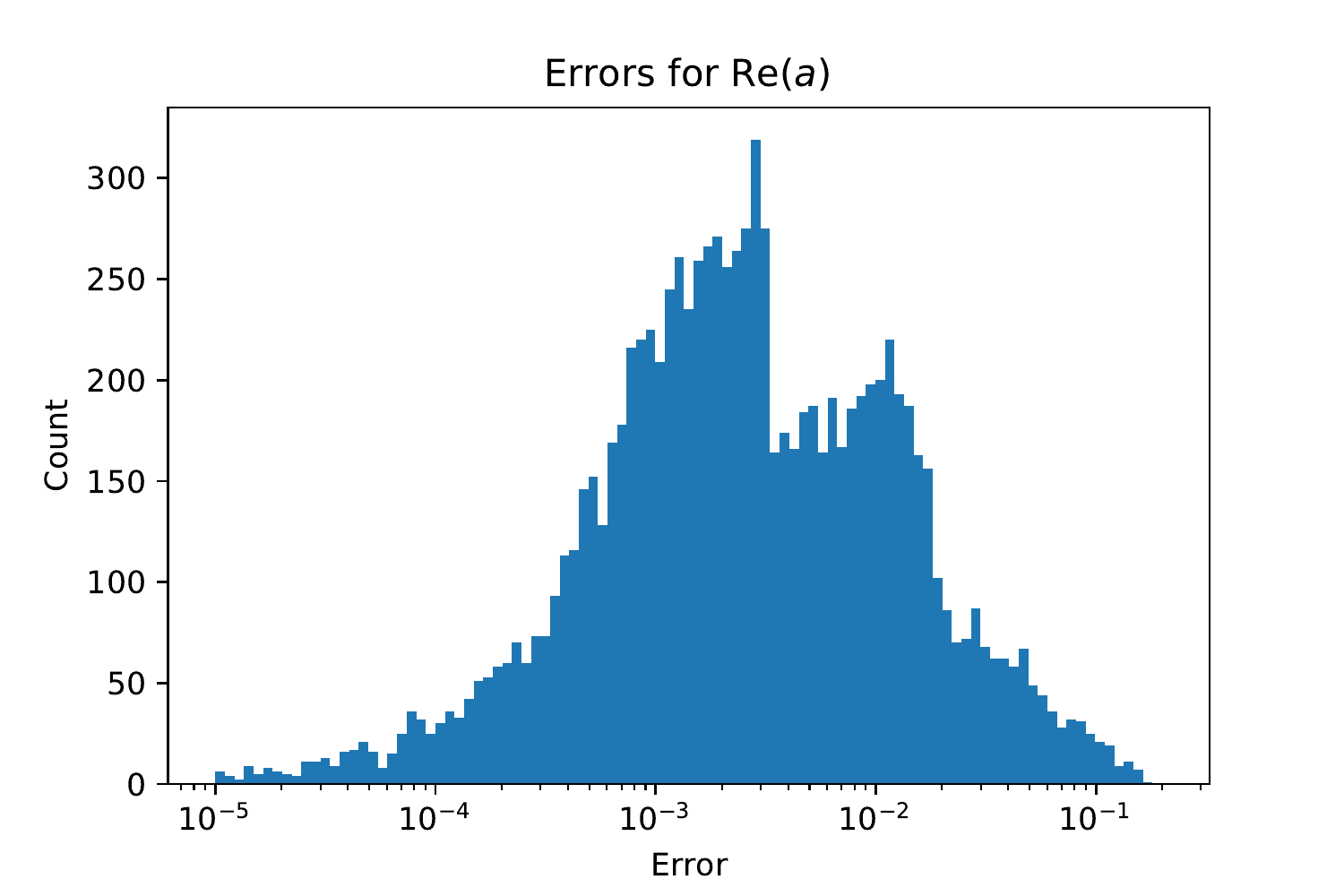}
	\includegraphics[width=0.49\textwidth]{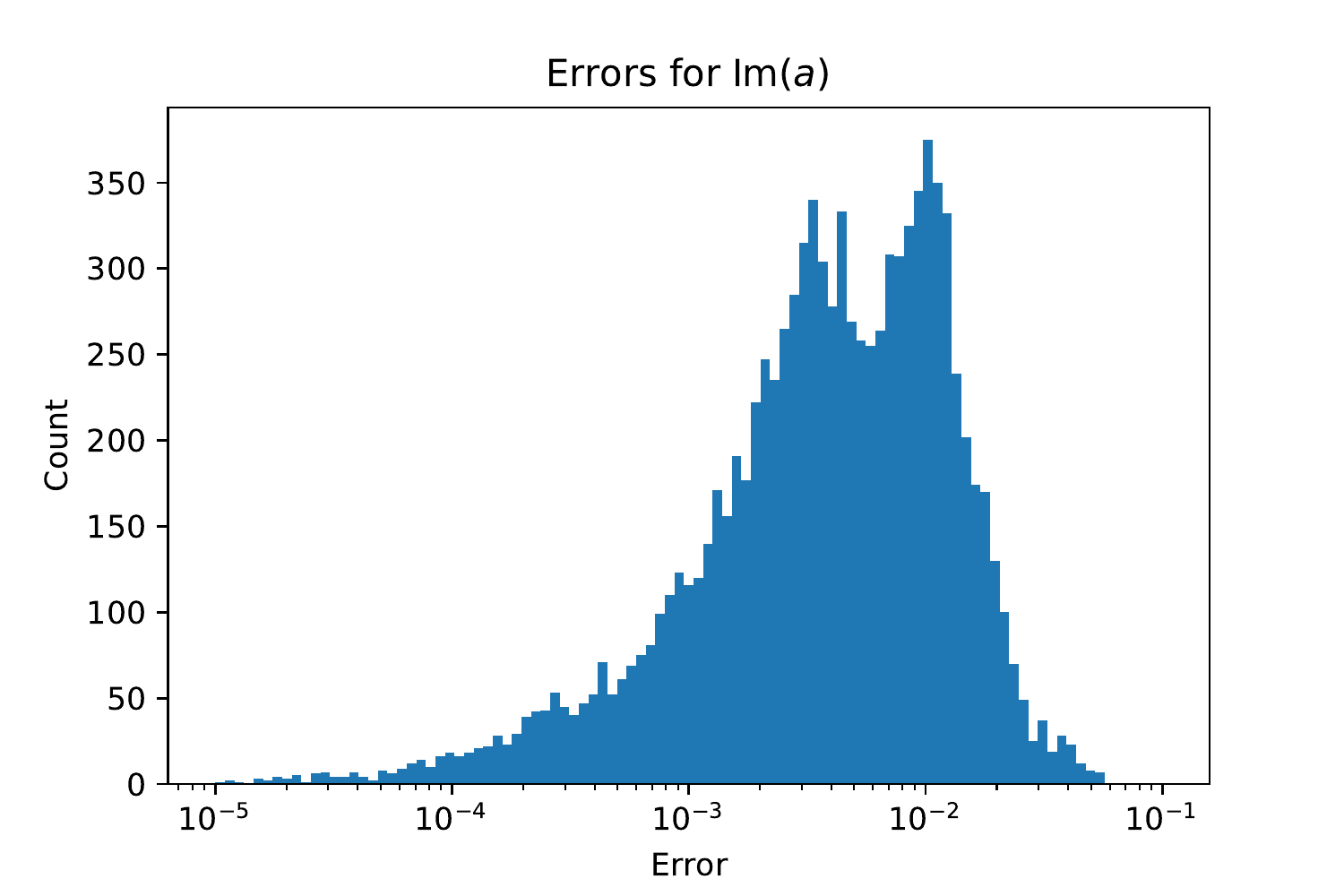}
	\caption{\label{fig:a_errors}The distribution of the absolute errors in the real and imaginary parts of the accessory parameter~\eqref{eq:a} for randomly sampled $10^4$ points in $\mathcal{M}_{0,4}$.}
\end{figure}

In the parametrization of~\cite{Erbin:2022rgx,Moeller:2004yy}, the accessory parameter is given by
\begin{align} \label{eq:a}
	a(\xi, \overline{\xi}) = 
	2 + \xi(\xi-1)\, c(\xi, \overline{\xi})
	=4 - {4 E(\xi) \over K(\xi)} - { \pi^2 q \, \widetilde{c}(q, \overline{q}) \over 2 K(\xi)^2 } \, .
\end{align}
Again, we compare this result with those obtained in~\cite{Erbin:2022rgx} by separately measuring the absolute errors for the real and imaginary parts of $a$. This is shown in figure~\ref{fig:a_errors}. The results are consistent with~\cite{Erbin:2022rgx}. We note that one can equivalently use~\eqref{eq:IntroAlt} to read off the accessory parameter. 

We emphasize that the accessory parameters entirely characterizes the Strebel differential $\varphi^{(S)} = \phi(z)dz^2$. Subsequently, it is possible to derive the local coordinates $h_i(w)$ from a unit disk to the uniformizing surfaces by solving the equation
\begin{align} \label{eq:StrebelLoc}
	\phi (h_i(w) ) \left( {d h_i(w) \over dw} \right)^2 = - {1 \over w^2} \, ,
\end{align} 
order-by-order in $w$, as explained in~\cite{Moeller:2004yy,Erbin:2022rgx}. The mapping radii associated with the individual punctures can be evaluated using the integral~\eqref{eq:map_rad}, see~\cite{Moeller:2004yy,Erbin:2022rgx}.

Alternatively, we can use the modulus to calculate them. This amounts to taking the WKB limit of the equation~\eqref{eq:connection} to realize
\begin{align}
	{\p \mathcal{S}_{0,n}^\ast(\xi_i; \alpha_i) \over \p \alpha_j} = 2 \alpha_j \log r_j[H_j] \, .
\end{align}
Let us compute the mapping radius $r_1$ associated with the puncture at $z=\xi$. We notice
\begin{align} \label{eq:indmap}
	{\p \mathcal{S}_{0,4}^\ast(\xi) \over \p \alpha_1}
	= 
	{\p \over \p \alpha_1} \left[
	\mathcal{S}_{0,3}^\ast(\alpha_1, 1, \alpha) 
	-2 f_{\alpha^2}
	\begin{bmatrix}
	1 & 1\\
	1& \alpha^2_1
	\end{bmatrix} (\xi)
	-2 \overline{f_{\alpha^2}}
	\begin{bmatrix}
	1 & 1\\
	1 & \alpha^2_1
	\end{bmatrix} (\overline{\xi})
	\right]_{\substack{\alpha = \alpha_s \\ \alpha_1 = 1}} \, ,
\end{align}
as the terms multiplying the derivative of $\alpha_s$ with respect to $\alpha_1$ add up to zero by the saddle point equation, just like for~\eqref{eq:IntroAlt}. A quick computation shows, for $0 \leq \alpha \leq 2$,
\begin{align} \label{eq:modder}
	{\p \mathcal{S}_{0,3}^\ast(\alpha, \alpha_1, 1)  \over \p \alpha_1}
	\bigg|_{ \alpha_1 = 1 }
	= \log \left[{16 \, (2-\alpha)^{-(2-\alpha)} \, (2+\alpha)^{-(2+ \alpha)} } \right] 
	= -{\alpha^2 \over 2}  - {\alpha^4 \over 48}- {\alpha^6 \over 480}  + \mathcal{O}(\alpha^8) \, ,
\end{align}
and the derivative of the classical conformal block with respect to $\alpha_1$ is
\begin{align} \label{eq:CCBder}
	 {\p f_{\alpha^2} \over \p \alpha_1}
	\begin{bmatrix}
	1 & 1\\
	1& \alpha^2_1
	\end{bmatrix} (\xi) \bigg|_{ \alpha_1 = 1} &=
	-{1 \over 2} \log \xi - {1 \over 2 } \log (1- \xi) - \log \left[{2 \over \pi} K(\xi) \right] \\
	&\hspace{0.5in} + {16 \over \alpha^2} \, q^2 + \left[  {48 \over \alpha^2}  -{576 \over \alpha^4} + {1280 \over \alpha^6} \right] q^4 + \cdots \, . \nonumber
\end{align} 
Then we see the mapping radius $r_1$ associated with the puncture $z=\xi$ is given by
\begin{align} \label{eq:argbelow}
	\log r_1 &= \log |\xi| + \log |1 - \xi| + 2 \log \left| {2 \over \pi} K(\xi) \right| + \\
	& |q| \left[ -e^2 - {4 \over e^2} \cos(2t) - {18 \over e^4} \cos(4t) + \cdots \right] \nonumber +|q|^2 \left[ {e^4 \over 6} - {4 \over 3} \cos(2t) + {26 \over 3e^4} \cos(4t) + \cdots \right] +\cdots \, .
\end{align}

\begin{figure}[h]
	\centering
	\includegraphics[width=0.75\textwidth]{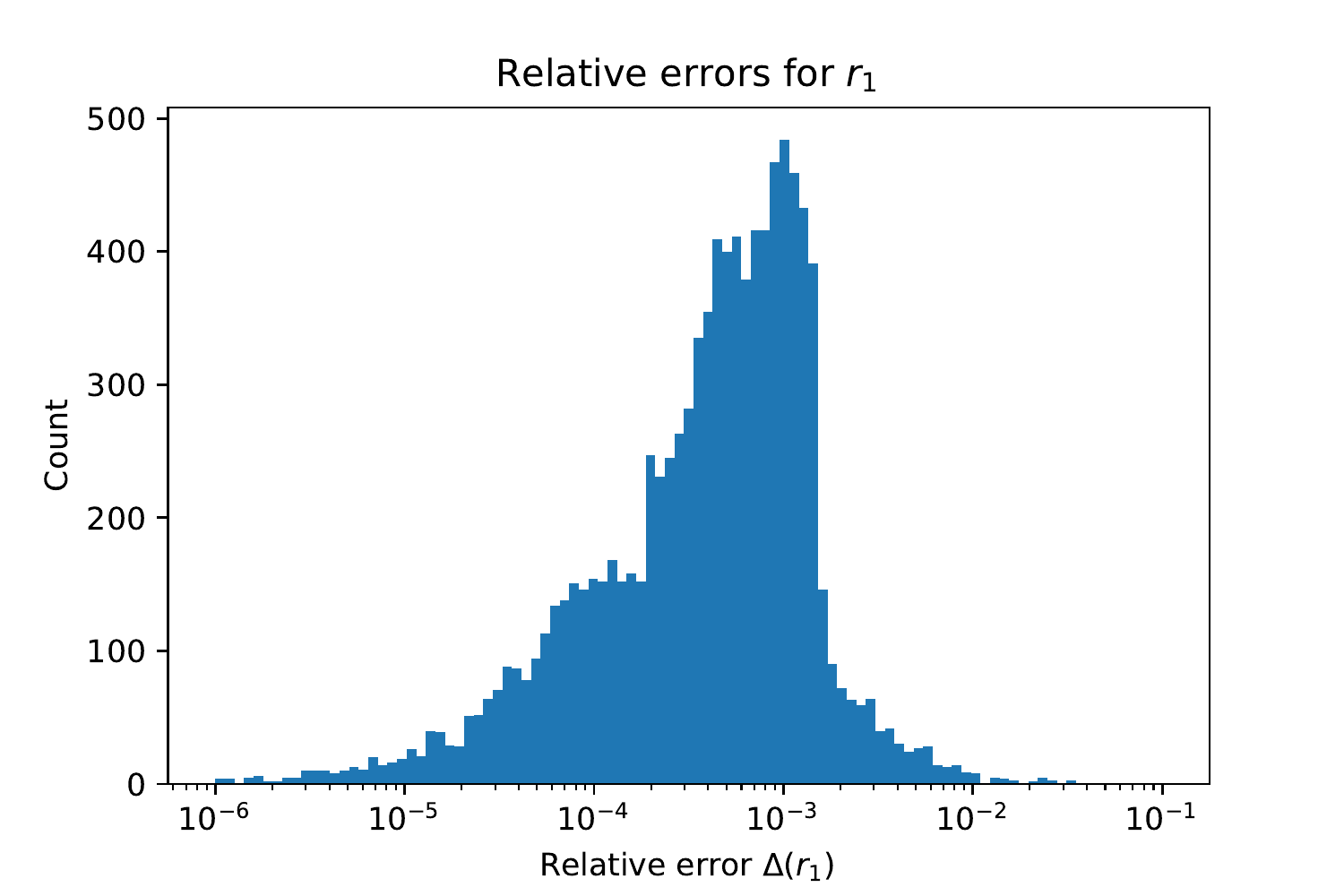}
	\caption{\label{fig:mod_errors_xi} The distribution of the relative errors in the mapping radius $r_1$ associated with the puncture at $z=\xi$ for randomly sampled $10^4$ points in $\mathcal{M}_{0,4}$.}
\end{figure}

In fact, solving for the remaining mapping radii is trivial. This is because the expression in~\eqref{eq:modder} stays the same, while~\eqref{eq:CCBder} only gets modified by the appearance/disappearance of the terms $\log |\xi|$ and $\log |1 - \xi|$ for $\log r_2, \log r_3, \log r_4$. The remaining parts of~\eqref{eq:CCBder} stays the same due to the symmetry of exchanging external weights. For $z =\xi$, both $\log |\xi|$ and $\log|1 - \xi|$ has appeared, but for $z=0 (1)$ only $\log |\xi|$ ($\log|1 - \xi|$) appears and for $z=\infty$ neither of them appears. Since the sum of the logarithms of the mapping radii adds up to $\mathcal{S}_{0,4}^\ast(\xi)$ by definition, we have
\begin{align}\label{eq:MapRel}
	\log r_1 &= {1 \over 4 }\mathcal{S}_{0,4}^\ast(\xi) + {1 \over 2} \log |\xi| + {1 \over 2}  \log|1- \xi| \, ,  \hspace{0.3in}
	\log r_2  = {1 \over 4 }\mathcal{S}_{0,4}^\ast(\xi)  + {1 \over 2} \log |\xi| - {1 \over 2}  \log|1- \xi|\, ,\\
	\log r_3 &= {1 \over 4 }\mathcal{S}_{0,4}^\ast(\xi)  - {1 \over 2} \log |\xi| + {1 \over 2}  \log|1- \xi|\, , \hspace{0.3in}
	\log r_4 = {1 \over 4 }\mathcal{S}_{0,4}^\ast(\xi)  - {1 \over 2} \log |\xi| - {1 \over 2}  \log|1- \xi| \, ,  \nonumber
\end{align}
which is manifestly true for~\eqref{eq:modulusq} and~\eqref{eq:argbelow}. Again, these are consistent with the results of~\cite{Erbin:2022rgx}. For example, the relative errors for $r_1$ is shown in figure~\ref{fig:mod_errors_xi}. We note that the relation~\eqref{eq:MapRel} leads to the following ratios:
\begin{align} \label{eq:ratios}
	{r_2 \over r_1} = {1 \over |1-\xi|}\, , \quad \quad
	{r_3 \over r_1} = {1 \over |\xi|}\, , \quad \quad
	{r_4 \over r_1} = {1 \over |\xi||1-\xi| } \, .
\end{align}
These are already argued from the $\mathbb{Z}_2 \times \mathbb{Z}_2$ symmetry of the symmetric quartic vertex in~\cite{Belopolsky:1994bj} and are consistent with them. One can check the mapping radii for $\xi \in \mathbb{R}$ given in~\eqref{eq:Rmap} satisfy~\eqref{eq:ratios}.

\subsection{The crossing symmetry} \label{sec:cross}

In this subsection we verify that there is a crossing symmetry among different decompositions of the modulus. More precisely, this entails checking the equations
\begin{align} \label{eq:crossing}
\mathcal{S}_{0,4}^\ast(\xi;\alpha_i) 
&=
{ \mathcal{S}_{0,3}^\ast (\alpha_s, \alpha_3, \alpha_4) }
+ { \mathcal{S}_{0,3}^\ast (\alpha_1, \alpha_2, \alpha_s) }  -2 f_{\alpha_s^2}
\begin{bmatrix}
\alpha_3^2 & \alpha^2_2\\
\alpha^2_4 & \alpha^2_1
\end{bmatrix} (\xi)
-2 \overline{f_{\alpha_s^2}}
\begin{bmatrix}
\alpha^2_3 & \alpha^2_2\\
\alpha^2_4 & \alpha^2_1
\end{bmatrix} (\overline{\xi})  \\
&\hspace{-0.5in}=
{ \mathcal{S}_{0,3}^\ast (\alpha_t, \alpha_1, \alpha_4) }
+ { \mathcal{S}_{0,3}^\ast (\alpha_3, \alpha_2, \alpha_t) } \nonumber
-2 f_{\alpha_t^2}
\begin{bmatrix}
\alpha_1^2 & \alpha^2_2\\
\alpha^2_4 & \alpha^2_3
\end{bmatrix} (1-\xi)
-2 \overline{f_{\alpha_t^2}}
\begin{bmatrix}
\alpha^2_1 & \alpha^2_2\\
\alpha^2_4 & \alpha^2_3
\end{bmatrix} (1-\overline{\xi}) \\
&\hspace{-0.5in}= 
{ \mathcal{S}_{0,3}^\ast (\alpha_u, \alpha_3, \alpha_1) }
+ { \mathcal{S}_{0,3}^\ast (\alpha_4, \alpha_2, \alpha_t) } 
 -2 f_{\alpha_u^2}
\begin{bmatrix}
\alpha_3^2 & \alpha^2_2\\
\alpha^2_1 & \alpha^2_4
\end{bmatrix} \left({1 \over \xi}\right)
-2 \overline{f_{\alpha_u^2}}
\begin{bmatrix}
\alpha^2_3 & \alpha^2_2\\
\alpha^2_1 & \alpha^2_4
\end{bmatrix} \left({1 \over \overline{\xi}}\right) 
+2 \alpha_2^2 \log |\xi|
\nonumber
\, ,
\end{align}
where $\alpha_s, \alpha_t$ and $\alpha_u$ are the solutions to the saddle point equation~\eqref{eq:SaddlePoint} and variations thereof. Note that for $u$-channel we have included an extra term to the modulus due to inversion. For equal external weights, it is easy to confirm that
\begin{align}
	\alpha_t (\xi) = \alpha_s(1-\xi), \hspace{0.5in}
	\alpha_u(\xi) = \alpha_s\left({1 \over \xi}\right) \, ,
\end{align}
by symmetry considerations, which implies that we can use the expansion~\eqref{eq:Resq} to compute $\alpha_t, \alpha_u$ by changing the argument of the elliptic nome $q$. 

\begin{figure}[h]
	\centering
	\includegraphics[width=0.49\textwidth]{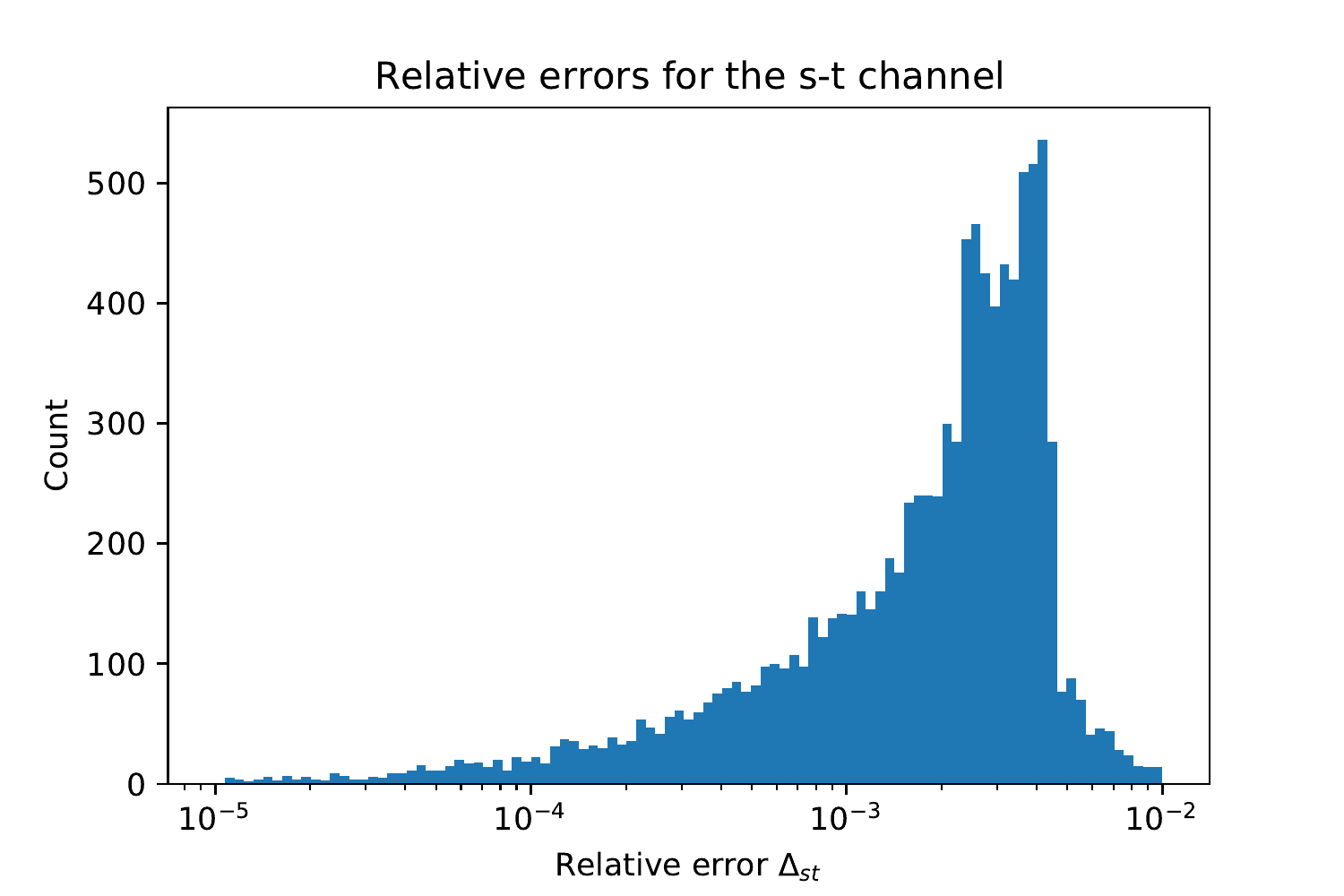}
	\includegraphics[width=0.49\textwidth]{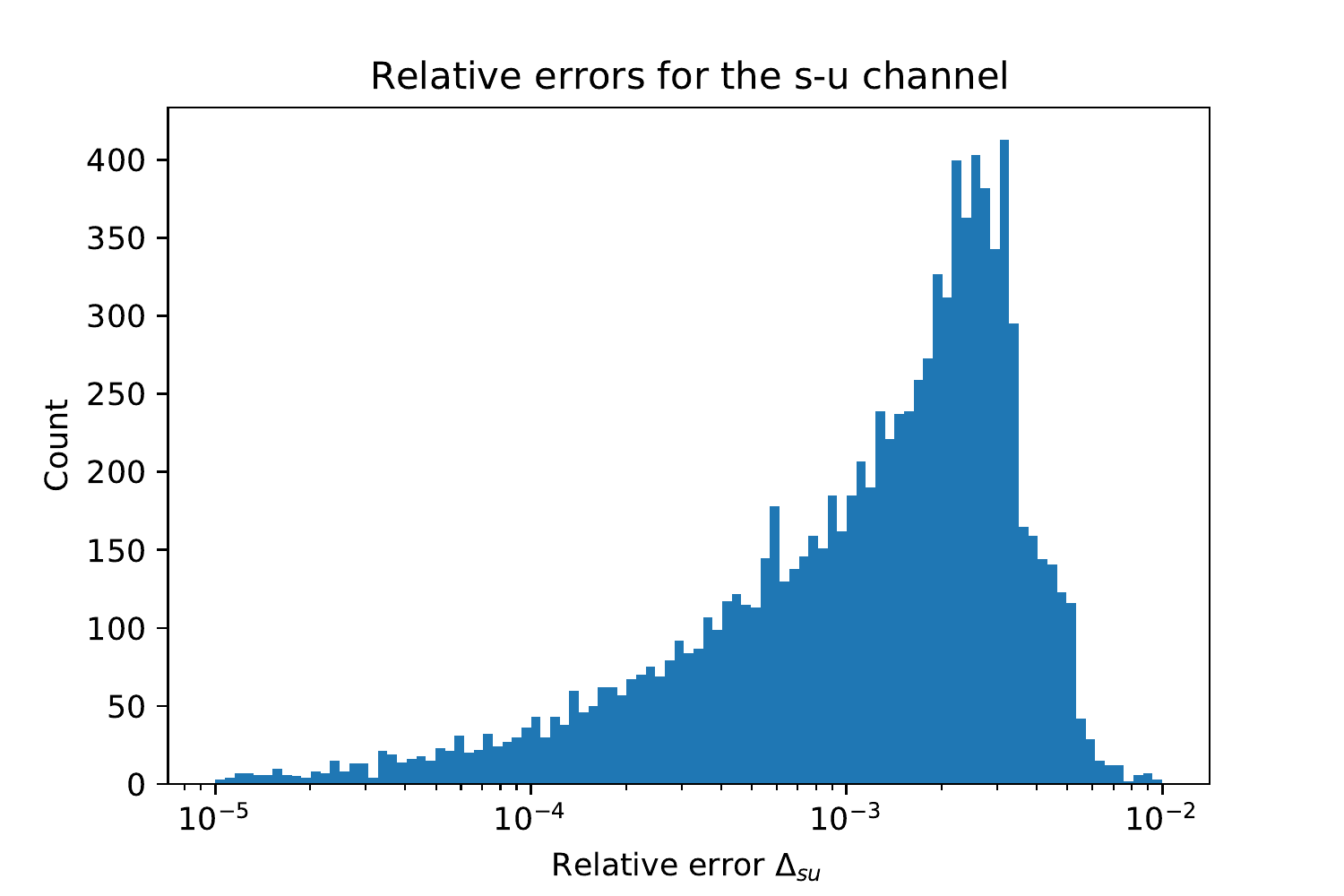}
	\caption{\label{fig:mod_errors_s}The distribution of the relative errors~\eqref{eq:cross} for randomly sampled $10^4$ points in $\mathcal{M}_{0,4}$.}
\end{figure}

Since we have constructed an operator formalism based on the DOZZ formula, it is expected that~\eqref{eq:crossing} is satisfied and we can use it to test the consistency of our framework. Like in the previous subsection, we set $\alpha_i = 1$ for convenience, but the results here would remain unchanged even if this is not the case. So, define the relative errors, similar to~\cite{Hadasz:2005gk},
\begin{align} \label{eq:cross}
	\Delta_{st} \equiv \left| 1 - {\mathcal{S}_{0,4}^{\ast, t}(\xi) \over \mathcal{S}_{0,4}^{\ast, s}(\xi)} \right|, 
	\hspace{0.5in}
	\Delta_{su} \equiv \left| 1 - {\mathcal{S}_{0,4}^{\ast, u}(\xi) \over \mathcal{S}_{0,4}^{\ast, s}(\xi)} \right| \, ,
\end{align}
to measure the deviation away from the crossing symmetry. The superscript on them indicates the channel the modulus is decomposed. These quantities should be zero all across the moduli space and this is what we numerically observe in figure~\ref{fig:mod_errors_s}.

\begin{figure}[h]
	\includegraphics[width=0.65\textwidth]{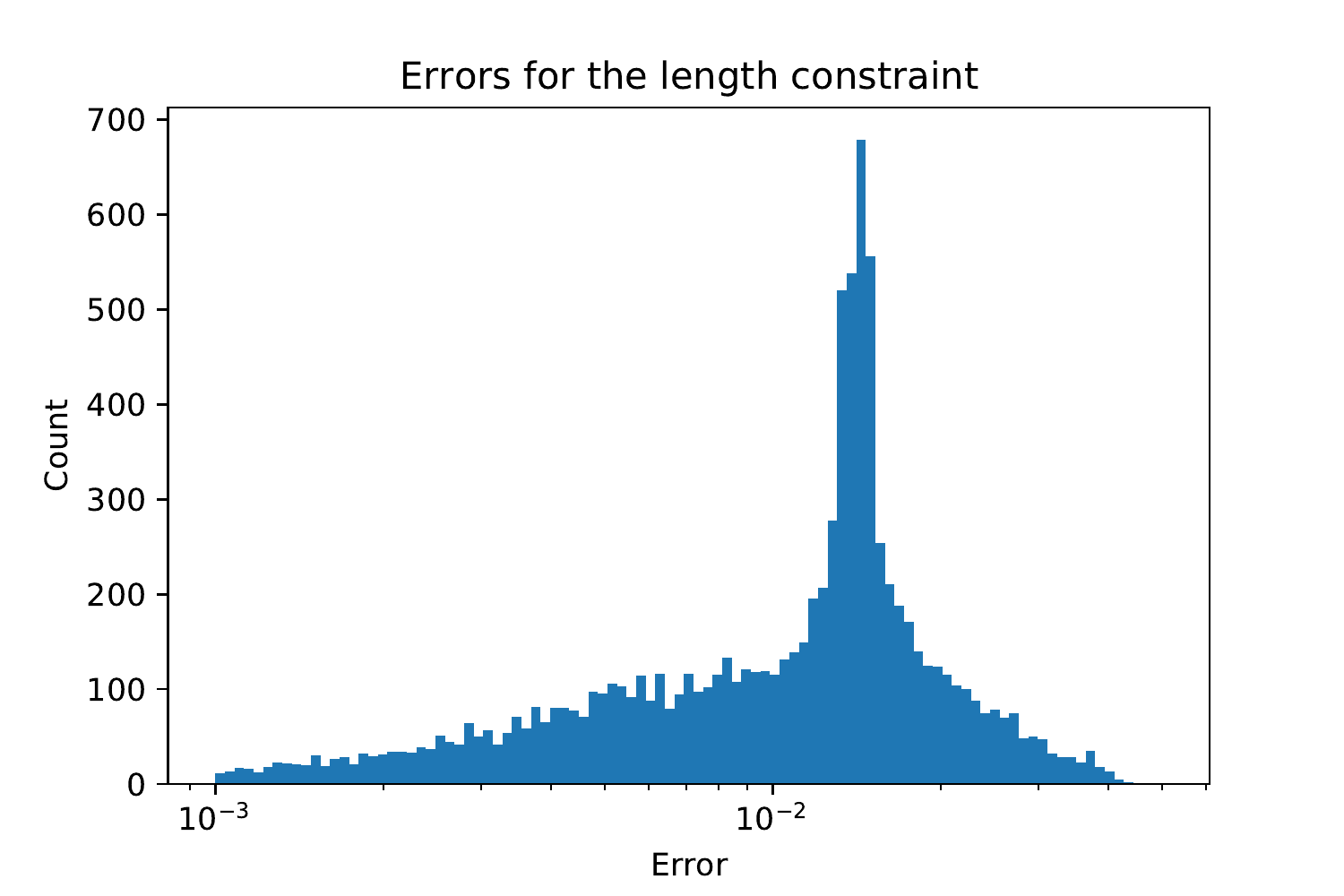}
	\parbox[t]{0.2\textwidth}{
		\vspace{-2in}
		\begin{tabular}{| l | r |} 
			\hline
			mean & 0.0114\\
			\hline
			std & 0.0076\\
			\hline
			median & 0.0121\\
			\hline
		\end{tabular}
	}
	\caption{\label{fig:len_sum_errors}The distribution of $|\alpha_s + \alpha_t + \alpha_u - 4|$ for randomly sampled $10^4$ points in $\mathcal{M}_{0,4}$ and its statistics.}
\end{figure}

Moreover, the geometry of the Strebel differentials with $\alpha_i = 1$ implies that 
\begin{align}
\alpha_s + \alpha_t + \alpha_u = 4 \, ,
\end{align}
over the moduli space $\mathcal{M}_{0,4}$. This is also shown to hold true in figure \ref{fig:len_sum_errors} by investigating its absolute errors. 

\subsection{Jenkins-Strebel differentials and the Feynman region} \label{sec:Feynman}

In this subsection we take $s>0$ to include the string propagators to the geometry, and correspondingly, to obtain the Jenkins-Strebel differentials. The saddle point equation~\eqref{eq:SaddlePoint} now reads
\begin{align}  \label{eq:SaddlePoints}
	s = {1 \over 2 \alpha_s} {\p \over \p \alpha} \, \Bigg[
	{\mathcal{S}_{0,3}^\ast (\alpha, \alpha_3, \alpha_4) }
	+ { \mathcal{S}_{0,3}^\ast (\alpha_1, \alpha_2, \alpha) } 
	-2 f_{\alpha^2}
	\begin{bmatrix}
	\alpha_3^2 & \alpha^2_2\\
	\alpha^2_4 & \alpha^2_1
	\end{bmatrix} (\xi)
	-2 \overline{f_{\alpha^2}}
	\begin{bmatrix}
	\alpha^2_3 & \alpha^2_2\\
	\alpha^2_4 & \alpha^2_1
	\end{bmatrix} (\overline{\xi})
	\Bigg]_{\alpha = \alpha_s} \, .
\end{align}
This indicates that fixing $\alpha_s$ determines $s$ and vice-versa, given $\xi$. Recall that $2 \pi \alpha_s$ is the circumference of the string propagator, while $s$ is its length, so~\eqref{eq:SaddlePoints} actually demonstrates that these quantities are not independent from each other. This is somewhat expected from the theory of quadratic differentials~\cite{strebel1984quadratic} and~\eqref{eq:SaddlePoints} provides an explicit realization of this feature. In the last subsection we set $s=0$ and related $\xi_i$ and $\alpha_s$.

We remark that the solutions with $s>0$ to~\eqref{eq:SaddlePoints} given $\alpha$ and $\xi$ may not always exist. The most famous example being $\xi \in \mathcal{V}_{0,4}$ with $\alpha =\alpha_i = 1$---the elementary quartic interactions of covariant CSFT cannot be covered with the Feynman diagrams of the symmetric cubic vertex~\cite{Sonoda:1989sj}.

\begin{figure}[h]
	\centering
	\includegraphics[height=2.9in,width=0.49\textwidth]{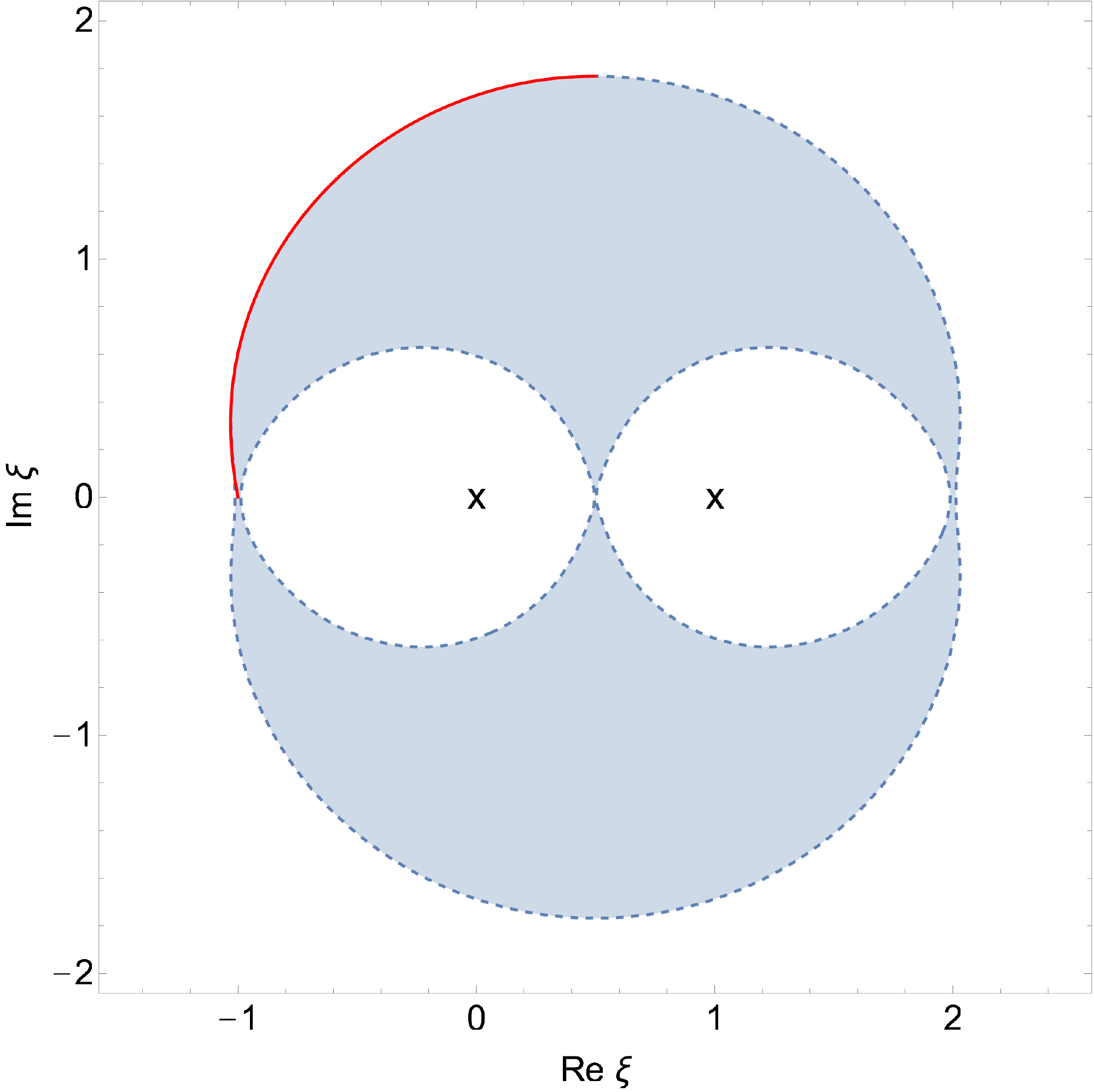}
	\includegraphics[height=3.0in,width=0.49\textwidth]{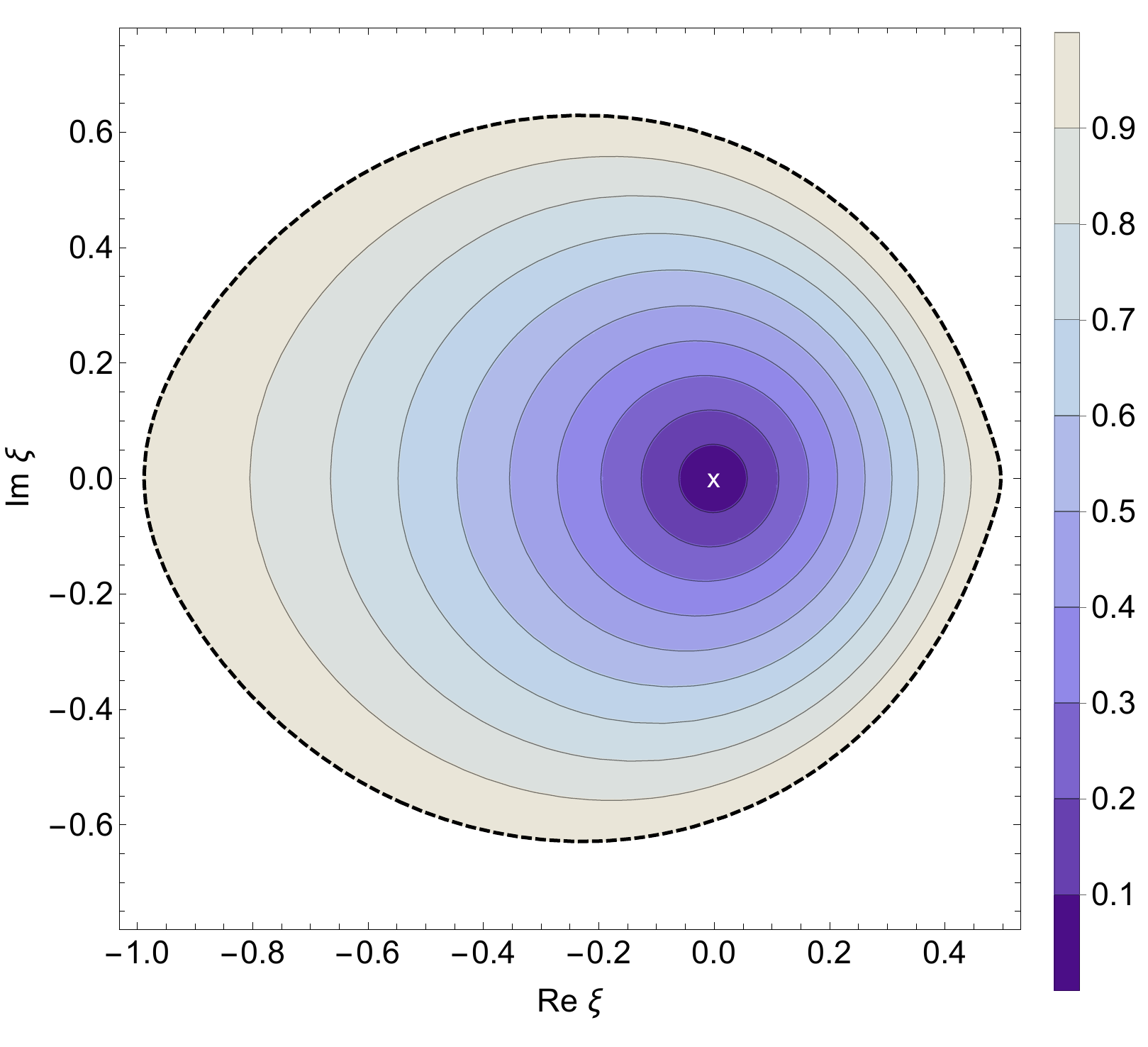}
	\caption{\label{fig:vertex}The vertex region $\mathcal{V}_{0,4}$ (left) and the constant $|\mathfrak{q}|$ contours on the Feynman region $\mathcal{F}_s$ (right). The blue dashed curve is $\p \mathcal{V}_{0,4}$ defined by~\eqref{eq:B} and the solid red curve is the numerical fit for the same curve given in~\cite{Moeller:2004yy} for the plot on the left.}
\end{figure}

The circumference of the string propagators in CSFT is $2 \pi$, so $\alpha_s =  1$. We also take $\alpha_i = 1$ and evaluate the resulting length of the cylinder by~\eqref{eq:SaddlePoints}. For this, we observe
\begin{align}
	{d \mathcal{S}_{0,3}^\ast (\alpha, 1,1) \over d \alpha} \bigg|_{\alpha = 1}
	= 2 \log {4 \over 3 \sqrt{3} } \, ,
\end{align}
and
\begin{align}
	{\p f_{\alpha^2} \over \p \alpha} (q)\bigg|_{\alpha = 1}
	&= {1 \over 2} \log 16 q - 32 q^2 -2400 q^4 -{1458560 \over 3} q^6 \\
	&\hspace{0.5in}- 131996640 q^8 - {209842963392 \over 5} q^{10}  
	- 14722716935552 q^{12} + \cdots \, , \nonumber
\end{align}
Using the fact that the Schwinger parameter of the string propagator is $\mathfrak{q} = e^{-s + i\theta}$, we evaluate the length $s$ of the string propagator to be
\begin{align} \label{eq:CylLen}
	s(\xi, \overline{\xi}) = - \log |\mathfrak{q}| &= 2 \log {4 \over 3 \sqrt{3} } 
	- \left[{\p f_{\alpha^2} \over \p \alpha} (q) \right]_{\alpha = 1}
	-\left[{\p \overline{f_{\alpha^2}} \over \p \alpha} (\overline{q}) \right]_{\alpha = 1} \\
	&= 2 \log {4 \over 3 \sqrt{3} } - \log 16 |q| + 64 \, \mathrm{Re}(q^2) + 4800 \, \mathrm{Re}(q^4) + \cdots \nonumber
	 \, ,
\end{align}
In particular, the curve $|\mathfrak{q}(\xi)| = 1$ describes the part of $\p \mathcal{V}_{0,4}$ separating the $s$-channel Feynman region $\mathcal{F}_s$ from the vertex region $\mathcal{V}_{0,4}$. This leads to the relation~\eqref{eq:DV}, which we report here again explicitly using~\eqref{eq:nome}:
\begin{align} \label{eq:B}
	{4 \over 3 \sqrt{3} } = \left| \exp {\p f_{\alpha^2} (q) \over \p \alpha}  \right|_{\alpha = 1} &=
	\left| 4q^{1/2} -128 q^{5/2} - 7552 q^{9/2} - 1659392 q^{13/2} + \cdots \right|
 	 \nonumber \\
 	&= \left|\xi^{1/2} + {\xi^{3/2} \over 4} + {\xi^{5/2} \over 128} - {35 \xi^{7/2} \over 512} 
 	-{ 8209 \xi^{9/2} \over 65536} + \cdots \right| \, .
\end{align}
Analogous curves for the $t$-and $u$ channels can be found by changing the argument of $q$ to $1- \xi$ and $1/\xi$. The comparison of these curves and the numerical fit provided in~\cite{Moeller:2004yy} is shown in figure~\ref{fig:vertex}.

The equation~\eqref{eq:CylLen} can be further used to obtain the Schwinger parameter $ \mathfrak{q}$ as a function of $\xi$. Recall that $\mathfrak{q}$ is a holomorphic function of $\xi$ from a unit disk $0 < |\mathfrak{q}| \leq 1$ to a given Feynman region~\cite{Zemba:1988rf,sonoda1990closed}. This map has to have a well-defined expansion and it should map $\mathfrak{q} =0$ to the degenerating surface (i.e. $\xi = 0$), while the relevant portion of $\p \mathcal{V}_{0,4}$ gets mapped by $| \mathfrak{q}(\xi)| = 1$. Given such constraints, there is a unique map $\mathfrak{q} = \mathfrak{q}(\xi)$ up to a phase\footnote{This is irrelevant for CSFT with level matching. Although it is expected that this would be important for CSFT without level matching, see~\cite{Okawa:2022mos, Erbin:2022cyb}.} by the Riemann mapping theorem and it is given by, using~\eqref{eq:nome},
\begin{align} \label{eq:Sch}
	\mathfrak{q} (\xi)= \left( {4 \over 3 \sqrt{3} } \right)^{-2} \exp \left[ 2\, {\p f_{\alpha^2} (q) \over \p \alpha} \right]_{\alpha=1} 
	&= \left( {4 \over 3 \sqrt{3} } \right)^{-2} \left( 16 q -1024 q^3 - 44032 q^5 + \cdots \right) \nonumber\\
	&=\left( {4 \over 3 \sqrt{3} } \right)^{-2} \left( \xi + {\xi^2 \over 2} + {5 \xi^3 \over 64 } - {17 \xi^4 \over 128} + \cdots \right)
	\, .
\end{align}
Using different arguments for $q$ produces the Schwinger parameters for other propagators. The constant $|\mathfrak{q} (\xi)|$ contours for the $s$-channel is shown in figure~\ref{fig:vertex}.

Now, let us return our attention back to $\mathcal{S}^\ast_{0,4}(\xi) $ for $s>0$ with $\alpha_s = \alpha_i = 1$. Denoting it by $\mathcal{S}_{0,4}^{\ast, (s>0)}(\xi) $ and using~\eqref{eq:mod} and~\eqref{eq:CylLen}, it reads
\begin{align} \label{eq:modulusJS}
	\mathcal{S}_{0,4}^{\ast, (s>0)}(\xi) 
	&= 4\log \left( {4 \over 3 \sqrt 3} \right) +
	 \left[{\p f_{\alpha^2} \over \p \alpha} (\xi) \right]_{\alpha = 1}
	+\left[{\p \overline{f_{\alpha^2}} \over \p \alpha} (\overline{\xi}) \right]_{\alpha = 1}
	-2 f_{1} (\xi)
	-2 \overline{f_{1}} (\overline{\xi}) \nonumber \\
	&= 
	4\log \left( {4 \over 3 \sqrt 3} \right) + 
	2 \log |\xi| + 2 \log|1-\xi| + 8 \log \left|{2 \over \pi} K(\xi)\right| \nonumber \\
	&\hspace{0.75in} -128 \; \mathrm{Re}(q^2) -6016 \;  \mathrm{Re}(q^4) -  
	{3401216 \over 3} \mathrm{Re}(q^6) + \cdots \, .
\end{align}
We first notice that this is an expansion in the real parts of $q^n$, unlike the double expansion of the modulus for the Strebel differentials~\eqref{eq:modulusq}. The choice of the argument for the elliptic nome $q$ determines the modulus for the $s,t,u$-channels. For $u$-channel one needs to include extra $2 \log |\xi|$ similar to~\eqref{eq:crossing}. The modulus of the previous section and~\eqref{eq:modulusJS} matches for $\xi \in \p \mathcal{V}_{0,4}$ by taking $s=0$ in~\eqref{eq:CylLen}.

We emphasize that~\eqref{eq:modulusJS} is \textit{not} the modulus of a Jenkins-Strebel differential: it just involves the sum of modulus for the punctures by~\eqref{eq:Stre}. On top of this, we have to include the modulus associated with the ring domain. So the modulus $\mathcal{S}_{0,4}^{\ast, (JS)}(\xi)$ for a Jenkins-Strebel differential is actually given by
\begin{align}
	\mathcal{S}_{0,4}^{\ast, (JS)}(\xi)  = \mathcal{S}_{0,4}^{\ast, (s>0)}(\xi)  + s 
	= 6\log \left( {4 \over 3 \sqrt 3} \right) 
	-2 f_{1} (\xi)
	-2 \overline{f_{1}} (\overline{\xi}) \, .
\end{align}

Given the modulus it is simple to derive the accessory parameters using~\eqref{eq:StrebelPolyakov1} and we find\footnote{This requires a slight modification to the Polyakov conjecture we used so far, see the discussion above~\eqref{eq:FlatPoly}. Clearly its WKB limit reduces to the modulus associated with Jenkins-Strebel differentials~\eqref{eq:Reducing}.}
\begin{align} \label{eq:JSacc}
	c^{(JS)} (\xi) = -4 {\p f_{1} (\xi) \over \p \xi} 
	= {2 \over \xi (1-\xi) } \left[{2E(\xi) \over K(\xi) } - 1\right]
	+ {\pi^2 q \,  \widetilde{c}^{(JS)} (q) \over 2 \xi (1-\xi)  K(\xi)^2 }
	 \, ,
\end{align}
with $\widetilde{c}^{JS}$ is given by, with $\mathcal{S}_{0,4}^{\ast, JS} $ representing the $q$-series in $\mathcal{S}_{0,4}^{\ast, (JS)}$,
\begin{align}
	\widetilde{c}^{(JS)}(q) \equiv  {\p \widehat{\mathcal{S}}_{0,4}^{\ast, JS} \over \p q} = 
	-{1 \over 2q}- 64 q - 2432 q^3 - 484096 q^5 - 127883008  q^7 + \cdots \, .
\end{align}
The accessory parameter for a Jenkins-Strebel differential is a holomorphic function of $\xi$ and its value along the boundary $\p \mathcal{V}_{0,4}$ matches with those of Strebel differential.

As a cross-check, let us investigate the degeneration behavior of the resulting Jenkins-Strebel differential. Expanding~\eqref{eq:JSacc} around $\xi = 0$ and upon using it in the quadratic differential, we find
\begin{align}
	c^{(JS)} (\xi)  = {1 \over \xi} + \mathcal{O} (\xi^0) 
	\implies
	\varphi^{(JS)} = 
	\left[ {-1 \over z^2} + {-1 \over (1-z)^2} + {-1 \over z} + {-1 \over 1-z} \right] dz^2 + \mathcal{O} (\xi) \, .
\end{align}
In the strict $\xi \to 0$ limit, this is the Strebel differential associated with the symmetric cubic vertex. This is as it should be: if~\eqref{eq:JSacc} is the accessory parameter resulting from sewing two symmetric cubic vertices, its associated quadratic differential must reduce to the Strebel differential above when two of the punctures collide~\cite{sonoda1990closed}.

Given the modulus and accessory parameters, the individual mapping radii and the local coordinates for the Feynman region can be derived similar to before. For brevity we don't report them. In principle these results should be consistent with sewing two cubic vertices, however obtaining the local coordinates through sewing remains open. Regardless, we now have a total analytic control as long as the geometry on 4-punctured spheres are concerned. In the next section we discuss the generalization to higher-string interactions.

\subsection{Higher-string interactions}

For the success of our framework, generalizing it to string interactions on $n-$punctured spheres with $n \geq 5$ is crucial. Unfortunately, the classical conformal blocks for $n$-point functions are not well-investigated and there is no good way to compute them efficiently, like in $n=4$~\cite{zamolodchikov1987conformal}. Therefore we only sketch the anticipated procedure in this subsection.

\begin{figure}[t]
	\centering
	\resizebox{15cm}{!}{%
		\begin{tikzpicture}[thick]
		\begin{scope}
		\node at (-1,0) {$\mathcal{S}_{0,n}^\ast (\xi_i)$};
		\node at (0,0) {$=$};
		\node at (1,0) {$0$};
		\draw[-] (1.5,0)--(10.5,0);
		\node at (11.1,0) {$\infty$};
		\draw[-] (2.5,0)--(2.5,1) node[above]{$\xi_1$};
		\draw[-] (4,0)--(4,1) node[above]{$\xi_2$};
		\node at (6,0.6) {$\cdots$};
		\draw[-] (8,0)--(8,1) node[above]{$\xi_{n-3}$};
		\draw[-] (9.5,0)--(9.5,1) node[above]{$1$};
		\node at (3.25,-0.5) {$(\alpha_1, s_1)$};
		\node at (8.75,-0.5) {$(\alpha_{n-3}, s_{n-3})$};
		\end{scope}
		\end{tikzpicture}
	}
	\caption{The decomposition of the modulus $\mathcal{S}_{0,n}^\ast (\xi_i)$ in a comb channel.}
	\label{fig:comb}
\end{figure}
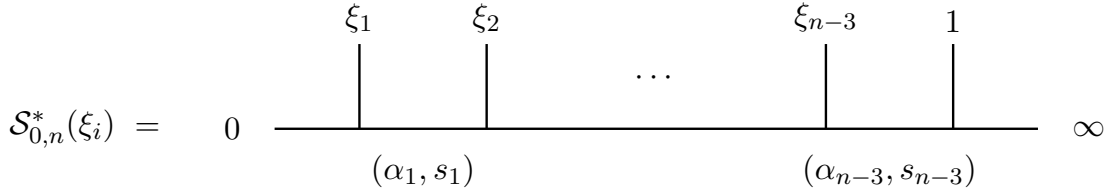

Repeating the arguments in previous subsections, we arrive at the following expression
\begin{align} \label{eq:nmodulus}
	\mathcal{S}_{0,n}^\ast(\xi_i) 
	&= \mathcal{S}_{0,3}^\ast (\alpha_1,1,1) +
	\sum_{k=1}^{n-4}\mathcal{S}_{0,3}^\ast(\alpha_{k+1},1, \alpha_k)
	+ \mathcal{S}_{0,3}^\ast (1,1, \alpha_{n-3}) 
	- \sum_{k=1}^{n-3} (\alpha_k^2 s_k) \nonumber \\
	&\hspace{2in}-2 f_{\{ \alpha_k^2 \}}( \xi_i ) -2 \overline{f_{\{ \alpha_k^2 \}}}( \overline{\xi}_i ) \, .
\end{align}
We have decomposed the $n$-punctured sphere in the comb channel shown in figure~\ref{fig:comb}. Different comb channels (or non-comb channels) give different decomposition of the surface, which are related by crossing-like symmetries similar to~\eqref{sec:cross}. Here $s_i$'s are the lengths of the string propagators if there is any. The functions $f_{\{ \alpha_k^2 \}}( \xi_i )$ are the classical conformal blocks for $n-$punctured spheres for equal external weights and internal weights $\alpha_k$ for $k=1,\cdots,n-3$. They depend on $n-3$ cross ratios. Here we assumed the conformal blocks for $n$-point functions have the exponentiation behavior akin to~\eqref{eq:ClassConfBlock} in the semi-classical limit.

Given~\eqref{eq:nmodulus}, it is possible to write $n-3$ saddle point equations, one for each $\alpha_k$. They would be coupled to each other via the classical conformal blocks $f_{\{ \alpha_k^2 \}}( \xi_i )$. Assuming $\alpha_{k,s}$ are the resulting saddle points, the lengths of the non-contractible geodesics in the comb channel shown in figure~\ref{fig:comb} are given by $2 \pi \alpha_{k,s}$. The modulus can be evaluated by setting $\alpha_k = \alpha_{k,s} (\xi_i, \overline{\xi_i})$ in~\eqref{eq:nmodulus}.

It is relatively straightforward to derive a simple expression for the accessory parameters in terms of classical conformal blocks
\begin{align}
	c_j (\xi_i, \overline{\xi_i})= -4 {\p f_{\{ \alpha_k^2 \}} (\xi_i) \over \p \xi_j} \bigg|_{\alpha_k = \alpha_{k,s}(\xi_i, \overline{\xi_i})}\, ,
\end{align}
analogous to~\eqref{eq:IntroAlt}. We can also find the individual mapping radii by differentiating the modulus with respect to external weights. With these, we can fully specify the local coordinates for the vertex region by solving~\eqref{eq:StrebelLoc}. Moreover, the boundary of the vertex region $\p \mathcal{V}_{0,n}$ as well as the Jenkins-Strebel differentials can be characterized using the saddle point equations after setting $\alpha_k = 1$ and $s_k = 0$ for one or multiple $k=1, \cdots, n-3$. Similarly, their local coordinates can be found.

With this note, our characterization of $n$-string contact interactions is complete for any $n \geq 3$ in principle. As we have argued, the only input was the modulus of the generalized cubic vertex~\eqref{eq:BaseCase1} and the classical conformal blocks at each order. The problem of specifying the background-independent geometric data of CSFT has transformed to finding the classical conformal blocks.

\section{Conclusion} \label{sec:conc}

In this paper, we have argued for the following:
\begin{enumerate}
	\item Hyperbolic string vertices can be generated by a properly regularized and modified on-shell Liouville action through the Polyakov conjecture~\eqref{eq:Polyakov1}. In particular, we have shown that this action~\eqref{eq:Lio5} reduces to the modulus $S_{0,n}^\ast$ of Strebel differentials~\eqref{eq:Reducing}, which we have used to establish a Polyakov-like conjecture for them~\eqref{eq:StrebelPolyakov1}. This has demonstrated that the modulus entirely determines the local coordinates for the classical minimal-area vertices. Since the modulus $S_{0,n}^\ast$ is related to the interactions of $n$ zero momentum tachyons~\eqref{eq:Tachyon}, this has also showed that the background-independent geometric data of classical CSFT is \textit{entirely} contained in the interactions of tachyons.
	
	\item The modulus $S_{0,n}^\ast$ can be constructed using $S_{0,3}^\ast(\alpha_1, \alpha_2,\alpha_3)$~\eqref{eq:BaseCase1} and the classical conformal blocks. This has been tested extensively for the symmetric quartic interactions and its vertex region is analytically characterized. An analogous construction for higher-string interactions works, but requires the knowledge of the higher-point classical conformal blocks.
	
\end{enumerate}

We find these results striking. It can be even said that the covariant closed string field theory is actually \textit{cubic} despite the famous no-go theorem in~\cite{Sonoda:1989sj}, but rather in an unconventional sense. We conclude the paper by listing some points for future investigations:

\begin{enumerate}
	\item The classical conformal blocks now appear to be one of the central ingredients for the hyperbolic CSFT. Understanding their behavior better is the essential next step for the success of this approach. Ideas in~\cite{Menotti:2016jut,Piatek:2021aiz} may help in this direction. Also the role of the Virasoro algebra in the semi-classical limit and its relation to hyperbolic geometry should be clarified, since it previously appeared in a similar context, see~\cite{Kravchuk:2021akc, Bonifacio:2021aqf, Mahanta:2022fvl}. It may be possible to address some of these questions in the context of AdS/CFT~\cite{Alkalaev:2015wia, Hijano:2015qja,Alkalaev:2015lca,Hijano:2015zsa,Chen:2016dfb,Alkalaev:2016rjl, Belavin:2017atm,Alkalaev:2018nik,Alkalaev:2019zhs} and/or AGT correspondence~\cite{ alday2010liouville,Alba:2010qc,Tai:2010ps,Piatek:2011tp,Ferrari:2012gc}. In passing, we point out Strebel differentials have found applications in the worldsheet approach to AdS/CFT, see~\cite{Gopakumar:2005fx,Gaberdiel:2020ycd,Bhat:2021dez,Knighton:2022ipy,Gopakumar:2022djw}. In particular, the on-shell Liouville action made a curious appearance in~\cite{Gaberdiel:2020ycd}. It may be helpful to flesh out its relation to our claims in this paper.
	
	\item Even though the parametrization of classical CSFT is entirely specified, the bootstrap program doesn't inform us how to perform the moduli integration over the vertex region. A hope here would be to generate a topological recursion for the effective tachyon potential in the lines of~\cite{mirzakhani2007simple, mirzakhani2007weil, andersen2017geometric, andersen2019topological}. An alternative possibility is to introduce a set of auxiliary string fields, together with their possibly non-standard propagators, to make the hidden cubic nature of CSFT manifest and eliminate the moduli integration all together.\footnote{The author thanks Harold Erbin for pointing out this possibility.}
	
	\item We have exclusively worked with classical CSFT, but it is possible to obtain the local coordinates and determined the vertex regions of quantum vertices using similar techniques. It would be wise to directly work with hyperbolic vertices whose grafted cylinders have circumference at most $2 \, \text{arcsinh} 1$ here~\cite{Costello:2019fuh}, but the rest of the procedure remains mostly unchanged. A slight difference is that the on-shell Liouville action no longer relates to the tachyon interactions directly, but it can be still used to read them via~\eqref{eq:connection}. The simplest quantum vertex is 1-punctured torus whose classical conformal blocks are known, see~\cite{Hadasz:2009db,Menotti:2012wq, Piatek:2013ifa, Menotti:2015gxa, Menotti:2018jsy}.
	
	\item The framework here can be generalized to open-closed hyperbolic string vertices~\cite{Cho:2019anu}. One has to consider Liouville theory on the upper half-plane with the FZZT boundary condition~\cite{Teschner:2000md,Fateev:2000ik} endowed with boundary and/or bulk hyperbolic singularities and its properly regularized on-shell action. Some direction towards it has been already undertaken for elliptic singularities in~\cite{Hadasz:2006vs}. Independently, it may be interesting to investigate the WKB limit of these vertices.
	
	\item It is an interesting question how our claims translate to supersymmetric SFTs. It is intuitive to imagine that a supersymmetric Liouville theory and its on-shell action should lead to superstring vertices, however (super-)geometric understanding of them is not sufficiently well-developed (although see~\cite{Belopolsky:1996cy,Belopolsky:1997bg,Belopolsky:1997jz,Witten:2012bh, Witten:2012ga, Witten:2012bg, Witten:2013cia, Ohmori:2017wtx,Takezaki:2019jkn}). In any case, we can still use the hyperbolic vertices as the bosonic input for the M\"unich construction~\cite{Erler:2014eba, Erler:2015lya} and/or WZW-like constructions~\cite{Okawa:2004ii,Berkovits:2004xh,Kunitomo:2021wiz}.
	
	\item Lastly, we observe that Liouville theory plays a central role in the framework sketched in this paper. One may wonder if this is more than a mere coincidence and actually an imprint of the uncoupled Liouville mode off-shell. There is already some indication that this might be relevant for CSFT~\cite{Ahmadain:2022tew, Ahmadain:2022eso}. Even if it is, we were mostly concerned with the semi-classical limit of Liouville theory and it is not clear to the author how such a situation may arise from the worldsheet of critical bosonic strings.
	
\end{enumerate}

\section*{Acknowledgments}
I am indebted to Harold Erbin and Barton Zwiebach for many enlightening discussions, as well as their constant encouragements and comments on the early draft. I also would like to thank Daniel Harlow and Manki Kim for valuable discussions. This material is based upon work supported by the U.S. Department of Energy, Office of Science, Office of High Energy Physics of U.S. Department of Energy under grant Contract Number  DE-SC0012567.

\appendix
\section{Deriving the Polyakov conjecture for hyperbolic singularities} \label{app:direct}

Here we review the original derivation of the Polyakov conjecture for hyperbolic singularities~\eqref{eq:Polyakov1}~\cite{Hadasz:2003kp}. In~\eqref{eq:GetRidofeps1}, we consider the terms in $S_{HJ}[\varphi]$ that may contribute to the accessory parameters via~\eqref{eq:Polyakov1}. Differentiating it and being careful about using the Leibniz rule we get
\begin{align} \label{eq:MA}
&{\p S_{HJ}[\varphi] \over \p \xi_i} 
= {1 \over 2 \pi} \int\limits_{R} d^2z \left[
\p \left({\p \varphi \over \p \xi_i} \right)  \overline{\p} \varphi 
+ \p \varphi  \, \overline{\p} \left({\p \varphi \over \p \xi_i} \right)
+ e^\varphi \left({\p \varphi \over \p \xi_i} \right)
\right] 
\nonumber\\
& \quad \quad
- {i \over 4 \pi}   \sum_{j=1}^n\int\limits_{\p H_j} \left(\p \varphi \, \overline{\p} \varphi + e^\varphi \right) \left( {\p \gamma_j \over \p \xi_i } d \overline{\gamma_j} - {\p \overline{\gamma_j} \over \p \xi_i } d\gamma_j\right) \nonumber
\\
& \quad \quad
+ \sum_{j=1}^n \Bigg[ {1 \over 2 \pi} \int\limits_{H_j^\epsilon} d^2z \left[
\p \left({\p \varphi \over \p \xi_i} \right)  \overline{\p} \varphi 
+ \p \varphi  \, \overline{\p} \left({\p \varphi \over \p \xi_i} \right)
\right]
+{i \over 4 \pi} \int\limits_{\p H_j}   \p \varphi \, \overline{\p} \varphi \left( {\p \gamma_j \over \p \xi_i } d \overline{\gamma_j} - {\p \overline{\gamma_j} \over \p \xi_i } d\gamma_j\right)  \Bigg] \nonumber\\
& \quad \quad
+{i \over 4 \pi} \int\limits_{|z-\xi_i|=\epsilon} d\overline{z} \, \p \varphi \, \overline{\p} \varphi
+ \sum_{j=1}^n { \lambda_j^2 \over r_j[H_j] } {\p r_j[H_j] \over \p \xi_i}  \, ,
\end{align}
leaving the $\epsilon \to 0$ limit implicit. This is involved, so let us explain what each line represents. The first line is due to differentiating the first term in~\eqref{eq:GetRidofeps1} under the integral sign and the second line is due to the change of $\p R$. Notice there is an extra minus sign for this term due to the orientation.

The curve $\gamma_j(t) : \left[0,1\right] \to \Sigma_{0,n}$ is defined to be the geodesic seams in the Thurston metric:
\begin{align}
\gamma_j(t) \equiv \rho_j^{-1} \left( \exp\left({- \pi / 2 \lambda_j + 2 \pi i t}\right)\right) \, .
\end{align}
Here $\rho_j^{-1}$ is the inverse of $\rho_j$ (whose branches have adjusted as described below~\eqref{eq:SeamLimit}), which is guaranteed to exist in $\rho_j(H_j)$. In order to find its expression, we note that $\rho_j$ has the series expansion
\begin{align} \label{eq:Exp}
\rho_j (z) = {e^{-\pi / 2 \lambda_j} \over r_j[H_j]} \left[ (z-\xi_j) + {c_j \over 2 \delta_j} (z-\xi_j)^2 +
\mathcal{O}((z-\xi_j)^3) 
\right] \, .
\end{align}
The part inside the square brackets comes from solving the Fuchsian equation order-by-order and using~\eqref{eq:ScaledRatio}, while the terms outside can be found by the relation between the mapping radius $r_j[H_j]$ and the overall scale for $\rho_j$, just as in~\cite{Firat:2021ukc}. So we have
\begin{align} \label{eq:InvExp}
	\rho_j^{-1} (\rho) = \xi_j + e^{\pi / 2 \lambda_j} \, r_j[H_j] \, \rho - {c_j \over 2 \delta_j} ( e^{\pi / 2 \lambda_j} \, r_j[H_j] \, \rho)^2 + \mathcal{O}(\rho^3) \, .
\end{align}

Returning back to~\eqref{eq:MA}, we see that the third line is similar to the first/second line, but for the region $H_j$. Here we note that $\varphi$ satisfies the sewing relations at the seam $\p H_j$ and this allows us to cancel some terms to get
\begin{align} \label{eq:MA1}
&{\p S_{HJ}[\varphi] \over \p \xi_i}
= {1 \over 2 \pi} \int\limits_{R} d^2z \left[
\p \left({\p \varphi \over \p \xi_i} \right)  \overline{\p} \varphi 
+ \p \varphi  \, \overline{\p} \left({\p \varphi \over \p \xi_i} \right)
+ e^\varphi \left({\p \varphi \over \p \xi_i} \right)
\right] 
\nonumber\\
& \quad \quad
+  \sum_{j=1}^n\Bigg[ -{i \over 4 \pi}   \int\limits_{\p H_j} e^\varphi \left( {\p \gamma_j \over \p \xi_i } d \overline{\gamma_j} - {\p \overline{\gamma_j} \over \p \xi_i } d\gamma_j\right) 
+ {1 \over 2 \pi} \int\limits_{H_j^\epsilon} d^2z \left[
\p \left({\p \varphi \over \p \xi_i} \right)  \overline{\p} \varphi 
+ \p \varphi  \, \overline{\p} \left({\p \varphi \over \p \xi_i} \right)
\right] \Bigg] 
\nonumber\\
& \quad \quad
+{i \over 4 \pi} \int\limits_{|z-\xi_i|=\epsilon} d\overline{z} \; \p \varphi \, \overline{\p} \varphi
+ \sum_{j=1}^n { \lambda_j^2 \over r_j[H_j] } {\p r_j[H_j] \over \p \xi_i}  \, .
\end{align}

The region $H_i$ has a cutoff around the puncture $\xi_i$, so there should be an additional term due to its change, which we have included in the fourth line of~\eqref{eq:MA}. While we are at it, we have assumed that the remaining regularization circles don't change. Lastly, there are terms due to the changes of mapping radii, which have also been included in the fourth line of~\eqref{eq:MA}.

Let us now focus on the ``bulk'' terms in~\eqref{eq:MA1}, and use integration-by-parts, the equation of motion~\eqref{eq:EOM}, and the sewing relations at the seam $\p H_j$ to evaluate them to
\begin{align}
&{1 \over 2 \pi} \int\limits_{R} d^2z \left[
\p \left({\p \varphi \over \p \xi_i} \right)  \overline{\p} \varphi 
+ \p \varphi  \, \overline{\p} \left({\p \varphi \over \p \xi_i} \right)
+ e^\varphi \left({\p \varphi \over \p \xi_i} \right)
\right] 
+ {1 \over 2 \pi} \int\limits_{H_j^\epsilon} d^2z \left[
\p \left({\p \varphi \over \p \xi_i} \right)  \overline{\p} \varphi 
+ \p \varphi  \, \overline{\p} \left({\p \varphi \over \p \xi_i} \right)
\right] \nonumber \\
& \quad \quad \quad
= {i \over 4 \pi} \sum_{j=1}^n \; \int\limits_{|z-\xi_j| = \epsilon} {\p \varphi \over \p \xi_i} 
\left( \overline{\p} \varphi d \overline{z} - \p \varphi d z \right) \, .
\end{align}
Then we see
\begin{align} \label{eq:MA3}
{\p S_{HJ}[\varphi]  \over \p \xi_i}
&= - {i \over 4 \pi} \sum_{j=1}^n\Bigg[ \, \int\limits_{\p H_j} e^\varphi \left( {\p \gamma_j \over \p \xi_i } d \overline{\gamma_j} - {\p \overline{\gamma_j} \over \p \xi_i } d\gamma_j\right) \Bigg]
+{i \over 4 \pi} \int\limits_{|z-\xi_i|=\epsilon} d\overline{z} \; \p \varphi  \, \overline{\p} \varphi  
\nonumber\\
& \quad \quad \quad
+{i \over 4 \pi} \sum_{j=1}^n \int\limits_{|z-\xi_j| = \epsilon} {\p \varphi \over \p \xi_i} 
\left( \overline{\p} \varphi d \overline{z} - \p \varphi d z \right) 
+ \sum_{j=1}^n  { \lambda_j^2 \over r_j[H_j] } {\p r_j[H_j] \over \p \xi_i}  \, .
\end{align}

Focus on the first term above and note that we have
\begin{align} \label{eq:ev}
e^\varphi d \overline{\gamma_j(t)} = 
\lambda_j^2 \left| {\p \rho_j(\gamma_j(t)) \over \rho_j(\gamma_j(t)) } \right|^2 d \overline{\gamma_j(t)} 
= {\lambda_j^2 \over e^{-\pi/\lambda_j} } {1 \over (\rho^{-1})' } d \overline{\rho} 
\bigg|_{\overline{\rho} = e^{ - \pi / 2 \lambda_j - 2 \pi i t}}\, .
\end{align}
There are few things happened here. First we noticed $e^\varphi d \overline{\gamma_j(t)} $ is the line element on the geodesic seam $\p H_j$ and we have used the expression for $e^\varphi$, see~\eqref{eq:Flat}. This gave the first equality. Then we have used~\eqref{eq:SeamLimit} at the geodesic seam and pulled-backed the metric to $\rho$-plane. Lastly, we have used the inverse function theorem to get to the last line. Using the inverse~\eqref{eq:InvExp} and evaluating~\eqref{eq:ev} we obtain
\begin{align}
e^\varphi d \overline{\gamma_j(t)} =  -2 \pi i \lambda_j^2 
\left( {1 \over r_j[H_j]} e^{-2 \pi i t} +  {c_j \over \delta_j} \right) dt
+ \mathcal{O}( e^{2 \pi i t} ) \, ,
\end{align}
while $e^\varphi d \gamma_j(t) $ is given by the complex conjugation. We also see
\begin{subequations}
	\begin{align}
	{\p \gamma_j \over \p \xi_i} &=
	{\p \over \p \xi_i} \left( \rho_j^{-1}(e^{-\pi / 2 \lambda_j + 2 \pi i t})\right)
	= \delta_{ij} +  {\p r_j \over \p \xi_i} e^{2 \pi i t} + \mathcal{O}(e^{4 \pi i t}) \, ,\\
	{\p \overline{\gamma_j} \over \p \xi_i} &=
	{\p \over \p \xi_i} \left( \overline{\rho_j}^{-1}(e^{-\pi / 2 \lambda_j + 2 \pi i t})\right)
	= {\p r_j \over \p \xi_i} e^{-2 \pi i t} + \mathcal{O}(e^{-4 \pi i t}) \, ,
	\end{align}
\end{subequations}
using the inverse~\eqref{eq:InvExp}. Substituting them to the first term of~\eqref{eq:MA3} we get
\begin{align} \label{eq:Long}
&-{i \over 4 \pi} \sum_{j=1}^n\Bigg[  \; \int\limits_{\p H_j} e^\varphi \left( {\p \gamma_j \over \p \xi_i } d \overline{\gamma_j} - {\p \overline{\gamma_j} \over \p \xi_i } d\gamma_j\right) \Bigg] 
= -{\lambda_i^2 c_i \over 2 \delta_i} -\sum_{j=1}^n {\lambda_j^2 \over r_j[H_j]} {\p r_j[H_j] \over \p \xi_i } \, .
\end{align}
after some algebra.

Now focus on the second and third terms of~\eqref{eq:MA3}. In order to evaluate them, we begin writing down the expansion of $\varphi$ in $H_j$. Recall that the flat metric on $H_j$ is given in~\eqref{eq:Flat}. Together with the expansion~\eqref{eq:Exp}, we are lead to the following for $\varphi$ in $H_j$
\begin{align} \label{eq:metexp}
\varphi(z, \overline{z}) = \log \lambda_j^2 -\log|z-\xi_i|^2 + {c_j \over 2\delta_j} (z-\xi_j)
+ {\overline{c_j} \over 2\delta_j} (\overline{z}-\overline{\xi_j}) + \mathcal{O}(|z- \xi_j|^2) \, . 
\end{align}
In particular we see
\begin{align} \label{eq:Coll}
	\p \varphi = - {1 \over z- \xi_j} + {c_j \over 2 \delta_j } + \mathcal{O}(z- \xi_j), 
	\hspace{0,5in}
	{\p \varphi \over \p \xi_i} = \delta_{ij} \left({1 \over z- \xi_j} - {c_j \over 2 \delta_j} \right) + \mathcal{O}(|z- \xi_j|) \, .
\end{align}
Upon using them and taking the orientation of the regularization circles into account, we get
\begin{align} \label{eq:short}
{i \over 4 \pi} \int\limits_{|z-\xi_i|=\epsilon} d\overline{z} \; \p \varphi \, \overline{\p} \varphi  
+{i \over 4 \pi} \sum_{j=1}^n \, \int\limits_{|z-\xi_j| = \epsilon} {\p \varphi \over \p \xi_i} 
\left( \overline{\p} \varphi d \overline{z} - \p \varphi d z \right) = - {c_i \over 2 \delta_i} \, .
\end{align}
Combining all the terms for~\eqref{eq:MA3} (i.e.~\eqref{eq:Long} and~\eqref{eq:short}), the final result is
\begin{align}
{\p S_{HJ}[\varphi] \over \p \xi_i}
= -{\lambda_i^2 c_i \over 2 \delta_i} -\sum_{j=1}^n {\lambda_j^2 \over r_j[H_j]} {\p r_j[H_j] \over \p \xi_i }  - {c_i \over 2 \delta_i} + \sum_{j=1}^n  { \lambda_j^2 \over r_j[H_j] } {\p r_j[H_j] \over \p \xi_i}  
= -c_i \, ,
\end{align}
upon using $\delta_i = (1+\lambda_i^2 ) / 2$. This is the original and direct derivation of the Polyakov conjecture~\eqref{eq:Polyakov1}. Having a mapping radii terms in $S_{HJ}[\varphi]$ was crucial in order to derive this. 

Finally, let us comment on the Polyakov conjecture associated with the Thurston metrics $\varphi$ that contain $m$ internal flat cylinders of length $s_k$ and circumference $2 \pi \lambda_k$ in their geometry. We can model them by sewing pairs of holes $H_k$ and $H'_k$ on different surfaces for which no internal flat cylinders exists in their geometry. As a result,~\eqref{eq:Polyakov1} gets modified by taking out pairs of moduli associated with $H_k$ and $H'_k$ and replacing them with the moduli for the internal flat cylinder in~\eqref{eq:Long}. Then the Polyakov conjecture modifies to become
\begin{align} \label{eq:FlatPoly}
	{\p \over \p \xi_i} \left[ S_{HJ}[\varphi] + \sum_{k = 1}^m \lambda_k^2 s_k \right] = - c_i \, .
\end{align}

\section{The on-shell HJ action with three hyperbolic singularities} \label{app:DOZZ}

In this appendix we review the derivation in~\cite{Hadasz:2003he} for the relation between the DOZZ three-point function and the on-shell HJ action with three hyperbolic singularities $S_{HJ}^{(3)}[\varphi]$ defined in~\eqref{eq:Lio5}. Let us begin by differentiating~\eqref{eq:Lio5} with respect to $\lambda_i$ for any number of punctures. This is given by
\begin{align} \label{eq:connection}
	{\p S_{HJ}[\varphi] \over \p \lambda_i}
	&=  \lim_{\epsilon \to 0}\bigg[
	{i \over 4 \pi}
	\sum_{j=1}^n \; 
	\int\limits_{\, |z-\xi_j| = \epsilon} {\p \varphi_j \over \p \lambda_i} (\overline{\p} \varphi_j d \overline{z} - \p \varphi_j d z)
	+ {1 \over 2 \pi \epsilon}\sum_{j=1}^n \; \int\limits_{\, |z-\xi_j| = \epsilon} |dz| {\p \varphi \over \p \lambda_i}
	\bigg] \nonumber \\
	&\hspace{2in} + 2 \lambda_i \log r_i[H_i] 
	= 2 \lambda_i \log r_i[H_i]\, .
\end{align}
The first term is coming from combining $R^{1/\epsilon}$ and $H_i^\epsilon$ bulk terms in~\eqref{eq:Lio5} upon using the equation of motion~\eqref{eq:EOM}. The second term is coming from the regularization term around $z=\xi_j$ and the final term is coming from the mapping radii contributions. The term $2 \lambda_i \log r_i[H_i]$ is the only surviving mapping radii term by~\eqref{eq:red} as
\begin{align}
	{1 \over 2\pi} \int\limits_{H_j^\epsilon} d^2 z \; {\p  e^\varphi \over \p \lambda_i} =
	{1 \over 2\pi} \int\limits_{w_j^{-1}(H_j^\epsilon)} d^2 w_j \; {\p \over \p \lambda_i} 
	\left[{\lambda^2_j \over|w_j|^2}\right]
	= 2 \delta_{i j} \lambda_i \left( \log r_i[H_i] - \log \epsilon \right) \, .
\end{align}
Note that the contribution due to the change of the region $H_j^\epsilon$ is already accounted by the first term in~\eqref{eq:connection}. Now we observe that the first and second terms cancel each other by~\eqref{eq:metexp} and~\eqref{eq:Coll}, which implies the second equality in~\eqref{eq:connection}.

In the case of three-hyperbolic singularities, the mapping radii is given by~\eqref{eq:3maprad} when three punctures are placed at $z=0,1,\infty$ and we have, for $\lambda_i > 0$,
\begin{align} \label{eq:LambdaDer0}
	{d S_{HJ}^{(3)}[\varphi] \over d \lambda_i} = \pi - 2 v_i \, ,
\end{align}
where $v_i$ are given in terms of~\eqref{eq:v_func}. We can integrate this equation to get
\begin{align} \label{eq:LambdaDer}
	S_{HJ}^{(3)}[\varphi] 
	= 2 \sum_{\s_2, \s_3 = \pm} F\left( {1 \over 2} + {i \lambda_1 \over 2} + \s_2 {i \lambda_2 \over 2}
	+ \s_3 {i \lambda_3 \over 2} \right) + 
	2\sum_{j=1}^3 \left[H(i \lambda_j) + {\pi \over 2} |\lambda_j| \right] 
	+ C \, ,
\end{align}
for $\lambda_i \in \mathbb{R}$ and $C$ is the integration constant. The functions $F$ and $H$ are defined by
\begin{align} \label{eq:int}
	F(x) \equiv \int_{1 \over 2}^x dy \log \gamma (y),
	\quad \quad \quad
	H(x) \equiv\int_{0}^x dy \log {\Gamma(-y) \over \Gamma(y) } \, ,
\end{align}
and one particularly useful identity they satisfy is
\begin{align} \label{eq:FHIdent}
F(1 + i x) &= F(0) - H(ix) +i x \, (\log|x| -1) -{1 \over 2} \pi |x| \, ,
\end{align}
for $x \in \mathbb{R}$. This can be derived using various properties of the gamma function and combining the integrals~\eqref{eq:int} suitably.

We remark that $S_{HJ}^{(3)}[\varphi]$ has to be invariant under $\lambda_i \to - \lambda_i$ by its construction in~\eqref{eq:Lio5}. This is called the \textit{reflection symmetry}. Clearly, $F$ and $H$ parts of~\eqref{eq:LambdaDer} respect the reflection symmetry, while we have included an extra absolute value for $\lambda_j$'s to force it for the remaining part. We can justify this modification by considering the derivative of $S_{HJ}^{(3)}[\varphi]$ with respect to $\lambda_i$ for $\lambda_i < 0$: the right-hand side of~\eqref{eq:LambdaDer0} would have extra $-2\pi$ in order to keep the choice of branches uniform for every $\lambda_i$~\cite{Hadasz:2003he}.

Now we shift gears and look at the DOZZ 3-point function~\cite{Zamolodchikov:1995aa}. This is given by
\begin{align} \label{eq:DOZZ}
	C (\lambda_1, \lambda_2, \lambda_3) &={
	\left[ \pi \mu \gamma(b^2) \, b^{2-2b^2} \right]^{(Q-\beta_1 - \beta_2 - \beta_3)/b}
	\Upsilon_0 \Upsilon (2 \beta_1) \Upsilon (2 \beta_2) \Upsilon (2 \beta_3) \over
	\Upsilon(\beta_1 + \beta_2 + \beta_3 - Q) \Upsilon(Q + \beta_1 - \beta_2 - \beta_3)
	\Upsilon(\beta_1 + \beta_2 - \beta_3) \Upsilon(\beta_1 - \beta_2 + \beta_3) 
	} \, .
\end{align}
We have defined $\beta_i$ like in~\eqref{eq:FinalWeights1}. Here the function $\Upsilon(x)$ is a particular special function that has the following integral representation
\begin{align}
	\log \Upsilon (x) = \int\limits_0^\infty {dt \over t} \left[
	\left({Q \over 2} - x\right)^2 e^{-t}
	- {
		\sinh^2\left({Q \over 2} - x\right){t \over 2} \over
		\sinh {b t \over 2} \sinh{t \over 2 b}
	 }
	\right] \, ,
\end{align}
for $0 < \mathrm{Re}(x) < Q$, which is always the case for us. We also defined
\begin{align}
	\Upsilon_0 \equiv {d \Upsilon(x) \over dx} \bigg|_{x=0} \, .
\end{align}

One can verify the function $\Upsilon(x)$ satisfies
\begin{align} \label{eq:UpsIden}
	\Upsilon\left({Q \over 2}\right) =1,
	\quad \quad \quad
	\Upsilon(x + b )= \gamma(b x) \, b^{1-2bx} \, \Upsilon(x) \, .
\end{align}
The second relation in particular shows that as $b \to 0$ we have
\begin{align}
	{d \log \Upsilon(x) \over dx} \approx {1 \over b} \left[
	\log \gamma(bx)
	+
	(1- 2 b x) \log b
	\right] \, .
\end{align}
Integrating this and using~\eqref{eq:UpsIden} to fix the integration constant, we get
\begin{align} \label{eq:FI}
	\log \Upsilon(x) \approx {1 \over b^2} \left[F(bx) + \left(bx \, (1-bx) - {1\over 4}\right) \log b \right] \, .
\end{align}
Using two equations above we further see
\begin{align}
	\log \Upsilon_0 \approx {1 \over b^2} \left[F(0)- {1 \over 4} \log b\right]  \, .
\end{align}

For the DOZZ formula in the $b \to 0$ limit, we need three more formulas. First of them is
\begin{align}
	\log \Upsilon\left( {1 + i x \over 2 b}\right)
	\approx {1 \over b^2} \left[ F\left({1 + i x \over 2}\right) + {x^2 \over 4} \log b\right] \, ,
\end{align}
for $x \in \mathbb{R}$ which is just a restatement of~\eqref{eq:FI}. We are going to use it for the functions in the denominator of~\eqref{eq:DOZZ}. For those in the numerator of~\eqref{eq:DOZZ} we instead use
\begin{align}
\log \Upsilon\left( {1 + i x \over b}\right)
\approx {1 \over b^2} \left[F(0) - H(ix) - {1 \over 2} \pi |x| + \left( x^2 - {1\over 4} \right) \log b\right]
+ {i x \over b^2} \left[\log{|x| \over b} -1 \right] \, ,
\end{align}
holding for $x \in \mathbb{R}$ as well. We have used~\eqref{eq:FI} here together with the identity~\eqref{eq:FHIdent}. Finally we notice
\begin{align}
	\log \left[ \pi \mu \gamma(b^2) b^{2-2b^2} \right]^{(Q - \beta_1 - \beta_2 - \beta_3)/b}
	\approx -{1 \over 2b^2} \left( 1 +i \lambda_1 + i\lambda_2 + i \lambda_3  \right) \log \pi \mu \, ,
\end{align}
as $b \to 0$. Recall that $\gamma(b^2) \approx b^{-2}$ as $b\to0$.

Now we truly consider the DOZZ formula as $b \to 0$. Combining the last three identities and after some cancellations, we obtain
\begin{align} \label{eq:DOZZClass}
	\log C(\lambda_1, \lambda_2, \lambda_3) &\approx 
	 -{1 \over 2 b^2} S_{HJ}^{(3)}[\varphi]
	- {i \over b^2} \sum_{j=1}^3 \lambda_j \left[1 - \log |\lambda_j| + {1 \over 2} \log(\pi \mu b^2)  \right]
	 \, ,
\end{align}
up to constant terms. We are going to comment on these constant terms at the end.

This is not quite complete, given that the DOZZ formula is not reflection symmetric whereas $S_{HJ}^{(3)}[\varphi]$ is. Recall that this was the invariance under $\lambda_i \to - \lambda_i$, or equivalently $\beta_i \to Q -\beta_i$. Under such transformations, the DOZZ formula~\eqref{eq:DOZZ} changes by~\cite{Zamolodchikov:1995aa}
\begin{align}
	C(-\lambda_1, \lambda_2, \lambda_3) 
	= S \left(i \beta_1 - i {Q \over 2} \right) C(\lambda_1, \lambda_2, \lambda_3) 
	= S\left( -{Q \lambda_1 \over 2} \right) C(\lambda_1, \lambda_2, \lambda_3)  \, ,
\end{align}
where
\begin{align} \label{eq:RefAmp}
	S(P) = - (\pi \mu \gamma(b^2) )^{-2 i P /b} \;
	{\Gamma(1+2i P/b) \Gamma(1+2iPb) \over \Gamma(1-2iP/b) \Gamma(1-2iPb)} \, ,
\end{align}
is the so-called reflection amplitude. Analogous expressions hold for $\lambda_2$ and $\lambda_3$. It is possible to construct a manifestly reflection symmetric three-point function by taking it to be
\begin{align} \label{eq:SymmDOZZ}
	\widetilde{C} (\lambda_1, \lambda_2, \lambda_3) =
	\left[ \prod_{i=1}^3 S\left(- {Q \lambda_i \over 2} \right) \right]^{1 \over 2} C(\lambda_1, \lambda_2, \lambda_3) \, . 
\end{align}
The overall factor here can be justified by demanding the canonical normalization for the two-point functions, see~\cite{Zamolodchikov:1995aa} and identities thereof.

So all we need to find is $b \to 0$ limit of the reflection amplitude~\eqref{eq:RefAmp}. Using~\eqref{eq:Stirling} in~\eqref{eq:RefAmp}, the reflection amplitude takes the form
\begin{align} \label{eq:RefClass}
	\log S\left(- {Q \lambda_i \over 2} \right) 
	\approx {2 i \lambda_i \over b^2} \left[1
	- \log |\lambda_i| 
	+ {1\over 2} \log \pi \mu b^2 \right] \, ,
\end{align}
as $b \to 0$. Combining~\eqref{eq:DOZZClass} and~\eqref{eq:RefClass} we then see
\begin{align} \label{eq:DOZZexp}
	\widetilde{C} (\lambda_1, \lambda_2, \lambda_3) \approx \exp\left[-{ S_{HJ}^{(3)}[\varphi]\over 2b^2} \right] \, ,
\end{align}
which was the main result of~\cite{Hadasz:2003he}. This relation motivates our bootstrap program for the Strebel differentials. We point out~\eqref{eq:DOZZexp} is true up to a multiplicative constant (that is, terms independent of $\lambda_i$). But, in the view of the WKB limit, they are subleading in $\lambda$, so we don't have to worry about them. In general, this constant is irrelevant, as we always consider the derivatives of ${ S_{HJ}^{(3)}[\varphi]}$.

\section{Classical conformal blocks} \label{app:ConfBlocks}

In this appendix we provide a brief introduction to classical conformal blocks and our conventions for them. Interested reader should consult deeper expositions in~\cite{Zamolodchikov:1995aa, zamolodchikov1987conformal,francesco2012conformal,zamolodchikov1984conformal}. The (Virasoro) conformal blocks are the functions that have an expansion of the form
\begin{align} \label{eq:ConfBlock1}
\mathcal{F}_{c , \Delta } 
\begin{bmatrix}
\Delta_3 & \Delta_2\\
\Delta_4 & \Delta_1
\end{bmatrix} 
(\xi)
&= \xi^{\Delta - \Delta_2 - \Delta_1}  \left[1 + \sum_{n=1}^\infty \xi^n 
\mathcal{F}_{c, \, \Delta }^n
\begin{bmatrix}
\Delta_3 & \Delta_2\\
\Delta_4 & \Delta_1
\end{bmatrix} 
\right] \\
&=
\xi^{\Delta - \Delta_1 - \Delta_2}\left[
1 + {(\Delta + \Delta_3  - \Delta_4) (\Delta + \Delta_2  - \Delta_1) \over 2 \Delta} \xi + \cdots 
\right] 
\nonumber 
\, ,
\end{align}
where $\mathcal{F}_{c, \Delta }^n$ are the coefficients that are completely specified by the Virasoro algebra.

In principle it is possible to find the conformal blocks just using the Virasoro algebra, but this is computationally expensive. A better method is given by Zamolodchikov who suggested the following recursion relation for the coefficients $\mathcal{F}_{c, \Delta }^n$~\cite{zamolodchikov1984conformal}:
\begin{align}
\hspace{-0.1in}
\mathcal{F}_{c, \Delta }^n
\begin{bmatrix}
\Delta_3 & \Delta_2\\
\Delta_4 & \Delta_1
\end{bmatrix} 
=
g_{\Delta }^n
\begin{bmatrix}
\Delta_3 & \Delta_2\\
\Delta_4 & \Delta_1
\end{bmatrix} 
+
\sum_{\substack{r \geq 2,s \geq 1 \\ n \geq rs \geq 2}}
{ 1 
	\over
	c - c_{rs}(\Delta)
}
\widetilde{R}_\Delta^{rs}
\begin{bmatrix}
\Delta_3 & \Delta_2\\
\Delta_4 & \Delta_1
\end{bmatrix} 
\mathcal{F}_{c_{rs} (\Delta), \, \Delta + rs }^{n-rs}
\begin{bmatrix}
\Delta_3 & \Delta_2\\
\Delta_4 & \Delta_1
\end{bmatrix} \, ,
\end{align}
where
\begin{subequations}
\begin{align}
	&_2F_1(\Delta + \Delta_2 - \Delta_1, \Delta + \Delta_3 - \Delta_4, 2 \Delta, z) =
	\sum_{n=0}^\infty
	g_{\Delta }^n
	\begin{bmatrix}
	\Delta_3 & \Delta_2\\
	\Delta_4 & \Delta_1
	\end{bmatrix} 
	z^n \, ,
	\\
	&c_{rs}(\Delta) = 13 - 6 \left[T_{rs} (\Delta) + {1 \over T_{rs} (\Delta)}\right] \, , \\
	&T_{rs} (\Delta) = {rs - 1 + 2 \Delta + \sqrt{(r-s)^2 + 4 (rs - 1) \Delta + 4 \Delta^2}  \over r^2 -1} \, .
\end{align}
and
\begin{align}
\widetilde{R}_\Delta^{rs}
\begin{bmatrix}
\Delta_3 & \Delta_2\\
\Delta_4 & \Delta_1
\end{bmatrix} 
&= - {\p c_{rs} ( \Delta ) \over \p \Delta } A_{rs} ( T_{rs} (\Delta) ) \\
&\hspace{0.5in} \times P_{rs} (T_{rs} (\Delta), \Delta_4  + \Delta_3, \Delta_4 - \Delta_3 ) \;
P_{rs} (T_{rs} (\Delta), \Delta_1  + \Delta_2, \Delta_1 - \Delta_2 ) \nonumber \, .
\end{align}
with
\begin{align} \label{eq:Afunc}
	A_{mn} (\alpha^2) = - {1 \over 2} \left( \prod_{k=1-m}^{m} \prod_{l=1-n}^{n} \right)'
	{1 \over k \alpha - {l \over \alpha} } \, ,
\end{align}
where the prime indicates that the terms with $(k,l) = (0,0)$ and $(k,l) = (m,n)$ are skipped in the product. Finally, we have
\begin{align} \label{eq:Pfunc}
	P_{mn} (\alpha^2, \Delta_1 + \Delta_2, \Delta_1 - \Delta_2)
	= \prod_{\substack{p = 1-m \\ p+m \; \text{is odd} }}^{m-1} 
	\;
	\prod_{\substack{q = 1-n \\ p+n \; \text{is odd} }}^{n-1}
	{\alpha_1 + \alpha_2 - p \alpha + {q \over \alpha} \over 2}
	{\alpha_1 - \alpha_2 - p \alpha + {q \over \alpha} \over 2} \, ,
\end{align}
with
\begin{align}
	\Delta_i = -{1 \over 4} \left[ \alpha - {1 \over \alpha} \right]^2 + {\alpha_i^2 \over 4 } \, .
\end{align}
\end{subequations}
Notice that the function $P_{mn}$ is odd in its second argument as a consequence of the last two equations. This is the reason behind the symmetry mentioned below~\eqref{eq:argbelow} for the equal external weights.

We are not going to report the entire conformal blocks, but only the classical conformal blocks that are obtained through exponentiation~\eqref{eq:ClassConfBlock} and taking the WKB limit~\eqref{eq:them} with $\alpha_i = 1$:
\begin{align} \label{eq:xi}
 f_{\alpha^2}
\begin{bmatrix}
1 & 1\\
1 & 1
\end{bmatrix} (\xi)
\equiv
f_{\alpha^2} (\xi)
&= \left( -{1 \over 2} + {\alpha^2 \over 4} \right) \log \xi 
+ {\alpha^2 \over 8} \xi
+ \left[{ 1 \over 16 \alpha^2} + {1 \over 32} + {13 \alpha^2 \over 256} \right] \xi^2 \nonumber \\
&\hspace{0.7in} + \left[ {1 \over 16 \alpha^2} + {1 \over 32} + {23 \alpha^2 \over 768}  \right] \xi^3 
+\mathcal{O}(\xi^4) \,. 
\end{align}
This series is expected to converge for $|\xi| < 1$. We have used the expansion up to $\mathcal{O}(\xi^5)$ in this work. We suppress the dependence on the external weights $\alpha_i$ if we set all of them equal to 1 and denote the block simply by $f_{\alpha^2}(\xi)$.

For our purposes, a better way of computing the expansion of classical conformal blocks is through the recursion involving the elliptic nome $q(\xi)$ defined by the complete elliptic integral of the first kind $K(\xi)$~\cite{zamolodchikov1987conformal}, which we already defined in~\eqref{eq:nome},
\begin{align} 
q(\xi) \equiv \exp \left[ - \pi {K(1-\xi) \over K(\xi) }\right],
\hspace{0.5in}
K(\xi) = \int_{0}^1 {dt \over \sqrt{(1-t^2)(1- \xi t^2) } } \, ,
\end{align}
as the conformal blocks are expected to converge for (nearly) every $\xi$ at much faster rate in this expansion. Here the elliptic nome $q$ maps the moduli space ($\mathbb{C} \setminus \{0,1\}$) to a unit disk $0 < |q| \leq 1$. The conformal blocks in this case are given by
\begin{align}
	\mathcal{F}_{c , \, \Delta } 
	\begin{bmatrix}
	\Delta_3 & \Delta_2\\
	\Delta_4 & \Delta_1
	\end{bmatrix} 
	(\xi) &=
	\xi^{{c-1 \over 24} - \Delta_1 - \Delta_2} (1-\xi)^{{c-1 \over 24} - \Delta_1 - \Delta_3} 
	\left[ {2 \over \pi} K(\xi) \right]^{{c-1 \over 4} - 2 (\Delta_1 + \Delta_2 + \Delta_3 + \Delta_4)}
	 \nonumber \\
	& \hspace{0.5in} \times
	(16 q )^{\Delta - {c-1 \over 24} }\left[ 1 +
	\sum_{n=1}^\infty \left( 16 q \right)^n
	H_{c, \, \Delta }^n
	\begin{bmatrix}
	\Delta_3 & \Delta_2\\
	\Delta_4 & \Delta_1
	\end{bmatrix} 
	\right]
	\, ,
\end{align}
with $H_{c, \, \Delta }^n$ for $n >0$ satisfying the recursion, with $H_{c, \, \Delta }^0 =1$,
\begin{align} \label{eq:rec}
	H_{c, \, \Delta }^n
	\begin{bmatrix}
	\Delta_3 & \Delta_2\\
	\Delta_4 & \Delta_1
	\end{bmatrix} 
	=
	\sum_{\substack{r \geq 1,s \geq 1 \\ n \geq rs \geq 1}}
	{ 1 
		\over
		\Delta - \Delta_{rs}(c)
	}
	R_c^{rs}
	\begin{bmatrix}
	\Delta_3 & \Delta_2\\
	\Delta_4 & \Delta_1
	\end{bmatrix} 
	H_{c, \, \Delta_{rs}(c) + rs }^{n-rs}
	\begin{bmatrix}
	\Delta_3 & \Delta_2\\
	\Delta_4 & \Delta_1
	\end{bmatrix} \, ,
\end{align}
where
\begin{subequations}
\begin{align}
	\Delta_{rs}(c) = {1 - r^2 \over 4} b^2 +
	{1 - rs \over 2} + {1 - s^2 \over 4} {1 \over b^2} \, ,
\end{align}
and
\begin{align}
		R_c^{rs}
	\begin{bmatrix}
	\Delta_3 & \Delta_2\\
	\Delta_4 & \Delta_1
	\end{bmatrix}  = A_{rs}(-b^2) 
	P_{rs} (-b^2, \Delta_1 + \Delta_2, \Delta_1 - \Delta_2)
	P_{rs} (-b^2, \Delta_3 + \Delta_4, \Delta_4 - \Delta_3) \, .
\end{align}
\end{subequations}
Notice the functions on the right-hand side are already defined in~\eqref{eq:Afunc}-~\eqref{eq:Pfunc}.

Again, we don't report the general block but only the ones we obtain after exponentiation~\eqref{eq:ClassConfBlock} and taking the WKB limit~\eqref{eq:them}:
\begin{align}
 f_{\alpha^2}
\begin{bmatrix}
\alpha_3^2 & \alpha_2^2\\
\alpha_4^2 & \alpha_1^2
\end{bmatrix} (\xi) 
&=-{1 \over 4} (\alpha_1^2 + \alpha_2^2) \log \xi 
-{1 \over 4} ( \alpha_1^2 + \alpha_3^2) \log (1 - \xi)  \\
&\hspace{0.5in} -{1 \over 2} (\alpha_1^2 + \alpha_2^2 + \alpha_3^2 + \alpha_4^2) \log \left[{2 \over \pi} K(\xi) \right] + 
 h_{\alpha^2}
\begin{bmatrix}
\alpha_3^2 & \alpha_2^2\\
\alpha_4^2 & \alpha_1^2
\end{bmatrix} (q)  \, , \nonumber
\end{align}
with $h$ is a function of the elliptic nome 
\begin{align}
	 h_{\alpha^2}
	\begin{bmatrix}
	\alpha_3^2 & \alpha_2^2\\
	\alpha_4^2 & \alpha_1^2
	\end{bmatrix} (q) =
	{\alpha^2 \over 4} \log q + \cdots \, .
\end{align}
The rest of the terms are too involved to report, but they can be obtained upon using the recursion~\eqref{eq:rec}. For example, the case for which the external weights are equal $\alpha_i = 1$ is given by
\begin{align} \label{eq:qexp}
 f_{\alpha^2}
\begin{bmatrix}
1 &  1\\
1 & 1
\end{bmatrix} (\xi)
&\equiv
f_{\alpha^2} (\xi)
= -{1 \over 2} \log \xi
- {1 \over 2} \log (1-\xi) 
-2 \log \left[{2 \over \pi} K(\xi)\right] 
+ { \alpha^2 \over 4} \log 16 q  \\
&\hspace{-0.5in}+{16 \over \alpha^2} q^2
+ \left[{48 \over \alpha^2} -{384 \over \alpha^4} + {640 \over \alpha^6}\right] q^4
+ \left[{64 \over \alpha^2} - {2048 \over \alpha^4} + {20480 \over \alpha^6} - {229376 \over \alpha^8} + {98304 \over \alpha^{10}} \right] q^6 + \mathcal{O}(q^8)\, . \nonumber
\end{align}
This is the classical conformal block relevant for the symmetric quartic vertex of CSFT. We use terms up to $\mathcal{O}(q^{12})$ in this work, but only showed the first few terms above for brevity.


\begin{thebibliography}{100}
	
	\bibitem{Zwiebach:1992ie}
	B.~Zwiebach, ``{Closed string field theory: Quantum action and the B-V master
		equation},'' {\em Nucl. Phys. B} {\bf 390} (1993) 33--152,
	\href{http://www.arXiv.org/abs/hep-th/9206084}{{\tt hep-th/9206084}}.
	
	\bibitem{deLacroix:2017lif}
	C.~de~Lacroix, H.~Erbin, S.~P. Kashyap, A.~Sen, and M.~Verma, ``{Closed
		Superstring Field Theory and its Applications},'' {\em Int. J. Mod. Phys. A}
	{\bf 32} (2017), no.~28n29, 1730021,
	\href{http://www.arXiv.org/abs/1703.06410}{{\tt 1703.06410}}.
	
	\bibitem{Erler:2019loq}
	T.~Erler, ``{Four Lectures on Closed String Field Theory},'' {\em Phys. Rept.}
	{\bf 851} (2020) 1--36, \href{http://www.arXiv.org/abs/1905.06785}{{\tt
			1905.06785}}.
	
	\bibitem{Erbin:2021smf}
	H.~Erbin, {\em {String Field Theory: A Modern Introduction}}, vol.~980 of {\em
		Lecture Notes in Physics}.
	\newblock 3, 2021.
	
	\bibitem{Pius:2014iaa}
	R.~Pius, A.~Rudra, and A.~Sen, ``{Mass Renormalization in String Theory:
		General States},'' {\em JHEP} {\bf 07} (2014) 062,
	\href{http://www.arXiv.org/abs/1401.7014}{{\tt 1401.7014}}.
	
	\bibitem{Pius:2014gza}
	R.~Pius, A.~Rudra, and A.~Sen, ``{String Perturbation Theory Around Dynamically
		Shifted Vacuum},'' {\em JHEP} {\bf 10} (2014) 070,
	\href{http://www.arXiv.org/abs/1404.6254}{{\tt 1404.6254}}.
	
	\bibitem{Sen:2015uoa}
	A.~Sen, ``{Supersymmetry Restoration in Superstring Perturbation Theory},''
	{\em JHEP} {\bf 12} (2015) 075,
	\href{http://www.arXiv.org/abs/1508.02481}{{\tt 1508.02481}}.
	
	\bibitem{Sen:2016gqt}
	A.~Sen, ``{One Loop Mass Renormalization of Unstable Particles in Superstring
		Theory},'' {\em JHEP} {\bf 11} (2016) 050,
	\href{http://www.arXiv.org/abs/1607.06500}{{\tt 1607.06500}}.
	
	\bibitem{Erler:2017pgf}
	T.~Erler, S.~Konopka, and I.~Sachs, ``{One Loop Tadpole in Heterotic String
		Field Theory},'' {\em JHEP} {\bf 11} (2017) 056,
	\href{http://www.arXiv.org/abs/1704.01210}{{\tt 1704.01210}}.
	
	\bibitem{DeLacroix:2018arq}
	C.~De~Lacroix, H.~Erbin, and A.~Sen, ``{Analyticity and Crossing Symmetry of
		Superstring Loop Amplitudes},'' {\em JHEP} {\bf 05} (2019) 139,
	\href{http://www.arXiv.org/abs/1810.07197}{{\tt 1810.07197}}.
	
	\bibitem{Zwiebach:1990ni}
	B.~Zwiebach, ``{Consistency of Closed String Polyhedra From Minimal Area},''
	{\em Phys. Lett. B} {\bf 241} (1990) 343--349.
	
	\bibitem{Zwiebach:1990nh}
	B.~Zwiebach, ``{How covariant closed string theory solves a minimal area
		problem},'' {\em Commun. Math. Phys.} {\bf 136} (1991) 83--118.
	
	\bibitem{ranganathan1992criterion}
	K.~Ranganathan, ``A criterion for flatness in minimal area metrics that define
	string diagrams,'' {\em Communications in mathematical physics} {\bf 146}
	(1992), no.~3, 429--445.
	
	\bibitem{Wolf:1992bk}
	M.~Wolf and B.~Zwiebach, ``{The Plumbing of minimal area surfaces},''
	\href{http://www.arXiv.org/abs/hep-th/9202062}{{\tt hep-th/9202062}}.
	
	\bibitem{Headrick:2018dlw}
	M.~Headrick and B.~Zwiebach, ``{Minimal-area metrics on the Swiss cross and
		punctured torus},'' {\em Commun. Math. Phys.} {\bf 377} (2020), no.~3,
	2287--2343, \href{http://www.arXiv.org/abs/1806.00450}{{\tt 1806.00450}}.
	
	\bibitem{Headrick:2018ncs}
	M.~Headrick and B.~Zwiebach, ``{Convex programs for minimal-area problems},''
	{\em Commun. Math. Phys.} {\bf 377} (2020), no.~3, 2217--2285,
	\href{http://www.arXiv.org/abs/1806.00449}{{\tt 1806.00449}}.
	
	\bibitem{Naseer:2019zau}
	U.~Naseer and B.~Zwiebach, ``{Extremal isosystolic metrics with multiple bands
		of crossing geodesics},'' \href{http://www.arXiv.org/abs/1903.11755}{{\tt
			1903.11755}}.
	
	\bibitem{Moosavian:2017qsp}
	S.~F. Moosavian and R.~Pius, ``{Hyperbolic geometry and closed bosonic string
		field theory. Part I. The string vertices via hyperbolic Riemann surfaces},''
	{\em JHEP} {\bf 08} (2019) 157,
	\href{http://www.arXiv.org/abs/1706.07366}{{\tt 1706.07366}}.
	
	\bibitem{Moosavian:2017sev}
	S.~F. Moosavian and R.~Pius, ``{Hyperbolic geometry and closed bosonic string
		field theory. Part II. The rules for evaluating the quantum BV master
		action},'' {\em JHEP} {\bf 08} (2019) 177,
	\href{http://www.arXiv.org/abs/1708.04977}{{\tt 1708.04977}}.
	
	\bibitem{Costello:2019fuh}
	K.~Costello and B.~Zwiebach, ``{Hyperbolic string vertices},'' {\em JHEP} {\bf
		02} (2022) 002, \href{http://www.arXiv.org/abs/1909.00033}{{\tt 1909.00033}}.
	
	\bibitem{Cho:2019anu}
	M.~Cho, ``{Open-closed Hyperbolic String Vertices},'' {\em JHEP} {\bf 05}
	(2020) 046, \href{http://www.arXiv.org/abs/1912.00030}{{\tt 1912.00030}}.
	
	\bibitem{Firat:2021ukc}
	A.~H. F\i{}rat, ``{Hyperbolic three-string vertex},'' {\em JHEP} {\bf 08}
	(2021) 035, \href{http://www.arXiv.org/abs/2102.03936}{{\tt 2102.03936}}.
	
	\bibitem{Wang:2021aog}
	P.~Wang, H.~Wu, and H.~Yang, ``{Connections between reflected entropies and
		hyperbolic string vertices},'' {\em JHEP} {\bf 05} (2022) 127,
	\href{http://www.arXiv.org/abs/2112.09503}{{\tt 2112.09503}}.
	
	\bibitem{Ishibashi:2022qcz}
	N.~Ishibashi, ``{The Fokker-Planck formalism for closed bosonic strings},''
	\href{http://www.arXiv.org/abs/2210.04134}{{\tt 2210.04134}}.
	
	\bibitem{buser2010geometry}
	P.~Buser, {\em Geometry and spectra of compact Riemann surfaces}.
	\newblock Springer Science \& Business Media, 2010.
	
	\bibitem{Saadi:1989tb}
	M.~Saadi and B.~Zwiebach, ``{Closed String Field Theory from Polyhedra},'' {\em
		Annals Phys.} {\bf 192} (1989) 213.
	
	\bibitem{Kugo:1989aa}
	T.~Kugo, H.~Kunitomo, and K.~Suehiro, ``{Nonpolynomial Closed String Field
		Theory},'' {\em Phys. Lett. B} {\bf 226} (1989) 48--54.
	
	\bibitem{strebel1984quadratic}
	K.~Strebel, ``Quadratic differentials,'' in {\em Quadratic Differentials},
	pp.~16--26.
	\newblock Springer, 1984.
	
	\bibitem{Hadasz:2003kp}
	L.~Hadasz and Z.~Jaskolski, ``{Polyakov conjecture for hyperbolic
		singularities},'' {\em Phys. Lett. B} {\bf 574} (2003) 129--135,
	\href{http://www.arXiv.org/abs/hep-th/0308131}{{\tt hep-th/0308131}}.
	
	\bibitem{Hadasz:2003he}
	L.~Hadasz and Z.~Jaskolski, ``{Classical Liouville action on the sphere with
		three hyperbolic singularities},'' {\em Nucl. Phys. B} {\bf 694} (2004)
	493--508, \href{http://www.arXiv.org/abs/hep-th/0309267}{{\tt
			hep-th/0309267}}.
	
	\bibitem{Hadasz:2005gk}
	L.~Hadasz, Z.~Jaskolski, and M.~Piatek, ``{Classical geometry from the quantum
		Liouville theory},'' {\em Nucl. Phys. B} {\bf 724} (2005) 529--554,
	\href{http://www.arXiv.org/abs/hep-th/0504204}{{\tt hep-th/0504204}}.
	
	\bibitem{kuz1997methodsI}
	G.~V. Kuz'mina, ``Methods of the geometric theory of functions. i,'' {\em
		Algebra i Analiz} {\bf 9} (1997), no.~3, 41--103.
	
	\bibitem{kuz1997methodsII}
	G.~V. Kuz’mina, ``Methods of the geometric theory of functions. ii,'' {\em
		Algebra i Analiz} {\bf 9} (1997), no.~5, 1--50.
	
	\bibitem{solynin1999moduli}
	A.~Y. Solynin, ``Moduli and extremal metric problems,'' {\em Algebra i Analiz}
	{\bf 11} (1999), no.~1, 3--86.
	
	\bibitem{solynin2009quadratic}
	A.~Y. Solynin, ``Quadratic differentials and weighted graphs on compact
	surfaces,'' in {\em Analysis and mathematical physics}, pp.~473--505.
	\newblock Springer, 2009.
	
	\bibitem{solynin2020fingerprints}
	A.~Y. Solynin, ``Fingerprints, lemniscates and quadratic differentials,'' {\em
		arXiv preprint arXiv:2011.03855} (2020).
	
	\bibitem{bakhtin2022generalized}
	A.~K. Bakhtin and I.~V. Denega, ``Generalized ma lavrentiev’s inequality,''
	{\em Journal of Mathematical Sciences} {\bf 262} (2022), no.~2, 138--153.
	
	\bibitem{Belopolsky:1994sk}
	A.~Belopolsky and B.~Zwiebach, ``{Off-shell closed string amplitudes: Towards a
		computation of the tachyon potential},'' {\em Nucl. Phys. B} {\bf 442} (1995)
	494--532, \href{http://www.arXiv.org/abs/hep-th/9409015}{{\tt
			hep-th/9409015}}.
	
	\bibitem{Belopolsky:1994bj}
	A.~Belopolsky, ``{Effective Tachyonic potential in closed string field
		theory},'' {\em Nucl. Phys. B} {\bf 448} (1995) 245--276,
	\href{http://www.arXiv.org/abs/hep-th/9412106}{{\tt hep-th/9412106}}.
	
	\bibitem{Moeller:2004yy}
	N.~Moeller, ``{Closed bosonic string field theory at quartic order},'' {\em
		JHEP} {\bf 11} (2004) 018,
	\href{http://www.arXiv.org/abs/hep-th/0408067}{{\tt hep-th/0408067}}.
	
	\bibitem{Moeller:2006cw}
	N.~Moeller, ``{Closed Bosonic String Field Theory at Quintic Order:
		Five-Tachyon Contact Term and Dilaton Theorem},'' {\em JHEP} {\bf 03} (2007)
	043, \href{http://www.arXiv.org/abs/hep-th/0609209}{{\tt hep-th/0609209}}.
	
	\bibitem{Moeller:2007mu}
	N.~Moeller, ``{Closed Bosonic String Field Theory at Quintic Order. II.
		Marginal Deformations and Effective Potential},'' {\em JHEP} {\bf 09} (2007)
	118, \href{http://www.arXiv.org/abs/0705.2102}{{\tt 0705.2102}}.
	
	\bibitem{Erbin:2022rgx}
	H.~Erbin and A.~H. F\i{}rat, ``{Characterizing 4-string contact interaction
		using machine learning},'' \href{http://www.arXiv.org/abs/2211.09129}{{\tt
			2211.09129}}.
	
	\bibitem{jenkins1954recent}
	J.~A. Jenkins, ``A recent note of kolbina,'' {\em Duke Mathematical Journal}
	{\bf 21} (1954), no.~1, 155--162.
	
	\bibitem{kuz1982problem}
	G.~Kuz'mina, ``Problem of the maximum of the product of the conformal radii of
	nonoverlapping domains,'' {\em Journal of Soviet Mathematics} {\bf 19}
	(1982), no.~6, 1715--1726.
	
	\bibitem{fedorov1982maximum}
	S.~Fedorov, ``Maximum of the product of the conformal radii of four
	nonoverlapping domains,'' {\em Journal of Soviet Mathematics} {\bf 19}
	(1982), no.~6, 1727--1741.
	
	\bibitem{emelyanov2004problem}
	E.~Emelyanov, ``On the problem of maximizing the product of powers of conformal
	radii nonoverlapping domains,'' {\em Journal of Mathematical Sciences} {\bf
		122} (2004), no.~6, 3641--3647.
	
	\bibitem{Zamolodchikov:1995aa}
	A.~B. Zamolodchikov and A.~B. Zamolodchikov, ``{Structure constants and
		conformal bootstrap in Liouville field theory},'' {\em Nucl. Phys. B} {\bf
		477} (1996) 577--605, \href{http://www.arXiv.org/abs/hep-th/9506136}{{\tt
			hep-th/9506136}}.
	
	\bibitem{zamolodchikov1987conformal}
	A.~B. Zamolodchikov, ``Conformal symmetry in two-dimensional space: recursion
	representation of conformal block,'' {\em Theoretical and Mathematical
		Physics} {\bf 73} (1987), no.~1, 1088--1093.
	
	\bibitem{francesco2012conformal}
	P.~Francesco, P.~Mathieu, and D.~S{\'e}n{\'e}chal, {\em Conformal field
		theory}.
	\newblock Springer Science \& Business Media, 2012.
	
	\bibitem{Belavin:1984vu}
	A.~A. Belavin, A.~M. Polyakov, and A.~B. Zamolodchikov, ``{Infinite Conformal
		Symmetry in Two-Dimensional Quantum Field Theory},'' {\em Nucl. Phys. B} {\bf
		241} (1984) 333--380.
	
	\bibitem{zamolodchikov1986two}
	A.~B. Zamolodchikov, ``Two-dimensional conformal symmetry and critical
	four-spin correlation functions in the ashkin-teller model,'' {\em Soviet
		Journal of Experimental and Theoretical Physics} {\bf 63} (1986), no.~5,
	1061.
	
	\bibitem{becsken2020semi}
	M.~Be{\c{s}}ken, S.~Datta, and P.~Kraus, ``Semi-classical virasoro blocks:
	proof of exponentiation,'' {\em Journal of High Energy Physics} {\bf 2020}
	(2020), no.~1, 1--16.
	
	\bibitem{Teschner:2001rv}
	J.~Teschner, ``{Liouville theory revisited},'' {\em Class. Quant. Grav.} {\bf
		18} (2001) R153--R222, \href{http://www.arXiv.org/abs/hep-th/0104158}{{\tt
			hep-th/0104158}}.
	
	\bibitem{Dorn:1994xn}
	H.~Dorn and H.~J. Otto, ``{Two and three point functions in Liouville
		theory},'' {\em Nucl. Phys. B} {\bf 429} (1994) 375--388,
	\href{http://www.arXiv.org/abs/hep-th/9403141}{{\tt hep-th/9403141}}.
	
	\bibitem{Sen:1999nx}
	A.~Sen and B.~Zwiebach, ``{Tachyon condensation in string field theory},'' {\em
		JHEP} {\bf 03} (2000) 002,
	\href{http://www.arXiv.org/abs/hep-th/9912249}{{\tt hep-th/9912249}}.
	
	\bibitem{mondello2011riemann}
	G.~Mondello, ``Riemann surfaces with boundary and natural triangulations of the
	teichm{\"u}ller space,'' {\em Journal of the European Mathematical Society}
	{\bf 13} (2011), no.~3, 635--684.
	
	\bibitem{takei2017wkb}
	Y.~Takei, ``Wkb analysis and stokes geometry of differential equations,'' in
	{\em Analytic, Algebraic and Geometric Aspects of Differential Equations:
		Bedlewo, Poland, September 2015}, pp.~263--304, Springer.
	\newblock 2017.
	
	\bibitem{seiberg1990notes}
	N.~Seiberg, ``Notes on quantum liouville theory and quantum gravity,'' {\em
		Progress of Theoretical Physics Supplement} {\bf 102} (1990) 319--349.
	
	\bibitem{Nakayama:2004vk}
	Y.~Nakayama, ``{Liouville field theory: A Decade after the revolution},'' {\em
		Int. J. Mod. Phys. A} {\bf 19} (2004) 2771--2930,
	\href{http://www.arXiv.org/abs/hep-th/0402009}{{\tt hep-th/0402009}}.
	
	\bibitem{Harlow:2011ny}
	D.~Harlow, J.~Maltz, and E.~Witten, ``{Analytic Continuation of Liouville
		Theory},'' {\em JHEP} {\bf 12} (2011) 071,
	\href{http://www.arXiv.org/abs/1108.4417}{{\tt 1108.4417}}.
	
	\bibitem{erbin2015notes}
	H.~Erbin, ``Notes on 2d quantum gravity and liouville theory.''
	{\url{https://www.lpthe.jussieu.fr/~erbin/files/liouville_theory.pdf}}, 2015.
	
	\bibitem{Ribault:2014hia}
	S.~Ribault, ``{Conformal field theory on the plane},''
	\href{http://www.arXiv.org/abs/1406.4290}{{\tt 1406.4290}}.
	
	\bibitem{Hadasz:2006rb}
	L.~Hadasz and Z.~Jaskolski, ``{Liouville theory and uniformization of
		four-punctured sphere},'' {\em J. Math. Phys.} {\bf 47} (2006) 082304,
	\href{http://www.arXiv.org/abs/hep-th/0604187}{{\tt hep-th/0604187}}.
	
	\bibitem{Rastelli:2000iu}
	L.~Rastelli and B.~Zwiebach, ``{Tachyon potentials, star products and
		universality},'' {\em JHEP} {\bf 09} (2001) 038,
	\href{http://www.arXiv.org/abs/hep-th/0006240}{{\tt hep-th/0006240}}.
	
	\bibitem{polchinski1998string}
	J.~G. Polchinski, {\em String theory, volume I: An introduction to the bosonic
		string}.
	\newblock Cambridge university press Cambridge, 1998.
	
	\bibitem{Harrison:2022frl}
	S.~M. Harrison, A.~Maloney, and T.~Numasawa, ``{Liouville Theory and the
		Weil-Petersson Geometry of Moduli Space},''
	\href{http://www.arXiv.org/abs/2210.08098}{{\tt 2210.08098}}.
	
	\bibitem{Sonoda:1989sj}
	H.~Sonoda and B.~Zwiebach, ``{COVARIANT CLOSED STRING THEORY CANNOT BE
		CUBIC},'' {\em Nucl. Phys. B} {\bf 336} (1990) 185--221.
	
	\bibitem{Zemba:1988rf}
	G.~Zemba and B.~Zwiebach, ``{Tadpole Graph in Covariant Closed String Field
		Theory},'' {\em J. Math. Phys.} {\bf 30} (1989) 2388.
	
	\bibitem{sonoda1990closed}
	H.~Sonoda and B.~Zwiebach, ``Closed string field theory loops with symmetric
	factorizable quadratic differentials,'' {\em Nuclear Physics B} {\bf 331}
	(1990), no.~3, 592--628.
	
	\bibitem{Okawa:2022mos}
	Y.~Okawa and R.~Sakaguchi, ``{Closed string field theory without the
		level-matching condition},'' \href{http://www.arXiv.org/abs/2209.06173}{{\tt
			2209.06173}}.
	
	\bibitem{Erbin:2022cyb}
	H.~Erbin and M.~M\'edevielle, ``{Closed string theory without level-matching at
		the free level},'' \href{http://www.arXiv.org/abs/2209.05585}{{\tt
			2209.05585}}.
	
	\bibitem{Menotti:2016jut}
	P.~Menotti, ``{Classical conformal blocks},'' {\em Mod. Phys. Lett. A} {\bf 31}
	(2016), no.~27, 1650159, \href{http://www.arXiv.org/abs/1601.04457}{{\tt
			1601.04457}}.
	
	\bibitem{Piatek:2021aiz}
	M.~R. Piatek, R.~G. Nazmitdinov, A.~Puente, and A.~R. Pietrykowski,
	``{Classical conformal blocks, Coulomb gas integrals and Richardson-Gaudin
		models},'' {\em JHEP} {\bf 04} (2022) 098,
	\href{http://www.arXiv.org/abs/2110.15009}{{\tt 2110.15009}}.
	
	\bibitem{Kravchuk:2021akc}
	P.~Kravchuk, D.~Mazac, and S.~Pal, ``{Automorphic Spectra and the Conformal
		Bootstrap},'' \href{http://www.arXiv.org/abs/2111.12716}{{\tt 2111.12716}}.
	
	\bibitem{Bonifacio:2021aqf}
	J.~Bonifacio, ``{Bootstrapping closed hyperbolic surfaces},'' {\em JHEP} {\bf
		03} (2022) 093, \href{http://www.arXiv.org/abs/2111.13215}{{\tt 2111.13215}}.
	
	\bibitem{Mahanta:2022fvl}
	R.~Mahanta and T.~Sengupta, ``{Modular linear differential equations for
		four-point sphere conformal blocks},''
	\href{http://www.arXiv.org/abs/2211.05158}{{\tt 2211.05158}}.
	
	\bibitem{Alkalaev:2015wia}
	K.~B. Alkalaev and V.~A. Belavin, ``{Classical conformal blocks via AdS/CFT
		correspondence},'' {\em JHEP} {\bf 08} (2015) 049,
	\href{http://www.arXiv.org/abs/1504.05943}{{\tt 1504.05943}}.
	
	\bibitem{Hijano:2015qja}
	E.~Hijano, P.~Kraus, E.~Perlmutter, and R.~Snively, ``{Semiclassical Virasoro
		blocks from AdS$_{3}$ gravity},'' {\em JHEP} {\bf 12} (2015) 077,
	\href{http://www.arXiv.org/abs/1508.04987}{{\tt 1508.04987}}.
	
	\bibitem{Alkalaev:2015lca}
	K.~B. Alkalaev and V.~A. Belavin, ``{Monodromic vs geodesic computation of
		Virasoro classical conformal blocks},'' {\em Nucl. Phys. B} {\bf 904} (2016)
	367--385, \href{http://www.arXiv.org/abs/1510.06685}{{\tt 1510.06685}}.
	
	\bibitem{Hijano:2015zsa}
	E.~Hijano, P.~Kraus, E.~Perlmutter, and R.~Snively, ``{Witten Diagrams
		Revisited: The AdS Geometry of Conformal Blocks},'' {\em JHEP} {\bf 01}
	(2016) 146, \href{http://www.arXiv.org/abs/1508.00501}{{\tt 1508.00501}}.
	
	\bibitem{Chen:2016dfb}
	B.~Chen, J.-q. Wu, and J.-j. Zhang, ``{Holographic Description of 2D Conformal
		Block in Semi-classical Limit},'' {\em JHEP} {\bf 10} (2016) 110,
	\href{http://www.arXiv.org/abs/1609.00801}{{\tt 1609.00801}}.
	
	\bibitem{Alkalaev:2016rjl}
	K.~B. Alkalaev, ``{Many-point classical conformal blocks and geodesic networks
		on the hyperbolic plane},'' {\em JHEP} {\bf 12} (2016) 070,
	\href{http://www.arXiv.org/abs/1610.06717}{{\tt 1610.06717}}.
	
	\bibitem{Belavin:2017atm}
	V.~A. Belavin and R.~V. Geiko, ``{Geodesic description of Heavy-Light Virasoro
		blocks},'' {\em JHEP} {\bf 08} (2017) 125,
	\href{http://www.arXiv.org/abs/1705.10950}{{\tt 1705.10950}}.
	
	\bibitem{Alkalaev:2018nik}
	K.~Alkalaev and M.~Pavlov, ``{Perturbative classical conformal blocks as
		Steiner trees on the hyperbolic disk},'' {\em JHEP} {\bf 02} (2019) 023,
	\href{http://www.arXiv.org/abs/1810.07741}{{\tt 1810.07741}}.
	
	\bibitem{Alkalaev:2019zhs}
	K.~B. Alkalaev and M.~Pavlov, ``{Four-point conformal blocks with three heavy
		background operators},'' {\em JHEP} {\bf 08} (2019) 038,
	\href{http://www.arXiv.org/abs/1905.03195}{{\tt 1905.03195}}.
	
	\bibitem{alday2010liouville}
	L.~F. Alday, D.~Gaiotto, and Y.~Tachikawa, ``Liouville correlation functions
	from four-dimensional gauge theories,'' {\em Letters in Mathematical Physics}
	{\bf 91} (2010), no.~2, 167--197.
	
	\bibitem{Alba:2010qc}
	V.~A. Alba, V.~A. Fateev, A.~V. Litvinov, and G.~M. Tarnopolskiy, ``{On
		combinatorial expansion of the conformal blocks arising from AGT
		conjecture},'' {\em Lett. Math. Phys.} {\bf 98} (2011) 33--64,
	\href{http://www.arXiv.org/abs/1012.1312}{{\tt 1012.1312}}.
	
	\bibitem{Tai:2010ps}
	T.-S. Tai, ``{Uniformization, Calogero-Moser/Heun duality and
		Sutherland/bubbling pants},'' {\em JHEP} {\bf 10} (2010) 107,
	\href{http://www.arXiv.org/abs/1008.4332}{{\tt 1008.4332}}.
	
	\bibitem{Piatek:2011tp}
	M.~Piatek, ``{Classical conformal blocks from TBA for the elliptic
		Calogero-Moser system},'' {\em JHEP} {\bf 06} (2011) 050,
	\href{http://www.arXiv.org/abs/1102.5403}{{\tt 1102.5403}}.
	
	\bibitem{Ferrari:2012gc}
	F.~Ferrari and M.~Piatek, ``{Liouville theory, N=2 gauge theories and accessory
		parameters},'' {\em JHEP} {\bf 05} (2012) 025,
	\href{http://www.arXiv.org/abs/1202.2149}{{\tt 1202.2149}}.
	
	\bibitem{Gopakumar:2005fx}
	R.~Gopakumar, ``{From free fields to AdS: III},'' {\em Phys. Rev. D} {\bf 72}
	(2005) 066008, \href{http://www.arXiv.org/abs/hep-th/0504229}{{\tt
			hep-th/0504229}}.
	
	\bibitem{Gaberdiel:2020ycd}
	M.~R. Gaberdiel, R.~Gopakumar, B.~Knighton, and P.~Maity, ``{From symmetric
		product CFTs to AdS$_{3}$},'' {\em JHEP} {\bf 05} (2021) 073,
	\href{http://www.arXiv.org/abs/2011.10038}{{\tt 2011.10038}}.
	
	\bibitem{Bhat:2021dez}
	F.~Bhat, R.~Gopakumar, P.~Maity, and B.~Radhakrishnan, ``{Twistor coverings and
		Feynman diagrams},'' {\em JHEP} {\bf 05} (2022) 150,
	\href{http://www.arXiv.org/abs/2112.05115}{{\tt 2112.05115}}.
	
	\bibitem{Knighton:2022ipy}
	B.~Knighton, ``{Classical geometry from the tensionless string},''
	\href{http://www.arXiv.org/abs/2207.01293}{{\tt 2207.01293}}.
	
	\bibitem{Gopakumar:2022djw}
	R.~Gopakumar and E.~A. Mazenc, ``{Deriving the Simplest Gauge-String Duality -
		I: Open-Closed-Open Triality},''
	\href{http://www.arXiv.org/abs/2212.05999}{{\tt 2212.05999}}.
	
	\bibitem{mirzakhani2007simple}
	M.~Mirzakhani, ``Simple geodesics and weil-petersson volumes of moduli spaces
	of bordered riemann surfaces,'' {\em Inventiones mathematicae} {\bf 167}
	(2007), no.~1, 179--222.
	
	\bibitem{mirzakhani2007weil}
	M.~Mirzakhani, ``Weil-petersson volumes and intersection theory on the moduli
	space of curves,'' {\em Journal of the American Mathematical Society} {\bf
		20} (2007), no.~1, 1--23.
	
	\bibitem{andersen2017geometric}
	J.~E. Andersen, G.~Borot, and N.~Orantin, ``Geometric recursion,'' {\em arXiv
		preprint arXiv:1711.04729} (2017).
	
	\bibitem{andersen2019topological}
	J.~E. Andersen, G.~Borot, S.~Charbonnier, V.~Delecroix, A.~Giacchetto,
	D.~Lewanski, and C.~Wheeler, ``Topological recursion for masur-veech
	volumes,'' {\em arXiv preprint arXiv:1905.10352} (2019).
	
	\bibitem{Hadasz:2009db}
	L.~Hadasz, Z.~Jaskolski, and P.~Suchanek, ``{Recursive representation of the
		torus 1-point conformal block},'' {\em JHEP} {\bf 01} (2010) 063,
	\href{http://www.arXiv.org/abs/0911.2353}{{\tt 0911.2353}}.
	
	\bibitem{Menotti:2012wq}
	P.~Menotti, ``{Accessory parameters for Liouville theory on the torus},'' {\em
		JHEP} {\bf 12} (2012) 001, \href{http://www.arXiv.org/abs/1207.6884}{{\tt
			1207.6884}}.
	
	\bibitem{Piatek:2013ifa}
	M.~Piatek, ``{Classical torus conformal block, $N = 2^*$ twisted superpotential
		and the accessory parameter of Lam\'e equation},'' {\em JHEP} {\bf 03} (2014)
	124, \href{http://www.arXiv.org/abs/1309.7672}{{\tt 1309.7672}}.
	
	\bibitem{Menotti:2015gxa}
	P.~Menotti, ``{The Polyakov relation for the sphere and higher genus
		surfaces},'' {\em J. Phys. A} {\bf 49} (2016), no.~19, 195203,
	\href{http://www.arXiv.org/abs/1507.04853}{{\tt 1507.04853}}.
	
	\bibitem{Menotti:2018jsy}
	P.~Menotti, ``{Torus classical conformal blocks},'' {\em Mod. Phys. Lett. A}
	{\bf 33} (2018), no.~28, 1850166,
	\href{http://www.arXiv.org/abs/1805.07788}{{\tt 1805.07788}}.
	
	\bibitem{Teschner:2000md}
	J.~Teschner, ``{Remarks on Liouville theory with boundary},'' {\em PoS} {\bf
		tmr2000} (2000) 041, \href{http://www.arXiv.org/abs/hep-th/0009138}{{\tt
			hep-th/0009138}}.
	
	\bibitem{Fateev:2000ik}
	V.~Fateev, A.~B. Zamolodchikov, and A.~B. Zamolodchikov, ``{Boundary Liouville
		field theory. 1. Boundary state and boundary two point function},''
	\href{http://www.arXiv.org/abs/hep-th/0001012}{{\tt hep-th/0001012}}.
	
	\bibitem{Hadasz:2006vs}
	L.~Hadasz and Z.~Jaskolski, ``{Semiclassical limit of the FZZT Liouville
		theory},'' {\em Nucl. Phys. B} {\bf 757} (2006) 233--258,
	\href{http://www.arXiv.org/abs/hep-th/0603164}{{\tt hep-th/0603164}}.
	
	\bibitem{Belopolsky:1996cy}
	A.~Belopolsky, ``{De Rham cohomology of the supermanifolds and superstring BRST
		cohomology},'' {\em Phys. Lett. B} {\bf 403} (1997) 47--50,
	\href{http://www.arXiv.org/abs/hep-th/9609220}{{\tt hep-th/9609220}}.
	
	\bibitem{Belopolsky:1997bg}
	A.~Belopolsky, ``{New geometrical approach to superstrings},''
	\href{http://www.arXiv.org/abs/hep-th/9703183}{{\tt hep-th/9703183}}.
	
	\bibitem{Belopolsky:1997jz}
	A.~Belopolsky, ``{Picture changing operators in supergeometry and superstring
		theory},'' \href{http://www.arXiv.org/abs/hep-th/9706033}{{\tt
			hep-th/9706033}}.
	
	\bibitem{Witten:2012bh}
	E.~Witten, ``{Superstring Perturbation Theory Revisited},''
	\href{http://www.arXiv.org/abs/1209.5461}{{\tt 1209.5461}}.
	
	\bibitem{Witten:2012ga}
	E.~Witten, ``{Notes On Super Riemann Surfaces And Their Moduli},'' {\em Pure
		Appl. Math. Quart.} {\bf 15} (2019), no.~1, 57--211,
	\href{http://www.arXiv.org/abs/1209.2459}{{\tt 1209.2459}}.
	
	\bibitem{Witten:2012bg}
	E.~Witten, ``{Notes On Supermanifolds and Integration},'' {\em Pure Appl. Math.
		Quart.} {\bf 15} (2019), no.~1, 3--56,
	\href{http://www.arXiv.org/abs/1209.2199}{{\tt 1209.2199}}.
	
	\bibitem{Witten:2013cia}
	E.~Witten, ``{More On Superstring Perturbation Theory: An Overview Of
		Superstring Perturbation Theory Via Super Riemann Surfaces},''
	\href{http://www.arXiv.org/abs/1304.2832}{{\tt 1304.2832}}.
	
	\bibitem{Ohmori:2017wtx}
	K.~Ohmori and Y.~Okawa, ``{Open superstring field theory based on the
		supermoduli space},'' {\em JHEP} {\bf 04} (2018) 035,
	\href{http://www.arXiv.org/abs/1703.08214}{{\tt 1703.08214}}.
	
	\bibitem{Takezaki:2019jkn}
	T.~Takezaki, ``{Open superstring field theory including the Ramond sector based
		on the supermoduli space},'' \href{http://www.arXiv.org/abs/1901.02176}{{\tt
			1901.02176}}.
	
	\bibitem{Erler:2014eba}
	T.~Erler, S.~Konopka, and I.~Sachs, ``{NS-NS Sector of Closed Superstring Field
		Theory},'' {\em JHEP} {\bf 08} (2014) 158,
	\href{http://www.arXiv.org/abs/1403.0940}{{\tt 1403.0940}}.
	
	\bibitem{Erler:2015lya}
	T.~Erler, S.~Konopka, and I.~Sachs, ``{Ramond Equations of Motion in
		Superstring Field Theory},'' {\em JHEP} {\bf 11} (2015) 199,
	\href{http://www.arXiv.org/abs/1506.05774}{{\tt 1506.05774}}.
	
	\bibitem{Okawa:2004ii}
	Y.~Okawa and B.~Zwiebach, ``{Heterotic string field theory},'' {\em JHEP} {\bf
		07} (2004) 042, \href{http://www.arXiv.org/abs/hep-th/0406212}{{\tt
			hep-th/0406212}}.
	
	\bibitem{Berkovits:2004xh}
	N.~Berkovits, Y.~Okawa, and B.~Zwiebach, ``{WZW-like action for heterotic
		string field theory},'' {\em JHEP} {\bf 11} (2004) 038,
	\href{http://www.arXiv.org/abs/hep-th/0409018}{{\tt hep-th/0409018}}.
	
	\bibitem{Kunitomo:2021wiz}
	H.~Kunitomo, ``{Type II superstring field theory revisited},'' {\em PTEP} {\bf
		2021} (2021), no.~9, 093B03, \href{http://www.arXiv.org/abs/2106.07917}{{\tt
			2106.07917}}.
	
	\bibitem{Ahmadain:2022tew}
	A.~Ahmadain and A.~C. Wall, ``{Off-Shell Strings I: S-matrix and Action},''
	\href{http://www.arXiv.org/abs/2211.08607}{{\tt 2211.08607}}.
	
	\bibitem{Ahmadain:2022eso}
	A.~Ahmadain and A.~C. Wall, ``{Off-Shell Strings II: Black Hole Entropy},''
	\href{http://www.arXiv.org/abs/2211.16448}{{\tt 2211.16448}}.
	
	\bibitem{zamolodchikov1984conformal}
	A.~B. Zamolodchikov, ``Conformal symmetry in two dimensions: an explicit
	recurrence formula for the conformal partial wave amplitude,'' {\em
		Communications in mathematical physics} {\bf 96} (1984), no.~3, 419--422.
	
\end{thebibliography}

\providecommand{\href}[2]{#2}\begingroup\raggedright\endgroup

\end{document}